\documentclass[12pt,preprint]{aastex}
 \slugcomment{ApJS accepted}
 \shorttitle{The Spectral Energy Distribution of Fermi Blazars}
 \shortauthors{Fan et al.}

\begin{document}
\title{The Spectral Energy Distributions of Fermi Blazars}

\author{J. H. Fan$^{1,2}~${\thanks{email:fjh@gzhu.edu.cn}}, J. H. Yang$^3$, Y. Liu$^{1,2}$, G. Y. Luo$^{4}$, C. Lin$^{1,2}$, Y.H. Yuan$^{1,2}$, H. B. Xiao$^{1,2}$, A. Y. Zhou$^{5}$, T. X. Hua$^{1,2}~$,  Z. Y. Pei$^{1,2}~$}
\affil{
1. Center for Astrophysics, Guangzhou University, Guangzhou 510006, China\\
2. Astronomy Science and Technology Research Laboratory of
   Department of Education of Guangdong Province, Guangzhou 510006,
   China\\
3.Department of Physics and Electronics Science, Hunan University
of Arts and Science, Changde 415000, China\\
4.Department of Electronics, School for Physics and Electronics Engineering, Guangzhou University, Guangzhou 510006, China\\
5. National Astronomical Observatories of China, Chinese Academy of China, Beijing, China}

\begin{abstract}
In this paper, multi-wavelength data are compiled for a sample
 of 1425 Fermi blazars to calculate their spectral energy distributions (SEDs).
A parabolic function, ${\rm log}(\nu F_{\nu}) = P_1({\rm log}\nu - P_2)^2 + P_3,$  is used for SED fitting. Synchrotron peak
frequency ($\rm log \nu_p$), spectral curvature ($\rm P_1$),  peak flux ($\rm \nu_p F_{\nu_p}$), and integrated flux ($\rm \nu F_{\nu}$)  are successfully obtained for 1392 blazars (461 flat spectrum radio quasars-FSRQs, 620 BL Lacs-BLs and 311 blazars of uncertain type-BCUs, 999 sources have known redshifts).  Monochromatic
luminosity at radio 1.4 GHz, optical R band, X-ray at 1 keV and
$\gamma$-ray at 1 GeV, peak luminosity, integrated luminosity and effective spectral indexes of radio to
optical ($\alpha_{\rm RO}$), and optical to X-ray ($\alpha_{\rm OX}$) are  calculated. The "Bayesian classification" is employed to log$\nu_{\rm p}$ in the rest frame for 999 blazars with available redshift and the results show that  3 components are enough to fit the log$\nu_{\rm p}$
 distribution, there is no ultra high peaked subclass. Based on the 3 components, the subclasses
 of blazars using the acronyms of Abdo et al. (2010a) are classified, and some mutual correlations are also studied.
 Conclusions are finally drawn as follows:
 (1) SEDs are successfully obtained for 1392 blazars.  The fitted peak frequencies are compared with common
 sources from samples available (
 Sambruna et al. 1996,
 Nieppola et al. 2006, 2008,
 Abdo et al. 2010a).
 (2)   Blazars are classified as low synchrotron peak sources (LSPs)  if log $\nu_{\rm p} (\rm Hz) \leq $ 14.0,
  intermediate synchrotron peak sources (ISPs)  if  $14.0 < \,\rm log\, \nu_p (\rm Hz) \leq 15.3 $, and
  high synchrotron peak sources (HSPs) if $ \rm log\, \nu_p (\rm Hz) > 15.3$.
 (3) $\gamma$-ray emissions are strongly  correlated with
 radio emissions. $\gamma$-ray luminosity is also correlated with synchrotron peak luminosity and integrated luminosity.
(4) There is an anti-correlation between peak frequency and peak luminosity
 within the whole blazar sample. However,
   there is a marginally positive correlation for HBLs, and  no correlations  for FSRQs or LBLs.
  (5) There are anti-correlations between the  monochromatic luminosities ($\gamma$-ray and radio bands) and the peak frequency
   within  the whole sample and BL Lacs.
 (6)  The optical to X-ray ($\alpha_{\rm OX}$) and radio
to optical ($\alpha_{\rm RO}$) spectral indexes are strongly anti-correlated with peak frequency (log
$\nu_{\rm p}$) within  the whole sample, but the correlations for subclasses of FSRQs, LBLs,
 and HBLs are different.

\end{abstract}


\section{Introduction}
The most powerful active galactic nuclei (AGNs) are the sources
referred to as blazars, which show rapid variability and high
luminosity, high and variable polarization, superluminal motions,
core-dominated non-thermal continuum and strong $\gamma$-ray
emissions, etc. ( Abdo et al. 2009a, 2010b,c;
 Acero et al. 2015;
 Ackermann et al. 2011a,b, 2015;
 Aller et al. 2011;
 Bai, et al. 1998;
 Bastieri et al. 2011;
 Chen et al., 2012;
 Fan \& Xie 1996;
 Fan et al. 2011;
 Ghisellini et al. 2010;
 Gu 2014;
 Gu \& Li 2013;
 Gupta 2011;
 Gupta et al. 2012;
 Hu, et al. 2006;
 Marscher et al. 2011;
 Nolan et al. 2012;
 Romero et al. 2002;
 Urry 2011;
 Wills et al. 1992;
 Wu, et al. 2007;
 Yang et al. 2010a,b, 2012a,b, 2014;
 You \& Cao, 2014).
  Quite recently, Massaro et al. (2015) published the largest blazar sample (The 5th edition of the Roma-BZCAT, see the BZCAT 5.0 (http://www.asdc.asi.it/bzcat/).  Blazars consist of two subclasses, namely BL Lacertae objects (BL Lacs) and flat
spectrum radio quasars (FSRQs),  both subclasses have  the common
continuum properties while their emission line features are quite  different,
namely FSRQs have strong emission lines while BL Lacs have no emission lines or very weak emission lines. The strong radio continuum is believed to be
produced via synchrotron process. This synchrotron radiation is
reflected in the blazar spectral energy distribution (SED) of the
${\rm log}(\nu F_\nu)$ vs ${\rm log}\nu$ as a bump in radio to
X-ray frequencies.  Another bump followed is often attributed to
the inverse Compton process.

BL Lacertae objects can be classified as radio selected BL Lacs (RBLs)
and X-ray selected BL Lacs (XBLs) from surveys. In 1995,  Giommi et al.
constructed  radio to X-ray energy distributions (${\rm log}(\nu F_\nu)$ vs ${\rm log}\nu$)
of a sample of 121 BL Lacs to investigate the difference
between XBLs and RBLs, and found that the synchrotron peak for
RBLs locates  at  the IR/optical band  while  for XBLs it locates
 at the UV/X-ray band. BL Lacs  were proposed to distinguish by the
ratio of $f_{x}/f_r$ as high-energy cut-off BL Lacs (HBLs) and
low-energy cut-off BL Lacs (LBLs) respectively when $f_x/f_r >
10^{-11}$ or  $f_x/f_r < 10^{-11}$ (where X-ray fluxes cover 0.3
$\sim$ 3.5 keV in units of erg/cm$^2$/s while the radio fluxes are
at 1.4 GHz in units of Jy) (Padovani \& Giommi 1995). In 1996,
with a parabola which was used by Landau
et al. (1986) to parameterize
the spectral flux distribution in ${\rm log}\, \nu F_{\nu} - {\rm log}
\, \nu$, Sambruna et al.  parameterized the power per decade energy
distribution by a logarithmic parabola form ${\rm log}\, \nu L_{\nu}
= A ({\rm log}\, \nu)^2 + B {\rm log}\, \nu + C$
 using a sample of blazars including  complete samples of RBLs and XBLs. They obtained integrated luminosity, peak frequency, and
the peak luminosity. The averaged peak frequency is
  $\left <{\rm log}\, \nu_{\rm p} \right > (\rm {Hz}) = 15.93 \pm 1.45$ for EMSS XBLs and
  $\left <{\rm log}\, \nu_{\rm p} \right >  (\rm {Hz}) = 13.91 \pm 0.09 $ for 1 Jy RBLs.
In 1996, Padovani \& Giommi classified  BL Lacs as HBLs (${\rm log}
\,\nu_{\rm p} (\rm {Hz}) > 15$) and LBLs (${\rm log} \,\nu_{\rm p}{\rm {Hz}})  < 15$).

Later, Fossati et al. (1998)  computed the average SED from radio to $\gamma$-rays using
  three complete samples of blazars (FSRQs, RBLs, and XBLs), and found that there is a clear
 continuity for more luminous blazars to have lower first peak
 frequency and for less luminous blazars to have higher first peak
 frequency.  However observations show high luminosity HBLs
 (Giommi et al. 2005) and low power LBLs
 (Padovani et al. 2003;
 Caccianiga \&  March$\tilde{a}$, 2004).

In 2002, we calculated  SEDs using a sample of $\gamma$-ray
blazars with available radio Doppler factors and estimated the
Doppler factors at $\gamma$-ray band
(Zhang, et al. 2002).
 Nieppola et al. (2006) calculated SEDs in the form of ${\rm log}\,\nu F_{\nu} - {\rm
log} \,\nu $ for 308 blazars, classified BL lacertae objects into
LBLs, IBLs, and HBLs,  and  set the boundaries as ${\rm log}
\,\nu_{\rm p} < 14.5$ for LBLs, $14.5 < {\rm log}\,\nu_{\rm p} < 16.5$
for IBLs, and ${\rm log} \,\nu_{\rm p}
> 16.5$ for HBLs. They also calculated peak frequencies for 135 AGNs by fitting $\rm {log (\nu F_{\nu})= A(log \nu)^2 + B (log \nu) + C}$ in 2008
(Nieppola, et al. 2008).
 Abdo et al. (2010a)
  calculated SEDs for 48 Fermi blazars, and proposed  an empirical parametrization for synchrotron peak frequency based on effective spectral indexes of  $\alpha_{ro}$ (radio-optical) and $\alpha_{ox}$ (optical-X-rays) using the available fitting peak frequencies and the effective spectral indexes. They also extended the definition to all types of non-thermal dominated AGNs using new acronyms: low synchrotron peaked blazars
(LSP, ${\rm log} \,\nu_{\rm p} < 14$ Hz), intermediate synchrotron
peaked blazars (ISP, $14~{\rm Hz} < {\rm log}\,\nu_{\rm p} < 15$
Hz), and high synchrotron peaked blazars (HSP, ${\rm log}\,\nu_{\rm
p} > 15$ Hz).  From the above work, it is noted that the criteria of the synchrotron peak
frequency boundaries used to assign a type of HSP, ISP and HSP, is not uniform.

For the 2nd bump in the plots of ${\rm log}\,\nu F_{\nu} - {\rm
log} \,\nu $ of blazars, the peak frequencies are in  the region of
GeV to TeV bands.  Blazars are strong $\gamma$-ray emitters.
EGRET/GRO has detected  about 60 high confidence $\gamma$-ray
bright blazars
 (Hartman et al. 1999).
 The second generation of $\gamma$-ray detector, Fermi/LAT,  has detected about 1800 blazars and unidentified blazars ( see
 Abdo et al 2010c,
 Ackermann, et al. 2011a,
 Nolan et al. 2012,
 Acero et al. 2015;
 Ackermann, et al. 2015).

From the 3FGL (Acero et al. 2015),  a sample of  Fermi detected blazars (classified as FSRQs, BLs and blazars of uncertain type-BCUs) is now available. In this paper, we  have  compiled  multiwavelength data from NED for a sample of 1425 Fermi blazars,  calculate their SEDs  and discuss the relationships between some relevant parameters. In Sect. 2, we  present our Fermi detected blazar sample, calculate
their SEDs, discuss their classification, and analyze the correlations. In Sect. 3,   correlation analysis results are given.
In Sect. 4, some discussions and in Sect. 5, some conclusions are both presented.

The spectral index $\alpha$ is defined as  $F_\nu \propto
\nu^{-\alpha}$, and all luminosities  $\nu L_\nu$ are  denoted simply by $L_{\nu}$.

\section{Sample and Results}

\subsection{Sample and SED Results}

In this paper, a sample of  1425 Fermi detected blazars (FSRQs, BLs, and blazars of uncertain type-BCUs)  are from the 3FGL (Acero et al. 2015). The multi-frequency data (radio to X-ray bands) collected from NED  are used to calculate their SEDs. In doing so, the infrared and optical data are corrected using $A_{\lambda}$ in NED for reddening/galactic absorption. Then the corrected infrared and optical magnitudes are transferred into flux densities, and afterwards, all the flux densities are K-corrected by $f_\nu = f_\nu^{\rm ob} (1+z)^ {(\alpha_\nu - 1)}$,
where
 $\alpha_\nu$ ($\alpha_\nu = \Gamma_\nu - 1$, $\Gamma$ is the photon spectral index for X-ray and $\gamma$-ray bands) is the spectral index at frequency $\nu$, and
 $z$ is the redshift.
 If the redshift and spectral index are not available, then we can  adopt averaged values of the sub-sample to replace them.
 For redshift, the following averaged values from our sample are obtained:
 $\left < z \right >  = 0.568 \pm 0.505 $ for BLs, and
 $\left < z \right >  = 0.524 \pm 0.628 $ for BCUs.
 For spectral indexes,
 we adopt $\alpha_{\rm R} = 0$ for radio band (Donato et al. 2001, Abdo et al. 2010a), while for
optical band, $\alpha$ = 0.5 for BLs and $\alpha$ = 1 for the rest of the sources
 as similar to what have been done by Donato et al. (2001):
 $\left < \alpha_X \right >  = 1.30 $ for BLs,
 $\left < \alpha_X \right >  = 0.78 $ for FSRQs, and
 $\left < \alpha_X \right >  = 1.05 $ for  BCUs.

\begin{figure}
    \centering
    \resizebox{\hsize}{!}{\includegraphics*{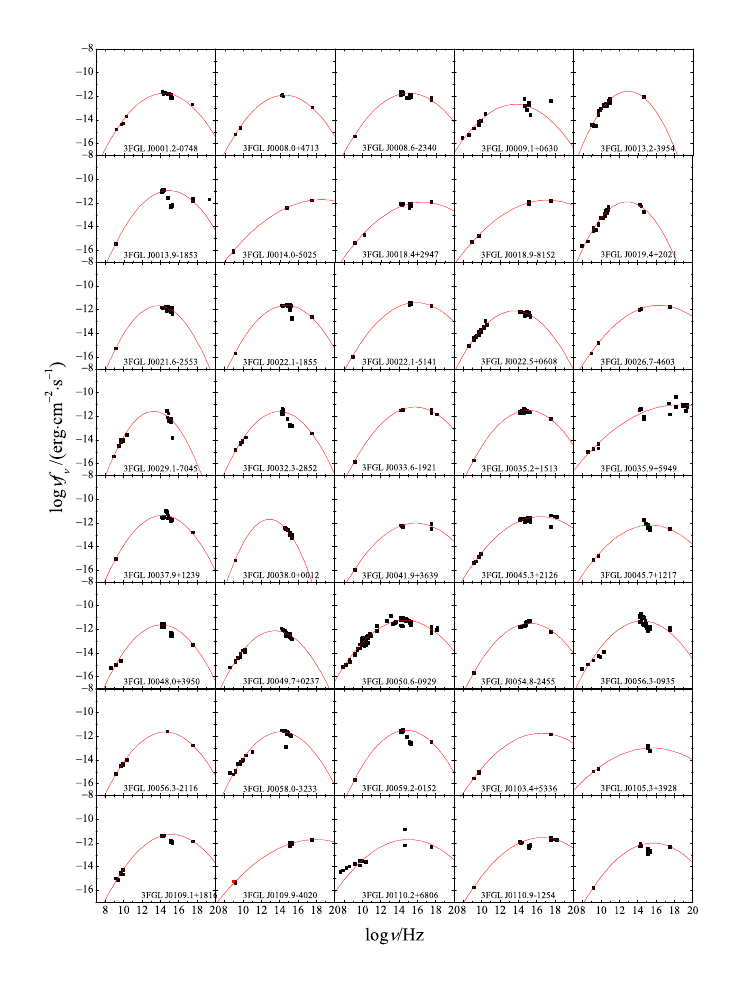}}
    \caption{SED Figures (see Appendix for all the SED figures).}
    \label{Fan-SED-Fig-B001-040}
\end{figure}

 The spectral energy distributions (SEDs)  are calculated by
fitting following relation with a least square fitting method,
$${\rm log}(\nu F_{\nu}) = P_1({\rm log}\nu - P_2)^2 + P_3,$$
where $P_1$, $P_2$ and $P_3$ are constants with
 $P_1$ being the spectral curvature,  $P_2$  the  peak frequency (${\rm log}\,\nu_{\rm p}$) and
 $P_3$  peak flux (${\rm log}\, (\nu_p F_{\nu_{\rm p}})$). The SED fitting figures of 1425 sources are shown in Fig \ref{Fan-SED-Fig-B001-040} and Appendix. From the fitting results,  it can be seen that SEDs  have been successfully fitted  only
   for 1392 sources with resulting  $P_1$, $P_2$, and $P_3$.  The fitting results (peak frequency, spectral curvature),  peak luminosity, integrated luminosity, monochromatic  luminosity, and  effective spectral indexes are listed in Table \ref{T1-samp}, where

Col. (1) gives the 3FGL name;

Col. (2) redshift from NED database at IPAC;

Col. (3) gives the SED  classification  by our method.
  HF stands for HSP FSRQ,
  IF for ISP FSRQ,
  LF for LSP FSRQ,
  HBL for HSP BL Lac,
  IBL for ISP BL Lac,
  LBL for LSP BL Lac,
  HU for HSP U-Blazar,
  IU for ISP U-Blazar, and
  LU for LSP U-Blazar;

Col. (4) radio luminosity ${\rm log}L_{\rm R}$ and its uncertainty at 1.4 GHz (${\rm
erg}\cdot {\rm s}^{-1}$);

Col. (5) optical R luminosity ${\rm log}L_{\rm O}$ and its uncertainty (${\rm
erg}\cdot {\rm s}^{-1}$) ;

Col. (6) X-ray luminosity ${\rm log}L_{\rm X}$ and its uncertainty at 1 keV (${\rm
erg}\cdot {\rm s}^{-1}$);

Col. (7)  $\gamma$-ray luminosity ${\rm log}L_\gamma$ and its uncertainty at
1 GeV (${\rm erg}\cdot {\rm s}^{-1}$);

Col. (8) and (9) give the effective spectral indices and the corresponding uncertainties of radio to
optical ($\alpha_{\rm RO}$) and optical to X-ray ($\alpha_{\rm
OX}$). They are calculated by a formula (Ledden \& O'Dell 1985),
$\alpha_{12} = -{\rm log}(f_1/f_2)/{\rm log}(\nu_1/\nu_2)$, where
$f_1$ and $f_2$ are the flux densities in frequencies $\nu_1$ and
$\nu_2$, respectively, and $f(\nu) \propto \nu^{-\alpha} $. In
this paper, $\nu_{\rm R} = 1.4$ GHz, $\nu_{\rm O} = 4.68 \times
10^{14}$ Hz and $\nu_{\rm X} = 2.416 \times 10^{17}$ Hz are adopted.

Col. (10) spectral curvature ($P_1$) and its uncertainty;

Col. (11) synchrotron peak frequency (${\rm log}\nu_{\rm p}$, Hz)  and its uncertainty;

Col. (12) peak luminosity (${\rm log}L_{\rm p}$, ${\rm erg}\cdot
{\rm s}^{-1}$)  and its uncertainty;

Col. (13) integrated (bolometric) luminosity (${\rm log}L_{\rm bol}$,
${\rm erg}\cdot {\rm s}^{-1}$)  and its uncertainty.

\begin{deluxetable}{ccccccccccccc}
\tabletypesize{\scriptsize}
 \rotate
  \tablecaption{Sample for blazars}
  \tablewidth{0pt}
 \tablehead{
  \colhead{3FGL name} &
   \colhead{$z$} &
  \colhead{C} &
  \colhead{$L_{\rm R}/\sigma_{L_{\rm R}}$} &
  \colhead{$L_{\rm O}/\sigma_{L_{\rm O}}$} &
  \colhead{$L_{\rm X}/\sigma_{L_{\rm X}}$} &
  \colhead{$L_\gamma/\sigma_{L_{\rm \gamma}}$} &
  \colhead{$\alpha _{\rm RO}/\sigma_{\alpha}$} &
  \colhead{$\alpha _{\rm OX}/\sigma_{\alpha}$} &
\colhead{$P_1/\sigma_{P_1}$} &
  \colhead{$\nu_{\rm p}/\sigma_{\nu_{\rm p}}$} &
  \colhead{$L_{\rm p}/\sigma_{L_{\rm p}}$} &
  \colhead{$L_{\rm bol}/\sigma_{L_{\rm bol}}$}
  }
 \startdata
(1)&(2)&(3)&(4)&(5)&(6)&(7)&(8)&(9)&(10)&(11)&(12)&(13)\\
\hline
J0001.2-0748    &       &   IB &   42.36/0.01    &   45.39/0.02    &               &   45.23/0.06    &   0.45/0.01    &               &      -0.12/0.01       &   14.37/0.12    &   45.35/0.03    &   45.71/0.05    \\
J0001.4+2120    &   1.106   &   HF  &   42.97/0.01    &               &               &   45.70/0.11    &               &               &      -0.05/0.00       &   16.79/0.28    &   45.70/0.03    &   46.32/0.04    \\
J0003.2-5246    &       &   HU  &               &               &   45.13/0.07    &   44.56/0.11    &               &               &      -0.05/0.01       &   17.89/0.81    &   45.15/0.14    &   45.76/0.14    \\
J0003.8-1151    &   1.310   &   LU  &   43.44/0.01    &   45.54/0.04    &               &   45.59/0.12    &   0.62/0.01    &               &      -0.12/0.01       &   13.06/0.14    &   45.57/0.11    &   46.01/0.15    \\
J0004.7-4740    &   0.880   &   IF  &               &   46.38/0.04    &   44.98/0.07    &   45.86/0.05    &               &   1.52/0.04    &      -0.12/0.01       &   14.14/0.09    &   46.20/0.06    &   46.59/0.09    \\
J0006.4+3825    &   0.229   &   IF  &   41.98/0.01    &   44.53/0.04    &   43.44/0.07    &   44.41/0.06    &   0.54/0.01    &   1.40/0.04    &      -0.11/0.01       &   14.03/0.12    &   44.65/0.10    &   45.08/0.14    \\
J0008.0+4713    &   0.280   &   IB &   41.18/0.01    &               &   43.51/0.07    &   44.87/0.03    &               &               &      -0.12/0.00       &   14.52/0.07    &   44.46/0.04    &   44.83/0.06    \\
J0008.6-2340    &   0.147   &   IB &   40.38/0.01    &               &   43.72/0.05    &   43.08/0.12    &               &               &      -0.10/0.01       &   15.09/0.19    &   44.01/0.05    &   44.40/0.07    \\
J0009.1+0630    &       &   LB &   42.43/0.02    &   44.97/0.04    &               &   45.14/0.07    &   0.54/0.01    &               &      -0.09/0.03       &   13.69/0.51    &   44.42/0.17    &   44.93/0.24    \\
J0009.6-3211    &   0.026   &   LU  &   39.87/0.01    &   44.48/0.04    &   41.53/0.13    &   41.91/0.10    &   0.17/0.01    &   2.09/0.06    &      -0.16/0.02       &   13.93/0.24    &   43.90/0.17    &   44.14/0.23    \\
J0013.2-3954    &       &   LB &   42.74/0.02    &   45.04/0.04    &               &   45.21/0.06    &   0.58/0.01    &               &      -0.19/0.01       &   12.95/0.14    &   45.53/0.09    &   45.79/0.13    \\
J0013.9-1853    &   0.095   &   IB &   39.90/0.02    &               &   43.72/0.03    &   42.88/0.11    &               &               &      -0.13/0.01       &   14.96/0.15    &   44.37/0.07    &   44.65/0.09    \\
J0014.0-5025    &       &   HB &               &               &   45.38/0.07    &   44.64/0.10    &               &               &      -0.05/0.00       &   18.55/0.33    &   45.38/0.06    &   45.94/0.07    \\
J0015.7+5552    &       &   HU  &   41.90/0.01    &               &               &   44.93/0.09    &               &               &      -0.10/0.00       &   15.82/0.10    &   45.95/0.03    &   46.32/0.04    \\
J0016.3-0013    &   1.577   &   IF  &   43.96/0.01    &   45.49/0.04    &   45.02/0.07    &   46.67/0.06    &   0.72/0.01    &   1.17/0.04    &      -0.09/0.01       &   13.58/0.10    &   45.58/0.04    &   46.12/0.06    \\
J0017.2-0643    &       &   IU  &   41.94/0.01    &   44.82/0.04    &               &   44.87/0.09    &   0.48/0.01    &               &      -0.10/0.01       &   14.64/0.37    &   44.79/0.06    &   45.21/0.09    \\
J0017.6-0512    &   0.227   &   IF  &   41.46/0.02    &   44.30/0.04    &   43.78/0.11    &   44.48/0.05    &   0.49/0.01    &   1.19/0.05    &      -0.11/0.01       &   14.48/0.13    &   44.63/0.15    &   45.02/0.21    \\
J0018.4+2947    &   0.100   &   HB &   40.00/0.01    &               &   43.54/0.07    &   42.84/0.13    &               &               &      -0.06/0.01       &   16.60/0.68    &   43.44/0.12    &   43.96/0.16    \\
J0018.9-8152    &       &   HB &               &               &   45.37/0.09    &   45.16/0.06    &               &               &      -0.05/0.01       &   17.16/0.46    &   45.33/0.07    &   45.90/0.07    \\
J0019.1-5645    &       &   LU  &               &               &               &   44.88/0.09    &               &               &      -0.13/0.01       &   13.35/0.10    &   44.04/0.06    &   44.41/0.10    \\
J0019.4+2021    &       &   LB &   43.04/0.01    &   44.42/0.04    &               &   44.91/0.10    &   0.75/0.01    &               &      -0.17/0.01       &   12.84/0.09    &   45.19/0.06    &   45.50/0.10    \\
J0021.6-2553    &       &   LB &   41.88/0.01    &   45.06/0.14    &               &   45.14/0.06    &   0.43/0.03    &               &      -0.17/0.02       &   13.77/0.17    &   45.43/0.08    &   45.67/0.12    \\
J0021.6-6835    &       &   IU  &               &               &   44.82/0.08    &   44.87/0.12    &               &               &      -0.09/0.01       &   14.90/0.13    &   45.47/0.04    &   45.92/0.05    \\
J0022.1-1855    &       &   IB &   41.39/0.02    &   45.60/0.02    &   44.56/0.11    &   45.13/0.05    &   0.24/0.01    &   1.38/0.05    &      -0.13/0.01       &   14.69/0.12    &   45.46/0.03    &   45.76/0.05    \\
J0022.1-5141    &       &   HB &               &               &   45.51/0.07    &   45.14/0.05    &               &               &      -0.09/0.00       &   15.86/0.16    &   45.69/0.03    &   46.07/0.05    \\
J0022.5+0608    &       &   LB &   42.57/0.01    &   44.64/0.04    &               &   45.68/0.03    &   0.63/0.01    &               &      -0.12/0.01       &   13.58/0.12    &   45.00/0.06    &   45.40/0.09    \\
...&...&...&...&...&...&...&...&...&...&...&...&...\\
...&...&...&...&...&...&...&...&...&...&...&...&...\\
\enddata
\label{T1-samp}
\end{deluxetable}

\subsection{Classification of Blazars}

\begin{figure}
    \centering
    \resizebox{\hsize}{!}{\includegraphics*{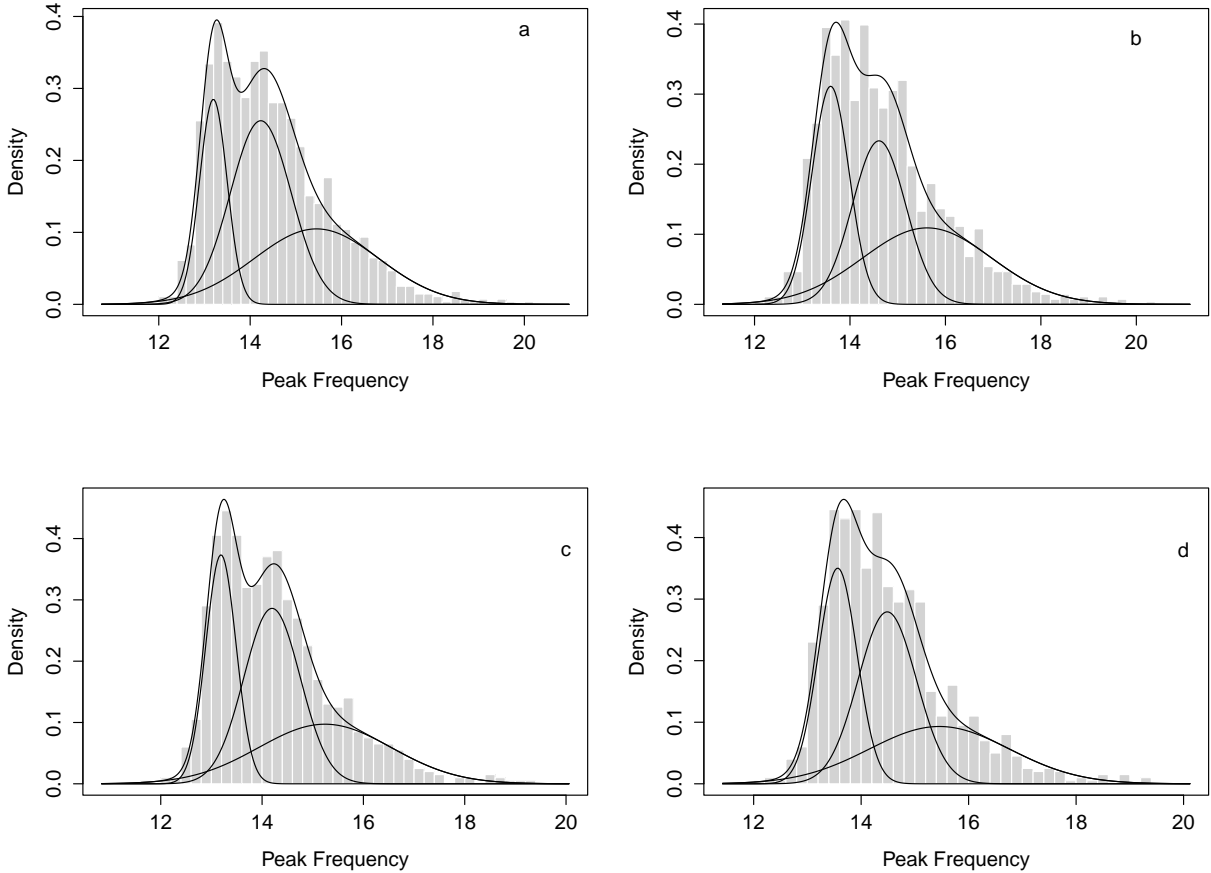}}
    \caption{Distribution and density of the peak frequencies with 3 Gaussian components.
    Fig $a$ is for the peak frequencies in the observer frame of the whole 1392 blazars;
    Fig $b$ is for the peak frequencies in the rest  frame of the whole 1392 blazars, the averaged values of redshift are used  for the sources with unknown redshift;
    Fig $c$ is for the peak frequencies in the observer frame of the 999 blazars with available redshift;
    Fig $d$ is for the peak frequencies in the rest frame of the 999 blazars with available redshift.
     }
    \label{fig-density-all}
\end{figure}

 In Table \ref{T1-samp}, the peak frequency, classification, luminosities, spectral indexes are listed for 1392 blazars, where 461
  are FSRQs, 620 are BLs and 311 are BCUs.  Amongst the 1392 blazars, 999 objects have available redshift. For the whole 1392 sources,  a distribution for logarithm of fitted peak frequencies is shown in Fig. \ref{fig-density-all}a. From  the distribution, it can be seen that  there are  4 peaks locating at about
 $\rm log\, \nu_p(Hz)\, =\, 13.3$,
 $\rm log\, \nu_p(Hz)\, =\, 14.3$,
 $\rm log\, \nu_p(Hz)\, =\, 15.9$, and
 $\rm log\, \nu_p(Hz)\, =\, 18.5$, and 3 concave points at about
  $\rm log\, \nu_p(Hz)\, =\, 13.9$,
 $\rm log\, \nu_p(Hz)\, =\, 15.5$, and
 $\rm log\, \nu_p(Hz)\, =\, 18.3$.

For the 999 blazars (463 BLs, 461 FSRQs, and 75 BCUs) with available redshifts, a distribution for the logarithm of fitted peak frequencies is shown in Fig. \ref{fig-density-all}c. From  the distribution, it can be seen that  there are  4 peaks locating at about
 $\rm log\, \nu_p(Hz)\, =\, 13.3$,
 $\rm log\, \nu_p(Hz)\, =\, 14.3$,
 $\rm log\, \nu_p(Hz)\, =\, 15.7$, and
 $\rm log\, \nu_p(Hz)\, =\, 18.5$, and 4 concave points at about
  $\rm log\, \nu_p(Hz)\, =\, 13.7$,
 $\rm log\, \nu_p(Hz)\, =\, 15.5$,
 $\rm log\, \nu_p(Hz)\, =\, 18.0$, and
  $\rm log\, \nu_p(Hz)\, =\, 18.3$.

When the peak frequencies are corrected to the rest frame, we have   $\rm{log \nu_p^{rest}}$ = $\rm{log \nu_p}$ + $\rm{log (1+z)}$. A distribution for the logarithm of corrected peak frequencies is shown in Fig. \ref{fig-density-all}d. From  the distribution, it can be seen that there are  many peaks locating at about
 $\rm log\, \nu_p(Hz)\, =\, 13.5-13.9$,
 $\rm log\, \nu_p(Hz)\, =\, 14.3$,
 $\rm log\, \nu_p(Hz)\, =\, 14.9$,
 $\rm log\, \nu_p(Hz)\, =\, 15.7$,
  $\rm log\, \nu_p(Hz)\, =\, 16.7$,
   $\rm log\, \nu_p(Hz)\, =\, 18.5$, and
 $\rm log\, \nu_p(Hz)\, =\, 18.9$, and many concave points at about
  $\rm log\, \nu_p(Hz)\, =\, 14.1$,
  $\rm log\, \nu_p(Hz)\, =\, 14.7$,
 $\rm log\, \nu_p(Hz)\, =\, 15.5$,
   $\rm log\, \nu_p(Hz)\, =\, 15.9$,
  $\rm log\, \nu_p(Hz)\, =\, 16.5$,
  $\rm log\, \nu_p(Hz)\, =\, 17.9$,
  $\rm log\, \nu_p(Hz)\, =\, 18.3$,
   $\rm log\, \nu_p(Hz)\, =\, 18.7$, and
 $\rm log\, \nu_p(Hz)\, =\, 18.9$.

\begin{figure}
    \centering
    \resizebox{\hsize}{!}{\includegraphics*{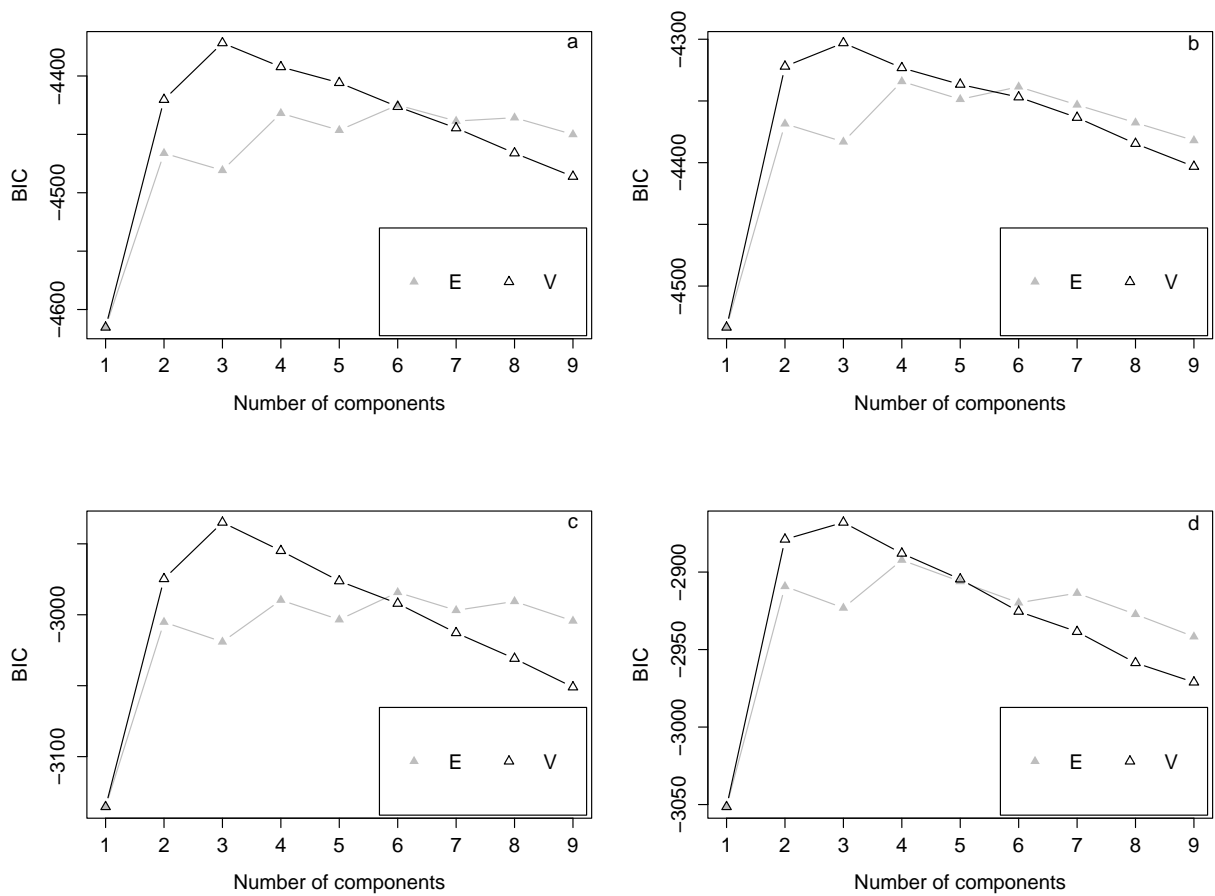}}
    \caption{BIC values from `mclust' for the models E and V with up to 9 clusters applied to the data of the peak frequencies.
    Fig $a$ is for the peak frequencies in the observer frame of the whole 1392 blazars;
    Fig $b$ is for the peak frequencies in the rest  frame of the whole 1392 blazars, the averaged values of redshift are used  for the sources with unknown redshift;
    Fig $c$ is for the peak frequencies in the observer frame of the 999 blazars with available redshift;
    Fig $d$ is for the peak frequencies in the rest frame of the 999 blazars with available redshift.
         }
    \label{fig-BIC-all}
\end{figure}

It is hard from the distributions to set boundary by eyes for different subclasses. To classify different subclass, a ¡¯normal mixture model¡¯ of Gaussian components is applied to the peak frequencies to confidently identify the peaks present.
 Then the existence of 3 vs. 2 or 4 components can be
quantitatively established using maximum likelihood estimation (via the expectation-maximization (EM)
Algorithm) and  Bayesian Information Criterion (BIC) for model
selection. The CRAN package `mclust' (Chris et al. 2012, Chris \& Adrian 2002) within the public domain R statistical software environment is used for the analysis. `mclust' provides iterative EM methods for maximum likelihood
clustering with parameterized Gaussian mixture models. First, density estimation via Gaussian finite mixture  modeling was conducted.  The BIC values for our 999 peak frequencies in the rest frame are shown in Fig \ref{fig-BIC-all}d. There are 2 kinds of models fitted by `mclust', one is the V (univariate, unequal variance) models, the other is the E (univariate, equal variance) models. An astro-oriented tutorial could be found in Sec. 9.92 of Modern Statistical Methods for Astronomy with R Applications (Feigelson \& Bau, 2012). Our analysis indicates that the better one is the V model with 3 components:
For the 1st component, it has a mean value of
 $\rm log\, \nu_p(Hz)$ = 13.56 with a variance of 0.129 and a clustering probability of 0.32;
 the 2nd one has a mean of
 $\rm log\, \nu_p(Hz)$ = 14.49 with a variance of 0.294 and a clustering probability of 0.38; and
 the 3rd one has
 $\rm log\, \nu_p(Hz)$ = 15.46 with a variance of 1.701 and a clustering probability of 0.30.
The density function with the 3 components is plotted also in Fig \ref{fig-density-all}d.
 As can be seen in Fig \ref{fig-density-all}d,  the crossing points of two adjacent Gaussian curves are
at ${\rm log \,\nu_{\rm p} (Hz)} = 13.98$ and ${\rm log \,\nu_{\rm p} (Hz)} = 15.30$.
 If we choose the frequencies at the jointing points as the boundaries for classification and
 follow  the acronyms of LSP, ISP, and HSP (Abdo et al. 2010a), the following classifications can be set:

 ${\rm log \,\nu_{\rm p} (Hz)} \leq 14.0$  for LSPs,

 $14.0 < {\rm log \,\nu_{\rm p} (Hz)}\leq 15.3$  for ISPs, and

${\rm log \,\nu_{\rm p} (Hz)} > 15.3$  for HSPs.

Based on the classifications, we obtain:
 38.6\% of the 999 blazar sample are LSPs,
 42.9\% are ISPs, and
 18.4\% are HSPs.

For the 999 peak frequencies at observer's frame, the same process is
 performed,  the following results are obtained:
 the 1st component has a mean value of
$\rm log\, \nu_p(Hz)$ = 13.19 with a variance of 0.090 and a clustering probability of 0.28; the 2nd one has a mean of
 $\rm log\, \nu_p(Hz)$ = 14.20 with a variance of 0.310 and a clustering probability of 0.40; the 3rd one has
 $\rm log\, \nu_p(Hz)$ = 15.24 with a variance of 1.724 and a clustering probability of 0.32.
 The  BIC values and the corresponding density function are shown in Fig \ref{fig-BIC-all}c and Fig \ref{fig-density-all}c respectively.

For the 1392 peak frequencies at observer's frame, we have:
the 1st component has a mean value of
 $\rm log\, \nu_p(Hz)$ = 13.19 with a variance of 0.091 and a clustering probability of 0.22; the 2nd one has a mean of
 $\rm log\, \nu_p(Hz)$  = 14.23 with a variance of 0.445 and a clustering probability of 0.43; and the 3rd one has
 $\rm log\, \nu_p(Hz)$ = 15.45 with a variance of 1.861 and a clustering probability of 0.36.
  The  BIC values and the corresponding density function are shown in Fig \ref{fig-BIC-all}a and Fig \ref{fig-density-all}a respectively.

For the 1392 objects,  we adopt the averaged redshift values for the unknown sources
($\left < z \right >  = 0.568 $ for BLs and  $\left < z \right >  = 0.524  $ for BCUs), then we obtain:
 the 1st component has a mean value of
 $\rm log\, \nu_p(Hz)$ = 13.59 with a variance of 0.151 and a clustering probability of 0.30; the 2nd one has a mean of
  $\rm log\, \nu_p(Hz)$ = 14.61 with a variance of 0.314 and a clustering probability of 0.33; and the 3rd one has
   $\rm log\, \nu_p(Hz)$ = 15.62 with a variance of 1.824 and a clustering probability of 0.37.
  The  BIC values and the corresponding density function are shown in Fig \ref{fig-BIC-all}b and Fig \ref{fig-density-all}b respectively.

 The classifications for 1392 blazars are shown in Col. (3) in Table
\ref{T1-samp}, where
 34.77\% of the whole sample are LSPs,
 40.09\% are ISPs, and
 25.14\% are HSPs. See Table \ref{ClassDis} for details.

\begin{deluxetable}{lccccc}
\tabletypesize{\scriptsize}
  \tablecaption{The statistics results of classification.}
  \tablewidth{0pt}
 \tablehead{
  \colhead{} &
  \colhead{$N$ for FSRQs } &
  \colhead{$N$ for BLs } &
  \colhead{$N$ for BCUs } &
  \colhead{$N$ for Sum  } &
  \colhead{percentage  }
    }
 \startdata
	&	FSRQs	&	BLs	&	BCUs	&	Sum	&	\%	\\
HSP	&	9	&	235	&	106	&	350	&	25.14	\\
ISP	&	180	&	271	&	107	&	558	&	40.09	\\
LSP	&	272	&	114	&	98	&	484	&	34.77	\\
Sum	&	461	&	620	&	311	&	1392	&	1	\\
percentage	&	33.12	&	45.54	&	22.34	&	1	&	---	\\
\enddata
\label{ClassDis}
\end{deluxetable}

\section{Correlations}
\subsection{ Correlations between $\gamma$-ray Luminosity and Other Luminosities}

In Table \ref{T1-samp},  monochromatic luminosities at
   1.4 GHz (${\rm log} L_{\rm R}$),
   optical R band (${\rm log} L_{\rm O}$),
   X-ray (${\rm log} L_{\rm X}$) at 1 KeV, and
   $\gamma$-ray (${\rm log} L_\gamma$) at 1 GeV are given.
 Here,  the  luminosity is calculated by a formulae $L = 4\pi d_L^2\nu f_\nu$,
 where, $d_L$ ($ = (1 + z) \cdot \frac{c}{{{H_0}}} \cdot
 \int_1^{{\rm{1 + }}z} {\frac{1}{{\sqrt {{\Omega _{{\rm{ M}}}}{x^3}
 + 1 - {\Omega _{{\rm{ M}}}}} }}} {\rm{ d}}x$)(Pedro \& Priyamvada
 2007) is luminosity distance and $f_\nu$ is the K-corrected flux
 density at the corresponding frequency $\nu$.

Following our pervious papers (Fan et al. 2013, 2014;
 Lin \& Fan 2016;
 Nie et al. 2014;
  Yang, et al. 2002b, 2014), the calculation of $\gamma$-ray  luminosity is further conducted.
  The luminosity  (${\rm log} L_{\rm
p}$) at synchrotron peak frequency and integrated luminosity (${\rm log} L_{\rm bol}$) are then
 obtained by $L = 4\pi d_L^2 (\nu f_\nu)$, where ($\nu f_\nu$)  is from  SED fittings.

\begin{figure}
    \centering
    \resizebox{\hsize}{!}{\includegraphics*{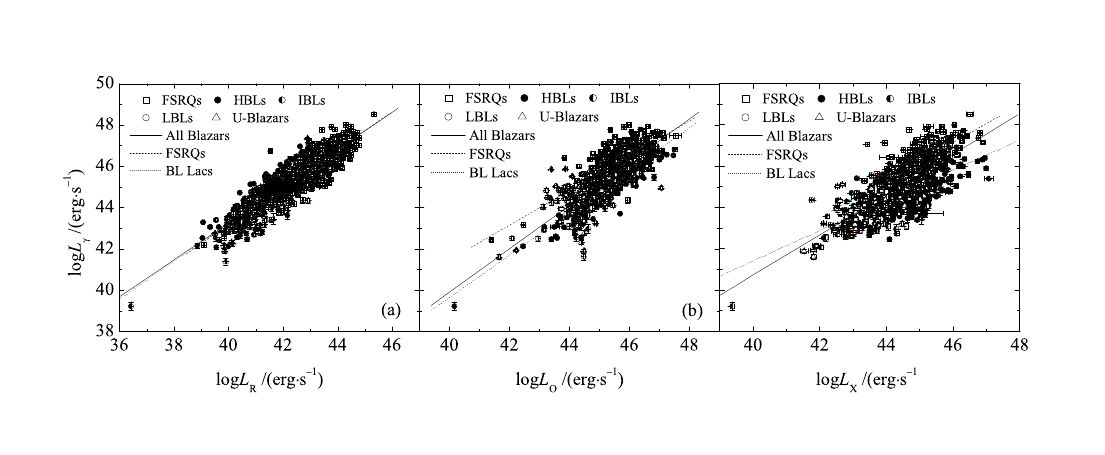}}
    \caption{The correlations between  $\gamma$-ray luminosity ($\rm log {L_{\gamma}}$ at 1 GeV) and
       (a) radio  luminosity ($\rm log {L_R}$ at 1.4 GHz),
       (b) optical  luminosity ($\rm log {L_O}$ at R band),
       (c) X-ray  luminosity ($\rm log {L_X}$ at 1 KeV).}
    \label{LG-LROX}
\end{figure}

\begin{figure}
    \centering
    \resizebox{\hsize}{!}{\includegraphics*{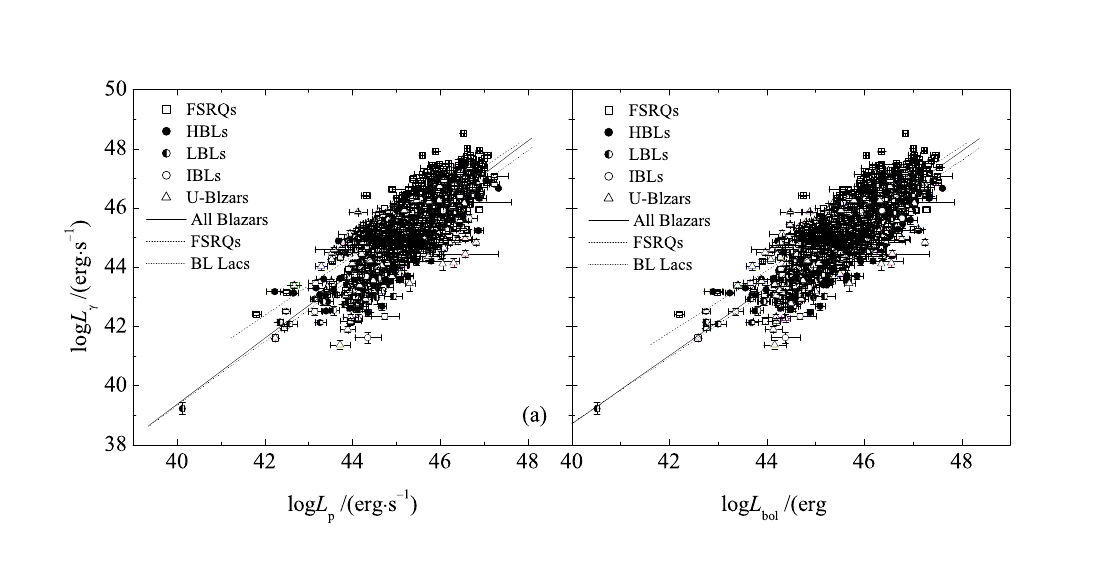}}
    \caption{The correlations between  $\gamma$-ray luminosity ($\rm log {L_{\gamma}}$ at 1 GeV) and (a) peak luminosity ($\rm log {L_p}$ ), (b) integrated luminosity ($\rm log {L_{bol}}$).}
    \label{LG-LpLbol}
\end{figure}

The correlations between $\gamma$-ray luminosity and lower energy
bands at radio, optical and X-ray are shown in Fig. \ref{LG-LROX}, and the
correlations between $\gamma$-ray luminosity and the peak and
integrated luminosities are shown in Fig. \ref{LG-LpLbol}. The linear regression
fitting results are shown in Table \ref{T2}, where the linear
regression fitting relation is expressed as
$y = (a \pm \Delta a) + (b \pm \Delta b)x$,
$r$ is a correlation coefficient,
$N$ is the number of sources in the corresponding sample (sub-sample),
$p$ is a chance probability,
$r_{LL,z}$ and $p_{LL,z}$ are correlation coefficient and the corresponding chance probability after removing redshift effect respectively.

\begin{deluxetable}{ccccccccc}
\tabletypesize{\scriptsize}
  \tablecaption{The correlations between $\gamma$-ray luminosity and other luminosities.}
  \tablewidth{0pt}
 \tablehead{
  \colhead{$y$ vs $x$} &
  \colhead{Sample} &
  \colhead{$a  \pm \Delta a$} &
  \colhead{$b  \pm \Delta b$ } &
  \colhead{$r$ } &
  \colhead{$N$ } &
  \colhead{$p$ } &
  \colhead{$r_{LL, z}$ } &
  \colhead{$p_{LL,z}$}
  }
 \startdata
${\rm log}L_\gamma $ vs ${\rm log}L_{\rm R}$    &   All Blazars &   7.53    $ \pm$  0.56    &   0.89    $ \pm$  0.01    &   0.90    &   1148    &   $<10^{-4}$  &   0.554   &   $<10^{-4}$  \\
    &   FSRQs   &   6.70    $ \pm$  1.42    &   0.91    $ \pm$  0.03    &   0.81    &   415 &   $<10^{-4}$  &   0.420   &   $<10^{-4}$  \\
    &   BL Lacs &   6.34    $ \pm$  0.90    &   0.92    $ \pm$  0.02    &   0.88    &   536 &   $<10^{-4}$  &   0.613   &   $<10^{-4}$  \\
    &   HBLs    &   3.16    $ \pm$  1.63    &   1.00    $ \pm$  0.04    &   0.88    &   198 &   $<10^{-4}$  &   0.535   &   $<10^{-4}$  \\
    &   IBLs    &   4.08    $ \pm$  1.29    &   0.98    $ \pm$  0.03    &   0.90    &   241 &   $<10^{-4}$  &   0.628   &   $<10^{-4}$  \\
    &   LBLs    &   6.49    $ \pm$  2.19    &   0.91    $ \pm$  0.05    &   0.88    &   97  &   $<10^{-4}$  &   0.536   &   $<10^{-4}$  \\
${\rm log}L_\gamma$ vs ${\rm log}L_{\rm O}$ &   All Blazars &   $-$1.98 $ \pm$  1.28    &   1.05    $ \pm$  0.03    &   0.78    &   867 &   $<10^{-4}$  &   0.257   &   $<10^{-4}$  \\
    &   FSRQs   &   7.84    $ \pm$  2.10    &   0.84    $ \pm$  0.05    &   0.69    &   368 &   $<10^{-4}$  &   0.325   &   $<10^{-4}$  \\
    &   BL Lacs &   $-$1.42 $ \pm$  1.37    &   1.03    $ \pm$  0.03    &   0.86    &   423 &   $<10^{-4}$  &   0.479   &   $<10^{-4}$  \\
    &   HBLs    &   $-$0.04 $ \pm$  1.85    &   0.99    $ \pm$  0.04    &   0.90    &   137 &   $<10^{-4}$  &   0.663   &   $<10^{-4}$  \\
    &   IBLs    &   $-$2.40 $ \pm$  1.89    &   1.05    $ \pm$  0.04    &   0.88    &   185 &   $<10^{-4}$  &   0.518   &   $<10^{-4}$  \\
    &   LBLs    &   $-$4.95 $ \pm$  3.70    &   1.11    $ \pm$  0.08    &   0.81    &   101  &   $<10^{-4}$  &   0.277   &   0.0150  \\
${\rm log}L_\gamma$ vs ${\rm log}L_{\rm X}$ &   All Blazars &   1.70    $ \pm$  1.54    &   0.98    $ \pm$  0.03    &   0.73    &   713 &   $<10^{-4}$  &   -0.134  &   0.0008  \\
    &   FSRQs   &   5.10    $ \pm$  2.52    &   0.91    $ \pm$  0.06    &   0.71    &   269 &   $<10^{-4}$  &   0.254   &   $<10^{-4}$  \\
    &   BL Lacs &   12.18   $ \pm$  1.76    &   0.73    $ \pm$  0.04    &   0.68    &   391 &   $<10^{-4}$  &   -0.082  &   0.1382  \\
    &   HBLs    &   6.87    $ \pm$  1.90    &   0.84    $ \pm$  0.04    &   0.83    &   180 &   $<10^{-4}$  &   0.265   &   0.0016  \\
    &   IBLs    &   $-$1.31 $ \pm$  2.49    &   1.05    $ \pm$  0.06    &   0.81    &   181 &   $<10^{-4}$  &   0.202   &   0.0105  \\
    &   LBLs    &   4.85    $ \pm$  5.54    &   0.93    $ \pm$  0.13    &   0.82    &   30  &   $<10^{-4}$  &   0.105   &   0.5302  \\
${\rm log}L_\gamma$ vs ${\rm log}L_{\rm p}$ &   All Blazars &   $-$5.09 $ \pm$  0.98    &   1.11    $ \pm$  0.02    &   0.81    &   1392    &   $<10^{-4}$  &   0.112   &   0.0004  \\
    &   FSRQs   &   0.30    $ \pm$  1.80    &   1.00    $ \pm$  0.04    &   0.77    &   461 &   $<10^{-4}$  &   0.372   &   $<10^{-4}$  \\
    &   BL Lacs &   $-$3.72 $ \pm$  1.20    &   1.08    $ \pm$  0.03    &   0.85    &   620 &   $<10^{-4}$  &   0.386   &   $<10^{-4}$  \\
    &   HBLs    &   1.78    $ \pm$  1.90    &   0.95    $ \pm$  0.04    &   0.83    &   235 &   $<10^{-4}$  &   0.380   &   $<10^{-4}$  \\
    &   IBLs    &   $-$7.62 $ \pm$  1.51    &   1.16    $ \pm$  0.03    &   0.90    &   271 &   $<10^{-4}$  &   0.516   &   $<10^{-4}$  \\
    &   LBLs    &   $-$3.04 $ \pm$  3.49    &   1.06    $ \pm$  0.08    &   0.80    &   114 &   $<10^{-4}$  &   0.170   &   0.1154  \\
${\rm log}L_\gamma$ vs ${\rm log}L_{\rm bol}$   &   All Blazars &   $-$7.35 $ \pm$  0.96    &   1.15    $ \pm$  0.02    &   0.83    &   1392    &   $<10^{-4}$  &   0.139   &   $<10^{-4}$  \\
    &   FSRQs   &   $-$2.26 $ \pm$  1.80    &   1.05    $ \pm$  0.04    &   0.78    &   461 &   $<10^{-4}$  &   0.376   &   $<10^{-4}$  \\
    &   BL Lacs &   $-$5.66 $ \pm$  1.18    &   1.11    $ \pm$  0.03    &   0.86    &   620 &   $<10^{-4}$  &   0.416   &   $<10^{-4}$  \\
    &   HBLs    &   $-$0.06 $ \pm$  1.91    &   0.98    $ \pm$  0.04    &   0.84    &   235 &   $<10^{-4}$  &   0.401   &   $<10^{-4}$  \\
    &   IBLs    &   $-$9.88 $ \pm$  1.40    &   1.20    $ \pm$  0.03    &   0.92    &   271 &   $<10^{-4}$  &   0.593   &   $<10^{-4}$  \\
    &   LBLs    &   $-$8.56 $ \pm$  3.03    &   1.18    $ \pm$  0.07    &   0.86    &   114 &   $<10^{-4}$  &   0.317   &   0.0030  \\

\enddata
\label{T2}
\end{deluxetable}

\subsection{ Correlations between Peak Frequency and Other
Parameters}

Now, we investigate  correlations between synchrotron peak frequency (${\rm log} \,\nu_{\rm p}$) and
 other parameters including monochromatic luminosity ($\gamma$-ray, X-ray, optical, and radio band),
 integrated luminosity,
 peak luminosity,
 spectral curvature ($P_1$),
 effective spectral indexes.
  The spectral curvature ($P_1$)  is from SED fitting, and the effective spectral
indexes are calculated as ${\alpha _{ij}} =  - \frac{{\log
({f_i}/{f_j})}}{{\log ({\nu _i}/{\nu _j})}}$ (Ledden \& O'Dell 1985).

\begin{figure}
    \centering
    \includegraphics[width=15cm,height=21cm]{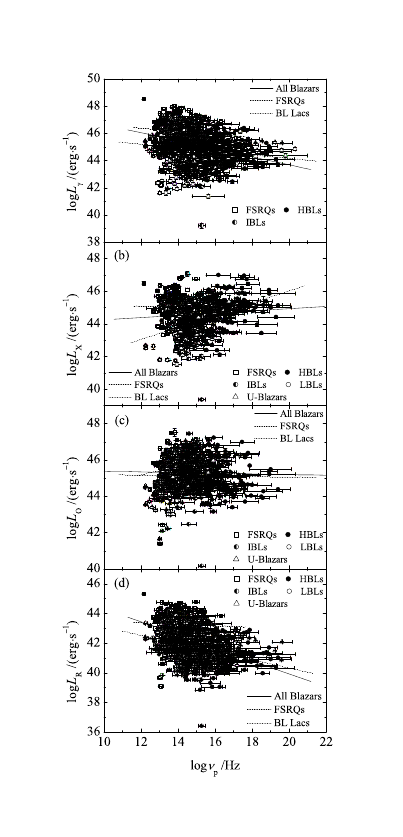}
    \caption{The correlation between monochromatic luminosity and peak frequency at rest frame.
     (a) $\gamma$-ray luminosity (1 GeV) and  peak frequency,
     (b) X-ray luminosity (1 KeV) and peak frequency,
     (c) optical luminosity and peak frequency, and
     (d) radio luminosity and peak frequency.}
    \label{vp-LROXG}
\end{figure}

\begin{figure}
    \centering
    \includegraphics[height=22cm]{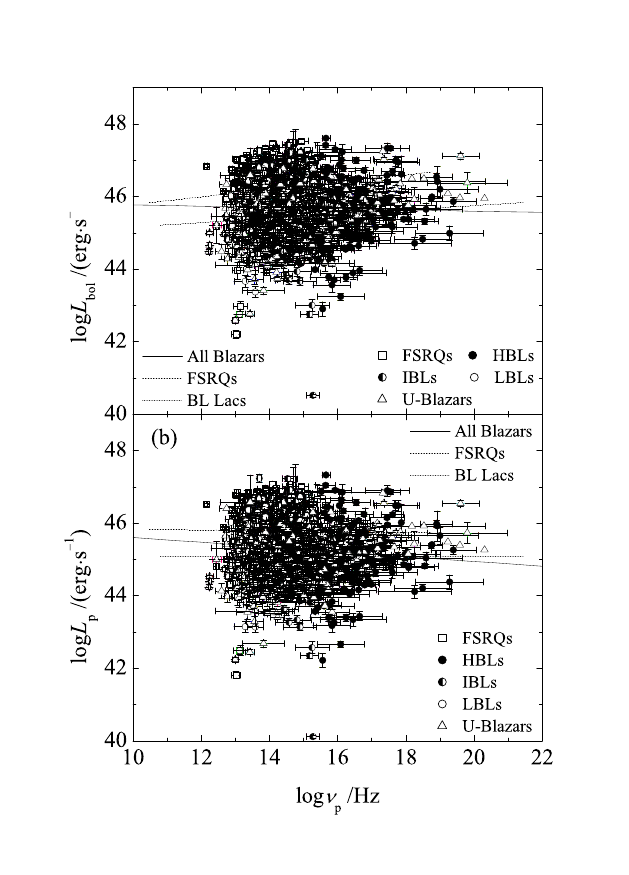}
    \caption{ (a) The correlation between peak frequency (${\rm log} \nu_{\rm p}$) and integrated luminosity  (${\rm log}
L_{\rm bol}$); (b) The correlation  between ${\rm log} \nu_{\rm p}$ and peak luminosity (${\rm log} L_{\rm p}$).}
    \label{vp-LpLb}
\end{figure}

\begin{figure}
    \centering
    \includegraphics[height=22cm]{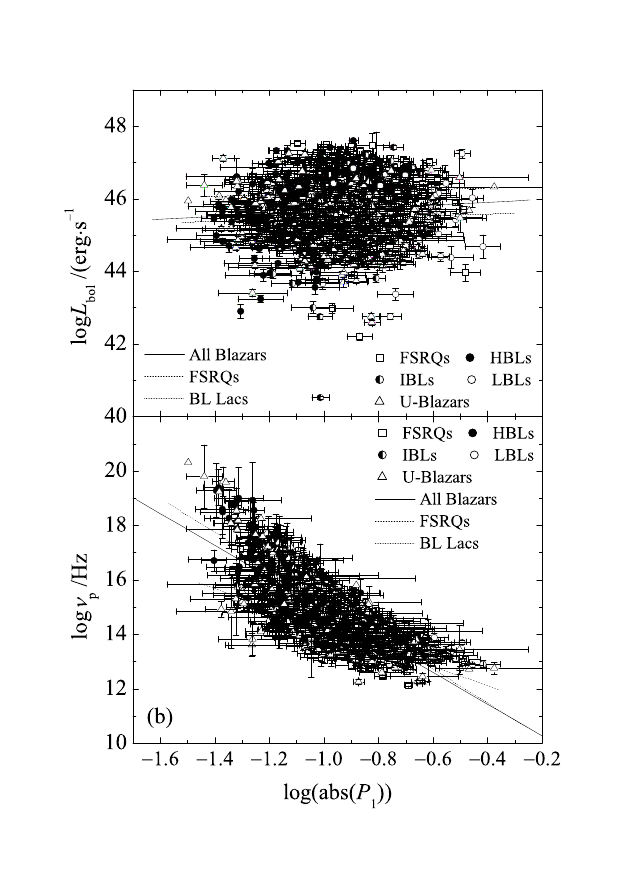}
    \caption{The correlation between spectral curvature (${\rm log} (|P_1|)$) and integrated luminosity  (${\rm log}
L_{\rm bol}$) (a), and that between ${\rm log} (|P_1|)$ and peak frequency (${\rm log} \nu_{\rm p}$) (b). }
    \label{P1-vpLb}
\end{figure}

\begin{figure}
    \centering
    \includegraphics[height=22cm]{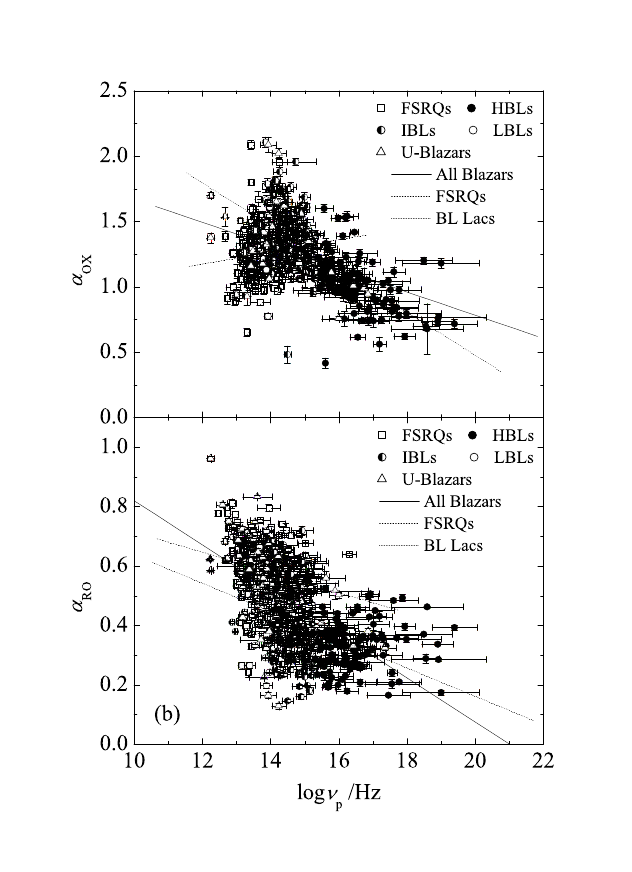}
    \caption{The correlation between peak frequency (${\rm log} \nu_{\rm p}$) and spectral index.
    (a) for ${\rm log} \nu_{\rm p}$ vs $\alpha_{\rm RO}$,
    (b) for ${\rm log} \nu_{\rm p}$ vs $\alpha_{\rm OX}$}.
    \label{vp-aRO-OX}
\end{figure}

 The relations between peak frequency (log $\nu_{\rm p}$) and monochromatic luminosities at
   1.4 GHz (${\rm log} L_{\rm R}$),
   optical R band (${\rm log} L_{\rm O}$),
   X-ray (${\rm log} L_{\rm X}$) at 1 KeV, and
   $\gamma$-ray (${\rm log} L_\gamma$) are shown in Fig. \ref{vp-LROXG}.
   The relations between the peak frequency (log $\nu_{\rm p}$) and the peak luminosity (log $L_{\nu_{\rm p}}$)/the integrated luminosity (log $L_{\rm bol}$) are shown in Fig. \ref{vp-LpLb}.
      The relations between the spectral curvature
   ($P_1$) and peak frequency (log $\nu_{\rm p}$)
   and those between the spectral curvature
   ($P_1$) and the integrated luminosity(log $L_{\rm bol}$) are shown in Fig. \ref{P1-vpLb}.
 The relations between  peak frequency and effective spectral indexes ($\alpha_{\rm RO}$, $\alpha_{\rm OX}$)
are shown in Fig. \ref{vp-aRO-OX}. The corresponding linear regression analysis
results are listed in Table \ref{T3}.

\begin{deluxetable}{ccccccc}
\tabletypesize{\scriptsize}
  \tablecaption{The linear regression analysis results for correlations between two parameters.}
  \tablewidth{0pt}
 \tablehead{
  \colhead{$y$ vs $x$} &
  \colhead{Sample} &
  \colhead{$a  \pm \Delta a$} &
  \colhead{$b  \pm \Delta b$ } &
  \colhead{$r$ } &
  \colhead{$n$ } &
  \colhead{$p$ }
    }
 \startdata
${\rm log} f_\gamma$ vs ${\rm log}f_{\rm R}$    &   All Blazars &   -11.75  $ \pm$  0.02    &   0.46    $ \pm$  0.02    &   0.61    &   1148    &   $<10^{-4}$  \\
    &   FSRQs   &   -11.73  $ \pm$  0.03    &   0.34    $ \pm$  0.04    &   0.39    &   415 &   $<10^{-4}$  \\
    &   BL Lacs &   -11.74  $ \pm$  0.03    &   0.48    $ \pm$  0.02    &   0.64    &   536 &   $<10^{-4}$  \\
    &   HBLs    &   -11.63  $ \pm$  0.07    &   0.54    $ \pm$  0.04    &   0.66    &   198 &   $<10^{-4}$  \\
    &   IBLs    &   -11.75  $ \pm$  0.05    &   0.47    $ \pm$  0.04    &   0.62    &   241 &   $<10^{-4}$  \\
    &   LBLs    &   -11.79  $ \pm$  0.07    &   0.49    $ \pm$  0.07    &   0.57    &   97  &   $<10^{-4}$  \\
${\rm log} f_\gamma$ vs ${\rm log}f_{\rm O}$    &   All Blazars &   -11.52  $ \pm$  0.09    &   0.17    $ \pm$  0.03    &   0.22    &   867 &   $<10^{-4}$  \\
    &   FSRQs   &   -11.15  $ \pm$  0.14    &   0.20    $ \pm$  0.04    &   0.27    &   368 &   $<10^{-4}$  \\
    &   BL Lacs &   -10.89  $ \pm$  0.12    &   0.42    $ \pm$  0.04    &   0.49    &   423 &   $<10^{-4}$  \\
    &   HBLs    &   -10.49  $ \pm$  0.17    &   0.59    $ \pm$  0.05    &   0.72    &   137 &   $<10^{-4}$  \\
    &   IBLs    &   -10.79  $ \pm$  0.18    &   0.44    $ \pm$  0.05    &   0.50    &   185 &   $<10^{-4}$  \\
    &   LBLs    &   -11.54  $ \pm$  0.33    &   0.19    $ \pm$  0.09    &   0.21    &   101  &   0.0357 \\
${\rm log} f_\gamma$ vs ${\rm log}f_{\rm X}$    &   All Blazars &   -12.70  $ \pm$  0.22    &   -0.08   $ \pm$  0.03    &   -0.09   &   713 &   0.0146  \\
    &   FSRQs   &   -10.49  $ \pm$  0.39    &   0.19    $ \pm$  0.05    &   0.21    &   269 &   0.0007  \\
    &   BL Lacs &   -12.11  $ \pm$  0.27    &   0.04    $ \pm$  0.04    &   0.04    &   391 &   0.3809  \\
    &   HBLs    &   -9.79   $ \pm$  0.40    &   0.44    $ \pm$  0.06    &   0.45    &   180 &   $<10^{-4}$  \\
    &   IBLs    &   -11.23  $ \pm$  0.46    &   0.15    $ \pm$  0.07    &   0.16    &   181 &   0.0282  \\
    &   LBLs    &   -10.62  $ \pm$  1.36    &   0.19    $ \pm$  0.19    &   0.18    &   30  &   0.3481  \\
${\rm log}L_{\rm R}$ vs ${\rm log} \nu_{\rm p}$ &   All Blazars &   48.70   $ \pm$  0.38    &   -0.44   $ \pm$  0.03    &   -0.45   &   1148    &   $<10^{-4}$  \\
    &   FSRQs   &   43.97   $ \pm$  0.81    &   -0.05   $ \pm$  0.06    &   -0.04   &   415 &   0.3950  \\
    &   BL Lacs &   45.83   $ \pm$  0.51    &   -0.27   $ \pm$  0.03    &   -0.33   &   536 &   $<10^{-4}$  \\
    &   HBLs    &   42.03   $ \pm$  1.07    &   -0.04   $ \pm$  0.07    &   -0.05   &   198 &   0.4981  \\
    &   IBLs    &   52.09   $ \pm$  2.64    &   -0.70   $ \pm$  0.18    &   -0.24   &   241 &   0.0001  \\
    &   LBLs    &   42.17   $ \pm$  4.22    &   0.00    $ \pm$  0.31    &   0.00    &   97  &   0.9913 \\
${\rm log}L_{\rm O}$ vs ${\rm log} \nu_{\rm p}$ &   All Blazars &   45.56   $ \pm$  0.38    &   -0.02   $ \pm$  0.03    &   -0.02   &   867 &   0.4876  \\
    &   FSRQs   &   42.50   $ \pm$  0.84    &   0.22    $ \pm$  0.06    &   0.19    &   368 &   0.0002  \\
    &   BL Lacs &   45.35   $ \pm$  0.53    &   -0.01   $ \pm$  0.04    &   -0.02   &   423 &   0.6930  \\
    &   HBLs    &   47.09   $ \pm$  1.44    &   -0.12   $ \pm$  0.09    &   -0.12   &   137 &   0.1632  \\
    &   IBLs    &   46.60   $ \pm$  2.74    &   -0.09   $ \pm$  0.19    &   -0.04   &   185 &   0.6312  \\
    &   LBLs    &   39.36   $ \pm$  3.36    &   0.41    $ \pm$  0.25    &   0.17    &   101  &   0.0954  \\
${\rm log}L_{\rm X}$ vs ${\rm log} \nu_{\rm p}$ &   All Blazars &   43.58   $ \pm$  0.44    &   0.07    $ \pm$  0.03    &   0.09    &   713 &   0.0217  \\
    &   FSRQs   &   45.36   $ \pm$  0.91    &   -0.02   $ \pm$  0.07    &   -0.02   &   269 &   0.7256  \\
    &   BL Lacs &   38.63   $ \pm$  0.61    &   0.37    $ \pm$  0.04    &   0.43    &   391 &   $<10^{-4}$  \\
    &   HBLs    &   40.39   $ \pm$  1.14    &   0.27    $ \pm$  0.07    &   0.28    &   180 &   0.0002  \\
    &   IBLs    &   38.68   $ \pm$  2.74    &   0.36    $ \pm$  0.19    &   0.14    &   181 &   0.0571  \\
    &   LBLs    &   87.14   $ \pm$  10.56   &   -3.13   $ \pm$  0.77    &   -0.63   &   30  &   0.0003  \\
${\rm log}L_\gamma$ vs ${\rm log} \nu_{\rm p}$  &   All Blazars &   49.58   $ \pm$  0.35    &   -0.29   $ \pm$  0.02    &   -0.32   &   1392    &   $<10^{-4}$  \\
    &   FSRQs   &   47.79   $ \pm$  0.88    &   -0.12   $ \pm$  0.06    &   -0.09   &   461 &   0.0661  \\
    &   BL Lacs &   46.80   $ \pm$  0.52    &   -0.13   $ \pm$  0.03    &   -0.15   &   620 &   0.0001  \\
    &   HBLs    &   45.74   $ \pm$  1.12    &   -0.07   $ \pm$  0.07    &   -0.06   &   235 &   0.3224  \\
    &   IBLs    &   50.88   $ \pm$  2.76    &   -0.41   $ \pm$  0.19    &   -0.13   &   271 &   0.0305  \\
    &   LBLs    &   40.12   $ \pm$  4.38    &   0.36    $ \pm$  0.32    &   0.11    &   114 &   0.2631  \\
${\rm log}L_{\rm p}$ vs ${\rm log} \nu_{\rm p}$ &   All Blazars &   46.26   $ \pm$  0.26    &   -0.07   $ \pm$  0.02    &   -0.10   &   1392    &   0.0003  \\
    &   FSRQs   &   45.95   $ \pm$  0.67    &   -0.01   $ \pm$  0.05    &   -0.01   &   461 &   0.8163  \\
    &   BL Lacs &   45.06   $ \pm$  0.41    &   0.00    $ \pm$  0.03    &   0.00    &   620 &   0.9558  \\
    &   HBLs    &   43.11   $ \pm$  0.97    &   0.12    $ \pm$  0.06    &   0.13    &   235 &   0.0436  \\
    &   IBLs    &   48.69   $ \pm$  2.15    &   -0.25   $ \pm$  0.15    &   -0.10   &   271 &   0.0921  \\
    &   LBLs    &   44.51   $ \pm$  3.29    &   0.05    $ \pm$  0.24    &   0.02    &   114 &   0.8471  \\
${\rm log}L_{\rm bol}$ vs ${\rm log} \nu_{\rm p}$   &   All Blazars &   45.94   $ \pm$  0.26    &   -0.02   $ \pm$  0.02    &   -0.03   &   1392    &   0.3290  \\
    &   FSRQs   &   44.77   $ \pm$  0.65    &   0.10    $ \pm$  0.05    &   0.10    &   461 &   0.0308  \\
    &   BL Lacs &   44.58   $ \pm$  0.41    &   0.06    $ \pm$  0.03    &   0.09    &   620 &   0.0292  \\
    &   HBLs    &   42.77   $ \pm$  0.94    &   0.17    $ \pm$  0.06    &   0.19    &   235 &   0.0036  \\
    &   IBLs    &   48.62   $ \pm$  2.12    &   -0.22   $ \pm$  0.14    &   -0.09   &   271 &   0.1330  \\
    &   LBLs    &   42.64   $ \pm$  3.20    &   0.20    $ \pm$  0.24    &   0.08    &   114 &   0.3851  \\
${\rm log} \nu_{\rm p}$ vs ${\rm log}({\rm abs}(P_1))$  &   All Blazars &   9.09    $ \pm$  0.12    &   -5.84   $ \pm$  0.12    &   -0.78   &   1392    &   $<10^{-4}$  \\
    &   FSRQs   &   10.68   $ \pm$  0.12    &   -3.57   $ \pm$  0.13    &   -0.79   &   461 &   $<10^{-4}$  \\
    &   BL Lacs &   8.97    $ \pm$  0.17    &   -6.27   $ \pm$  0.17    &   -0.83   &   620 &   $<10^{-4}$  \\
    &   HBLs    &   9.18    $ \pm$  0.38    &   -6.43   $ \pm$  0.34    &   -0.78   &   235 &   $<10^{-4}$  \\
    &   IBLs    &   13.35   $ \pm$  0.21    &   -1.39   $ \pm$  0.22    &   -0.35   &   271 &   $<10^{-4}$  \\
    &   LBLs    &   12.71   $ \pm$  0.15    &   -1.10   $ \pm$  0.18    &   -0.50   &   114 &   $<10^{-4}$  \\
${\rm log} L_{\rm bol}$ vs ${\rm log}({\rm abs}(P_1))$  &   All Blazars &   46.06   $ \pm$  0.13    &   0.39    $ \pm$  0.13    &   0.08    &   1392    &   0.0031  \\
    &   FSRQs   &   46.40   $ \pm$  0.20    &   0.24    $ \pm$  0.21    &   0.05    &   461 &   0.2580  \\
    &   BL Lacs &   45.66   $ \pm$  0.20    &   0.20    $ \pm$  0.20    &   0.04    &   620 &   0.3178  \\
    &   HBLs    &   46.30   $ \pm$  0.54    &   0.69    $ \pm$  0.48    &   0.09    &   235 &   0.1508  \\
    &   IBLs    &   46.94   $ \pm$  0.53    &   1.60    $ \pm$  0.56    &   0.17    &   271 &   0.0054  \\
    &   LBLs    &   45.80   $ \pm$  0.42    &   0.48    $ \pm$  0.52    &   0.09    &   114 &   0.3602  \\
$\alpha_{\rm RO}$ vs ${\rm log} \nu_{\rm p}$    &   All Blazars &   1.57    $ \pm$  0.05    &   -0.07   $ \pm$  0.00    &   -0.59   &   770 &   $<10^{-4}$  \\
    &   FSRQs   &   1.06    $ \pm$  0.10    &   -0.03   $ \pm$  0.01    &   -0.25   &   336 &   $<10^{-4}$  \\
    &   BL Lacs &   1.11    $ \pm$  0.07    &   -0.05   $ \pm$  0.00    &   -0.47   &   374 &   $<10^{-4}$  \\
    &   HBLs    &   0.24    $ \pm$  0.13    &   0.01    $ \pm$  0.01    &   0.07    &   116 &   0.4815  \\
    &   IBLs    &   1.94    $ \pm$  0.35    &   -0.11   $ \pm$  0.02    &   -0.32   &   171 &   $<10^{-4}$  \\
    &   LBLs    &   1.40    $ \pm$  0.56    &   -0.07   $ \pm$  0.04    &   -0.17   &   87  &   0.1100  \\
$\alpha_{\rm OX}$ vs ${\rm log} \nu_{\rm p}$    &   All Blazars &   2.56    $ \pm$  0.12    &   -0.09   $ \pm$  0.01    &   -0.43   &   525 &   $<10^{-4}$  \\
    &   FSRQs   &   0.61    $ \pm$  0.30    &   0.05    $ \pm$  0.02    &   0.14    &   230 &   0.0292  \\
    &   BL Lacs &   3.78    $ \pm$  0.16    &   -0.16   $ \pm$  0.01    &   -0.70   &   273 &   $<10^{-4}$  \\
    &   HBLs    &   2.94    $ \pm$  0.31    &   -0.12   $ \pm$  0.02    &   -0.51   &   115 &   $<10^{-4}$  \\
    &   IBLs    &   4.50    $ \pm$  0.65    &   -0.21   $ \pm$  0.04    &   -0.39   &   128 &   $<10^{-4}$  \\
    &   LBLs    &   -2.38   $ \pm$  2.80    &   0.27    $ \pm$  0.20    &   0.26    &   30  &   0.1624  \\
$\alpha_{\rm RO}$ vs $\alpha_{\rm OX}$  &   All Blazars &   0.57    $ \pm$  0.03    &   -0.07   $ \pm$  0.03    &   -0.13   &   478 &   0.0055  \\
    &   FSRQs   &   0.93    $ \pm$  0.03    &   -0.27   $ \pm$  0.02    &   -0.61   &   215 &   $<10^{-4}$  \\
    &   BL Lacs &   0.37    $ \pm$  0.03    &   0.01    $ \pm$  0.03    &   0.02    &   248 &   0.7566  \\
    &   HBLs    &   0.52    $ \pm$  0.03    &   -0.19   $ \pm$  0.03    &   -0.54   &   98  &   $<10^{-4}$  \\
    &   IBLs    &   0.50    $ \pm$  0.07    &   -0.08   $ \pm$  0.05    &   -0.15   &   120 &   0.1077  \\
    &   LBLs    &   0.75    $ \pm$  0.12    &   -0.17   $ \pm$  0.08    &   -0.38   &   30  &   0.0410  \\

\enddata
\label{T3}
\end{deluxetable}

\subsection{Effective spectral index correlation}

\begin{figure}
    \centering
    \resizebox{\hsize}{!}{\includegraphics*{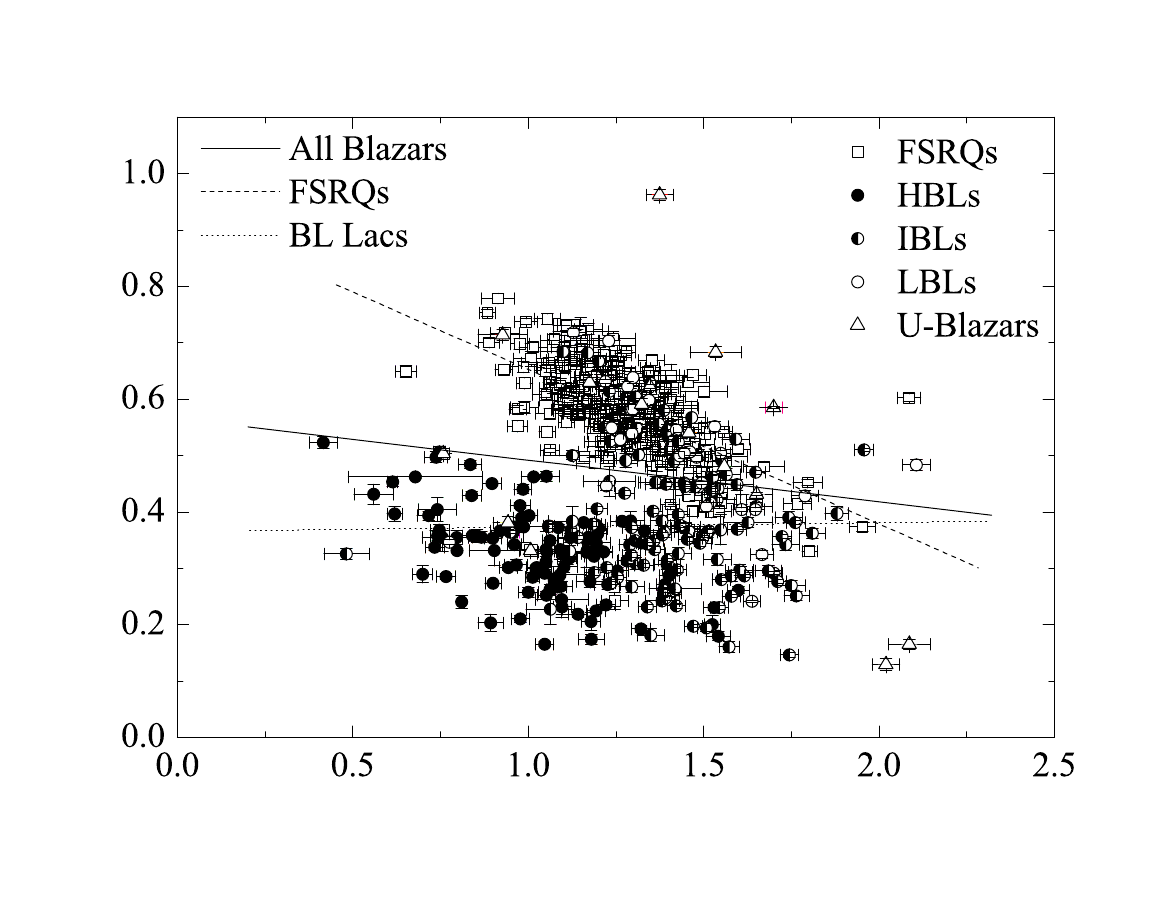}}
    \caption{The correlation between $\alpha_{\rm RO}$ and $\alpha_{\rm
OX}$.}
\label{aRO-OX}
\end{figure}

From the calculated effective spectral indexes, the scattering diagram between $\alpha_{\rm RO}$ and $\alpha_{\rm
OX}$ is plotted in Fig. \ref{aRO-OX}, and the linear regression analysis results are listed in Table \ref{T3}.

\section{Discussions}

As a special subclass of AGNs, blazars show many extreme
observational properties, which are associated with a beaming
effect. Blazars can be divided into  BL Lacertae objects (BL Lacs) and flat  spectrum radio quasars (FSRQs)
 by their emission line features. They are  the major population of detected sources in the
 Fermi missions (
 Abdo et al 2010b,c;
 Ackermann, et al. 2011a,b;
 Nolan et al. 2012;
 Acero et al. 2015;
 Ackermann, et al. 2015).
The Fermi detected  blazars provide us with a good opportunity to study
the emission mechanism  and beaming effects in $\gamma$-rays.
 The spectral energy distributions (SEDs) are available for some blazars and studied in the literatures (see
 Sambruana et al. 1996;
 Zhang, et al. 2002;
 Nieppola et al. 2006, 2008;
 Abdo et al. 2010a).   At the present work,  the multiwavelength data is compiled for
  a sample of 1425 Fermi blazars from the 3FGL (Acero et al. 2015) and
   their SEDs are calculated. SEDs for 1392 blazars have successfully been achieved, and their
   monochromatic luminosities  at radio, optical, X-ray and $\gamma$-ray, and effective spectral indexes
  are also calculated.

BL Lacertae objects can be divided into radio selected BL Lac
objects (RBLs) and X-ray selected BL Lacs objects (XBLs) from
surveys, or low frequency peaked BL Lacertae objects (LBLs, log
$\nu_{\rm p} < 15$ Hz) and high frequency peaked BL Lacertae
objects (HBLs, log $\nu_{\rm p} > 15$ Hz) from SEDs. Generally
RBLs correspond to LBLs while XBLs  to HBLs (
Padovani \& Giommi, 1995, 1996; Urry \& Padovani, 1995). Nieppola,
et al. (2006) calculated SEDs for a sample of BL Lacertae objects
and set the boundaries for different subclasses: ${\rm log}\,
\nu_{\rm p} < 14.5$ Hz for LBLs, 14.5 Hz $< {\rm log}\, \nu_{\rm
p} < 16.5$ Hz for IBLs, and ${\rm log} \, \nu_{\rm p} > 16.5$ Hz
for HBLs. The classification was extended to all non-thermal
dominated AGNs as
  low synchrotron peaked blazars-LSP (${\rm log}\, \nu_{\rm p} < 14$ Hz),
  intermediate synchrotron peaked blazars-ISP (${\rm log}\, \nu_{\rm p} = 14 \sim 15$ Hz), and
  high synchrotron peaked blazars-HSP (${\rm log}\, \nu_{\rm p} >15$ Hz) by Abdo et al. (2010a).

Ghisellini (1999)  proposed that there is a subclass of BL Lacs with their synchrotron peak
frequencies being higher than that of conventional HBLs, $\nu_{\rm p} > 10^{19}$ Hz. These objects can be called ultra-high-energy
synchrotron peak BL Lacs (UHBLs) (Giommi et al. 2001). In the work presented by Nieppola  et al. (2006), there are 22
objects with ${\rm log} \, \nu_{\rm p}
>19$, of which 9 objects have  ${\rm log} \,\nu_{\rm p}
>20$.  They also found  the appearance of
several low-radio-luminosity LBLs, which could even reach  lower radio luminosities than any of the HBLs.

\subsection{Peak Frequency}

At the present work, SEDs have successfully been calculated for 1392 Fermi blazars
 from  their  multiwavelength data. The peak frequencies obtained  in the range of
$ \rm{log}\nu_{\rm p}$ (Hz) = 11.60 $\sim$ 20.12 in the observer frame are
listed in Col. (12) of  Table 1.
 Out of the 1392 Fermi blazars, 999 blazars have  available redshifts.

For those 999 blazars,  the rest peak frequencies obtained  are in the range of
 $ \rm{log}\nu_{\rm p}^{rest}$ (Hz) = 12.15 $\sim$ 19.39 in the rest frame (frequency in the discussions below is assumed to be in the rest frame if it is not specifically stated). There are 3 sources, whose logarithm of peak frequencies is  greater
  than 19.0 ($ \rm{log}\nu_{\rm p}^{rest} (Hz) \geq 19.0$).
  When  FSRQs, BLs, and BCUs are considered separately,
  we have: for FSRQs, their peak frequencies $ \rm{log}\nu_{\rm p}^{rest}$ (Hz) are in the range of
 12.15 $\sim$ 17.11 with an averaged value of
  $ \rm {\left< log\, \nu_p^{rest}(Hz) \right> \, = \,13.94 \pm 0.65}$;
  for BL Lacs, $\rm{log \, \nu_p^{rest}} (Hz)$ are in the range of
  12.87 $\sim$ 19.39 with $ \rm {\left< log\, \nu_p^{rest}(Hz) \right> \,= 15.07 \pm 1.19}$; while for BCUs,
  $\rm{log \, \nu_p^{rest}} (Hz)$ are in the range of
  12.25 $\sim$ 17.36 with $ \rm {\left< log\, \nu_p^{rest}(Hz) \right> \,= 14.33 \pm 1.00}$.
The distribution of  peak frequencies in the rest frame for the 999 blazars is shown in Fig. \ref{fig-density-all}d. It can be seen that the distribution can be fitted using 3 components when the ¡±Bayesian classification¡± analysis method is employed. When the jointing points of two adjacent Gaussian curves are used to set the boundaries for different classes and the acronyms of Abdo et al. (2010a) are used,
 we obtain the following classifications:

 LSPs: ${\rm log \,\nu_{\rm p} (Hz)} \leq 14.0$,

 ISPs: $14.0 < {\rm log \,\nu_{\rm p} (Hz)} \leq 15.3$, and

 HSPs: ${\rm log \,\nu_{\rm p} (Hz)} > 15.3$ Hz.

For the whole sample of 1392 blazars, if the averaged values of redshift, $\left < z \right > \, =\, 0.568$ and  $\left < z \right > \, =\, 0.524$ are used to BLs and BCUs without redshifts, the rest peak
frequencies can be  obtained in the range of $ \rm{log}\nu_{\rm p}^{rest}$ (Hz) = 12.15 $\sim$ 20.31. In this case, there are 8
 sources,  whose logarithm of peak frequencies is greater than 19.0 ($ \rm{log}\nu_{\rm p}^{rest} (Hz) \geq 19.0$). Their distributions and statistical analysis results using the "Bayesian classification"  method are shown in Fig. \ref{fig-density-all}b and Fig. \ref{fig-BIC-all}b respectively. Based on the above-mentioned classification criteria, the classification (HSP, ISP, and LSP) obtained is
 listed in Col. 3 of Table  \ref{T1-samp}.
   From Fig. \ref{fig-density-all}b, it can be seen that the peak frequencies corresponding to the jointing points are higher than those in Fig. \ref{fig-density-all}d.
 This implies that the averaged values of redshifts used to get the rest frequencies are over estimated for most sources without redshifts.

 From the rest frame peak frequencies of the whole sample and our criteria, we have:
 25.14\% of them belong to HSP,
 40.09\% belong to ISP, and
 34.77\% belong to LSP.
 When FSRQs, BLs, and BCU are considered separately, we have:
 9 are HSP FSRQs,
 180 are ISP FSRQs, and
 272 are LSP FSRQs for 461 FSRQs; and
 235 are HSP BLs,
 272 are ISP BLs, and
 114 are LSP BLs for the 620 BLs. See  Table \ref{ClassDis} for details.

  As shown in Table \ref{T1-samp},  there are 22  sources, whose logarithm of the rest peak frequency  is greater
   than 18 ( $\rm{log \, \nu_p^{rest} (Hz) \geq 18.0}$)
   and the other 8 sources, $\rm{log \, \nu_p^{rest} (Hz) \geq 19.0}$, however statistical analysis does not come up with an ultra-high synchrotron peak component for the  999 sources with available redshift or the whole 1392 sources (See Fig. \ref{fig-density-all}d, \ref{fig-density-all}b) for the rest frequencies. We do not have such an ultra-high  synchrotron  peak component for the observer frequency either for the 999 sources with available redshift or for the whole sample of 1392 sources (See Fig. \ref{fig-density-all}c, \ref{fig-density-all}a).   In Fig. 27 of Abdo et al. (2010a), it is stated that there is no population of ultra high energy peaked (UHBLs) blazars. Our statistical analysis confirms the results.

 For the classifications, Abdo et al. (2010a) proposed,
  ${\rm log \,\nu_{\rm p} (Hz) }< 14.0$  for LSPs,
 $14.0 \leq {\rm log\, \nu_{\rm p} (Hz) } \leq 15.0$ for ISPs, and
${\rm log \nu_{\rm p} (Hz) } > 15.0$  for HSPs.
From Fig. \ref{fig-density-all}d at the present work,  two jointing points at $\rm {log \, \nu_p^{rest} (Hz)}$ = 13.98 and $\rm {log \, \nu_p^{rest} (Hz)}$ = 15.30 can be obtained,  and this results in that
 the boundaries are  at $\rm {log \, \nu_p^{rest} (Hz)}$ = 14.0 and $\rm {log \, \nu_p^{rest} (Hz)}$ = 15.3.  From Table 1,  the uncertainties for $\rm log\,\nu_p $  from
 $\rm \Delta log\,\nu_p $ = 0.02 to
 $\rm \Delta log\,\nu_p $ = 1.39 with an averaged value of $< \rm \Delta log\,\nu_p > = 0.24\pm0.21$ can be obtained.   Considering the uncertainties in $\rm {log \, \nu_p}$, we believe that our classifications are consistent with those by  Abdo et al. (2010a).

 For comparison, we analyze  our synchrotron peak frequencies (${\rm log} \nu_{\rm p}^{\rm TW}$) with those (${\rm log} \nu_{\rm p}^{\rm other}$)
known for common sources in the literatures by Sambruna et al. (1996), Nieppola et
al. (2006, 2008) and Abdo et al. (2010b). First we calculate the difference between our estimation and others',
$({\rm log} \nu_{\rm p}^{\rm TW}-{\rm log} \nu_{\rm p}^{\rm other})$, then investigate the relationship between the differences and our estimations.

\begin{figure}
    \centering
    \resizebox{\hsize}{!}{\includegraphics*{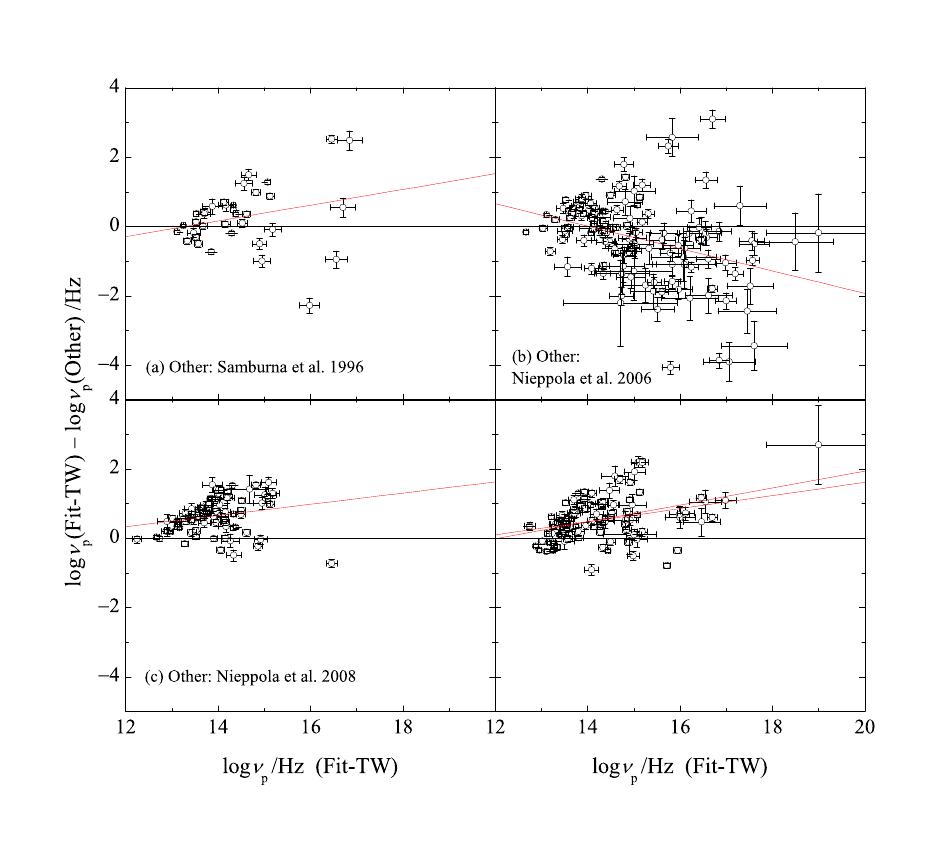}}
    \caption{The plot of the difference between fitted peak frequency ($\rm log \nu_p^{\rm{TW}}$)  and that obtained from other literatures,
    $\rm log \nu_p ^{Other}$, against that from SED fitting, $\rm log \nu_p^{\rm{TW}}$. The y = 0 line stands for $\rm log \nu_p^{\rm{TW}}$ = $\rm log \nu_p ^{Other}$, another line stands for the best fitting result.
     (a) Sambruna, et al. 1996,
     (b) Nieppola, et al. 2006,
     (c) Nieppola, et al. 2008,
     (d) Abdo, et al. 2010a.}
    \label{vp-vp}
\end{figure}

There are   35 sources in common with Sambruna et al. (1996), and we have
$({\rm log} \nu_{\rm p}^{\rm TW}-{\rm log} \nu_{\rm p}^{\rm S96})  =   (0.23 \pm0.15){\rm log} \nu_{\rm p}^{\rm TW}\,\, -3.02 \pm	2.20 $ with a correlation coefficient  $r = $ 0.25 and
 a chance probability  14.5\%. Our estimations tend to be larger than those by Sambruna et al. (1996) for some sources. It is also found that two sources with difference being larger than 2.0 are HSPs with $\rm log \nu_p(Hz) > 15 $ (See Fig. \ref{vp-vp}a).

There are 129 sources in common with Nieppola et al. (2006),
 $({\rm log} \nu_{\rm p}^{\rm TW}-{\rm log} \nu_{\rm p}^{\rm N06})  =   -(0.32 \pm 0.08){\rm log} \nu_{\rm p}^{\rm TW}\,\, + 4.52 \pm1.16 $ with
$r = $ -0.35 and  $ p < 10^{-4}$. There is a clear anti-correlation between them, indicating that our estimations are smaller than those by Nieppola et al.  (2006) for  most common sources. Fig. \ref{vp-vp}b shows that our estimations of 4 source  are greatly larger  than
 those by Nieppola et al. (2006) and that the sources with large deviation are the sources with $\rm log \nu_p(Hz) > 15$.

There are  82 sources in common with Nieppola et al. (2008),
 $({\rm log} \nu_{\rm p}^{\rm TW}-{\rm log} \nu_{\rm p}^{\rm N08})  =   (0.16 \pm 0.08){\rm log} \nu_{\rm p}^{\rm TW}\,\, - 1.61 \pm1.12 $ with
$r = $ 0.22 and  $ p = 4.7\%$. There is a marginal positive correlation between them, indicating that  our estimations are larger than those by Nieppola et al. (2008) (See Fig. \ref{vp-vp}c).

There are 102 sources in common with Abdo et al. (2010a),
 $({\rm log} \nu_{\rm p}^{\rm TW}-{\rm log} \nu_{\rm p}^{\rm A10})  =   (0.24 \pm 0.06){\rm log} \nu_{\rm p}^{\rm TW}\,\, - 2.95 \pm 0.80 $ with
$r = $ 0.40 and  $ p < 10^{-4}$.
 If the right hand corner point is excluded, we have
 $({\rm log} \nu_{\rm p}^{\rm TW}-{\rm log} \nu_{\rm p}^{\rm A10})  =   (0.19 \pm 0.06){\rm log} \nu_{\rm p}^{\rm TW}\,\, - 2.18 \pm 0.88 $ with
$r = $ 0.29 and  $ p =2.9\times10^{-3}$.  The analysis indicates  that our estimations are larger than those by Abdo et al. (2010a) (See Fig. \ref{vp-vp}d).
All the comparisons conducted and the best fitting results are listed in Table \ref{T4}.

  From the  literatures (Sambruna et al. 1996;
 Pian et al. 1998;
 Nieppola et al. 2006, 2008;
 Abdo et al. 2010a;
 Giommi et al. 2000) and the present work, it is noted that the peak frequency (in observer frame) for a certain source is different from one calculation to another based on different multiwavelength observations.
  For some sources, the different spectral shapes may be attributed to the variability and different data sets. The comparisons suggest that there is no much difference  between our estimations and those by Sambruan et al. (1996) and Nieppola et al. (2006), but ours are over estimated than those by Nieppola et al. (2008) and Abdo et al. (2010a).

\begin{deluxetable}{ccccccc}
\tabletypesize{\scriptsize}
  \tablecaption{The linear regression analysis results for correlations between two parameters.}
  \tablewidth{0pt}
 \tablehead{
  \colhead{$y$ vs $x$} &
  \colhead{$a  \pm \Delta a$} &
  \colhead{$b  \pm \Delta b$ } &
  \colhead{$r$ } &
  \colhead{$N$ } &
  \colhead{$p$ }
    }
 \startdata
$({\rm log} \nu_{\rm p}^{\rm TW}-{\rm log} \nu_{\rm p}^{\rm S96})$  vs ${\rm log} \nu_{\rm p}^{\rm TW}$
	&	-3.02 	$\pm$	2.20 	&	0.23 	$\pm$	0.15 	&	0.25 	&	35	&	0.145	 \\
$({\rm log} \nu_{\rm p}^{\rm TW}-{\rm log} \nu_{\rm p}^{\rm N06})$  vs ${\rm log} \nu_{\rm p}^{\rm TW}$	&	4.52 	$\pm$	1.16 	&	-0.32 	$\pm$	 0.08 	&	-0.35 	&	129	&	 $<10^{-4}$	\\
$({\rm log} \nu_{\rm p}^{\rm TW}-{\rm log} \nu_{\rm p}^{\rm N08})$  vs ${\rm log} \nu_{\rm p}^{\rm TW}$	&	-1.61 	$\pm$	1.12 	&	0.16 	$\pm$	 0.08 	&	0.22 	&	82	&	 0.047	\\
$({\rm log} \nu_{\rm p}^{\rm TW}-{\rm log} \nu_{\rm p}^{\rm A10})$  vs ${\rm log} \nu_{\rm p}^{\rm TW}$$^{\dag}$ 	&	-2.95 	$\pm$	0.80 	&	 0.19 	 $\pm$	0.06 	&	0.29 	&	 101	&	$<10^{-4}$	\\
$({\rm log} \nu_{\rm p}^{\rm TW}-{\rm log} \nu_{\rm p}^{\rm A10})$  vs ${\rm log} \nu_{\rm p}^{\rm TW}$ 	&	-2.18 	$\pm$	0.88 	&	 0.24 	 $\pm$	0.06 	&	0.40 	&	101	&	 $2.9\times10^{-3}$	\\

(${\rm \alpha_{ro}^{TW}-\alpha_{ro}^{A10}}$) vs ${\rm \alpha_{ro}^{TW}}$
	&	-0.02 	$\pm$	0.01 	&	-0.12 	$\pm$	0.03 	&	-0.17 	&	724	&	 $<10^{-4}$	\\
(${\rm \alpha_{ox}^{TW}-\alpha_{ox}^{A10}}$) vs ${\rm \alpha_{ox}^{TW}}$
	&	-0.33 	$\pm$	0.04 	&	0.29 	$\pm$	0.03 	&	0.36 	&	498	&	 $<10^{-4}$	\\

(${\rm \alpha_{ro}^{TW}-\alpha_{ro}^{A11}}$) vs ${\rm \alpha_{ro}^{TW}}$
	&	-0.01 	$\pm$	0.02 	&	-0.17 	$\pm$	0.03 	&	-0.22 	&	514	&	 $<10^{-4}$	\\
(${\rm \alpha_{ox}^{TW}-\alpha_{ox}^{A11}}$) vs ${\rm \alpha_{ox}^{TW}}$
	&	-0.33 	$\pm$	0.05 	&	0.29 	$\pm$	0.04 	&	0.35 	&	376	&	 $<10^{-4}$	\\
\enddata
${\dag}: \rm {Excluded\,  the\, up\, right\, corner\, point}$.
\label{T4}
\end{deluxetable}

 Abdo et al. (2010a) and Ackermann et al. (2011a)  obtained the spectral indexes, $\alpha_{ox}$ and  $\alpha_{ro}$ for  Fermi blazars. So comparisons can be conducted between their spectral indexes and ours.

\begin{figure}
    \centering
    \resizebox{\hsize}{!}{\includegraphics*{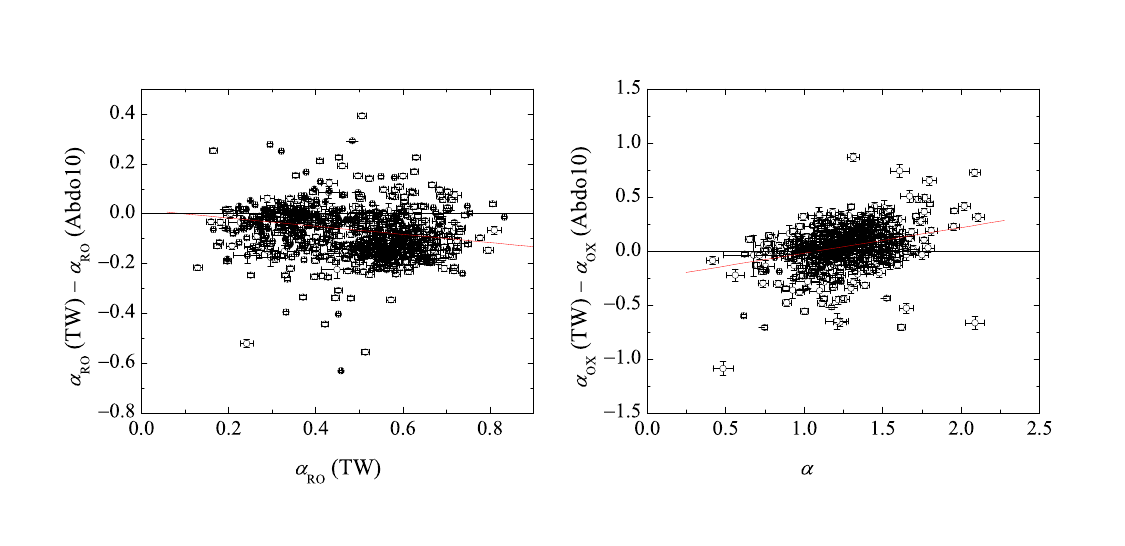}}
    \caption{
    The plot of the difference between our spectral index and that from Abdo et al. (2010a), against our spectral index.
    The y = 0 line stands for our spectral index is the same as that from Abdo et al. (2010a), another line stands for the best fitting result.
    Left panel is for $\alpha_{ro}$; Right panel is for $\alpha_{ox}$.}
    \label{A-A-A10}
\end{figure}

For spectral index, $\alpha_{ro}$, there are 724  sources in common with Abdo et al. (2010a),
 we have
${\rm \alpha_{ro}^{TW}-\alpha_{ro}^{A10}} =  -(0.12 	\pm	0.03){\rm \alpha_{ro}^{TW}} - (0.02 	\pm	0.01) $ with a correlation coefficient  $r = $ -0.17 and a chance probability  $p <10^{-4}$
 (Fig. \ref{A-A-A10}a);  while
 for $\alpha_{ox}$, there are 498 common sources, which follow
  ${\rm \alpha_{ox}^{TW}-\alpha_{ox}^{A10}} = (0.29 \pm	0.03){\rm \alpha_{ox}^{TW}} - (0.33 \pm 0.04)$ with $r = 0.36$ and $p <10^{-4}$
(Fig. \ref{A-A-A10}b).  The marginal positive correlation that we observe in Fig. \ref{vp-vp}d is likely due to the positive correlation we  observe here in Fig. \ref{A-A-A10}b.

\begin{figure}
    \centering
    \resizebox{\hsize}{!}{\includegraphics*{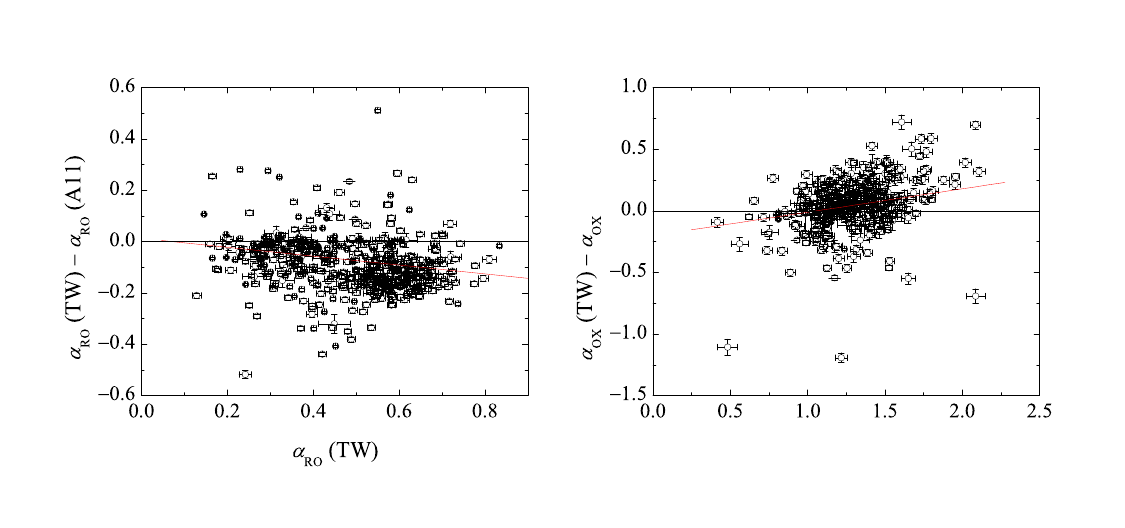}}
    \caption{
    The plot of the difference between our spectral index and that from Ackermann et al. (2011a), against our spectral index.
    The y = 0 line stands for our spectral index is the same as that from Ackermann et al. (2011a), another line stands for the best fitting result.
    Left panel is for $\alpha_{ro}$; Right panel is for $\alpha_{ox}$.}
    \label{A-A-A11}
\end{figure}

  For spectral indexes in Ackermann et al. (2011a), there are 514 $\alpha_{ro}$ for common sources,
 we have ${\rm \alpha_{ro}^{TW}-\alpha_{ro}^{A11}} =  -(0.17 	\pm	0.03){\rm \alpha_{ro}^{TW}} - (0.01 	\pm	0.02) $ with a correlation coefficient $r = $-0.22 and a chance probability  $p <10^{-4}$;
  while for $\alpha_{ox}$, there are 376 common sources, which follow
  ${\rm \alpha_{ox}^{TW}-\alpha_{ox}^{A11}} = (0.29 \pm	0.04){\rm \alpha_{ox}^{TW}} - (0.33 \pm 0.05)$ with $r = 0.35$ and $p <10^{-4}$. Results are shown in Fig. \ref{A-A-A11}.
 As shown in Fig. \ref{A-A-A10} and Fig. \ref{A-A-A11}, the differences in  our estimate and those by Abdo et al. (2010a) and Ackermann et al. (2011a)  scatter more or less around 0. Some of our $\alpha_{ro}$'s are slightly smaller than those by Abdo et al. (2010a) and Ackermann et al. (2011a) and some of our $\alpha_{ox}$'s are slightly larger than those by Abdo et al. (2010a) and Ackermann et al. (2011a), but no strong systematics appear.  Because the flux densities used for the calculations of spectral indexes by Abdo et al. (2010a) and Ackermann et al. (2011a) are different from those at the present work, the calculation results of the spectral indexes are different as well.

 In 2010, Abdo et al. (2010a) presented an empirical relation to estimate the synchrotron peak frequency, $\nu_p$ from effective spectral indexes $\alpha_{ox}\, \rm{and}\, \alpha_{ro}$.
 Following their work, we obtain an empirical relation to estimate the synchrotron peak frequency, $\nu_p^{Eq.}$ from effective spectral indexes $\alpha_{ox}\, \rm{and}\, \alpha_{ro}$ as
 \begin{equation}
\rm  log \nu_p^{Eq.} =\{
\begin{array}{l}
16+4.238X \qquad  \qquad X < 0\,\,  \\
16+4.005Y\qquad \qquad X > 0
\end{array}
,
\label{empirical-nu}
\end{equation}
where $X = 1.0 - 1.262\alpha_{ro} - 0.623\alpha_{ox}$, and
      $Y = 1.0 + 0.034\alpha_{ro} - 0.978 \alpha_{ox}$.

\begin{figure}
    \centering
    \resizebox{\hsize}{!}{\includegraphics*{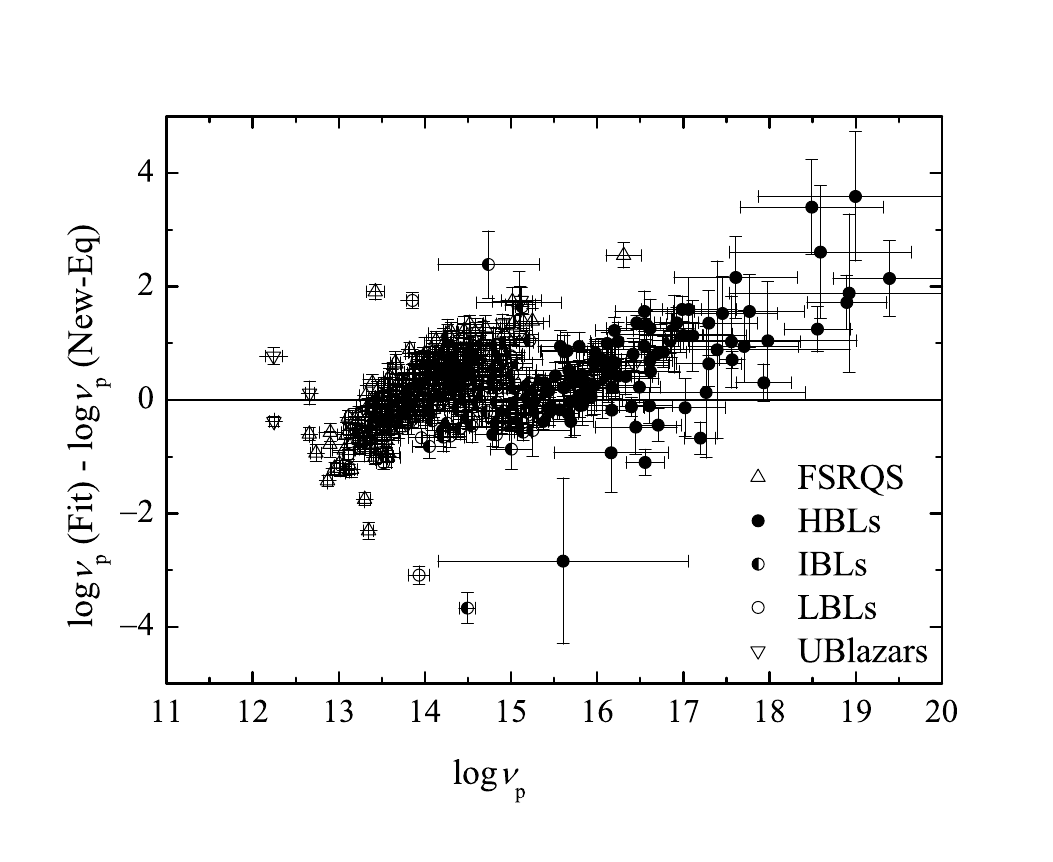}}
    \caption{The plot of the difference between fitted peak frequency ($\rm log \nu_p^{Fit}$)  and that estimated from the empirical equation (\ref{empirical-nu}), $\rm log \nu_p^{Eq}$, against that from SED fitting, $\rm log \nu_p^{Fit}$. The y = 0 line stands for
     $\rm log \nu_p^{Fit}$ = $\rm log \nu_p ^{Eq}$.
   }
    \label{emp-nu-nu}
\end{figure}

Based on Eq. (\ref{empirical-nu}) and the effective spectral indexes in the present paper,   $\rm {log \, \nu_p^{Eq}} $ can be derived, then we compare the relationship between the derived peak frequency ($\rm log \nu_p^{Eq}$) and fitted peak frequency ($\rm log \nu_p^{Fit}$).
 Similar to the above, we investigate the relationship between the difference ($\rm {log \, \nu_p^{Fit}} - \rm log \nu_p^{Eq}$) and $\rm {log \, \nu_p^{Fit}}$. The result is shown in Fig. \ref{emp-nu-nu}. The differences in our fitted peak frequency and the derived peak frequency for most sources scatter more or less around 0, but the derived peak frequency for high frequency source is underestimated as compared with the fitted peak frequency.

\subsection{Correlations}

\subsubsection{Luminosity vs Luminosity}

The correlations between  $\gamma$-ray luminosity (${\rm
{log}}\,L_\gamma$) and other luminosities including radio
(${\rm {log}}\,L_{\rm R}$), optical  (${\rm {log}}\,L_{\rm O}$), and
X-ray emission (${\rm {log}}\,L_{\rm X}$) have been discussed
 (Dondi \& Ghisellini 1995; Fan et al. 1998; Fossati
et al. 1998; Cheng et al. 2000). It is found that the correlation
between $\gamma$-ray and radio is the closest. At the present
work, we revisit those correlations using a large sample of
Fermi blazars. The relationship between $\gamma$-ray
luminosity and synchrotron peak luminosity (${\rm {log}}\,L_{\rm
p}$), and the relationship between $\gamma$-ray luminosity and integrated
luminosity (${\rm {log}}\,L_{\rm bol}$) are both investigated for
the whole blazar sample and subclasses of FSRQs, BL Lacs (HBLs, IBLs and
LBLs). The results are shown in Figs. \ref{LG-LROX}, \ref{LG-LpLbol} and Table  \ref{T2},
 and  demonstrate  very good correlations. The correlation coefficients are all greater than 0.68 with all chance
probabilities being less than $10^{-4}$.

It is known that luminosities are all correlated with redshift ($z$), so  luminosity-luminosity correlation
may be from a redshift effect (Kendall \& Stuart, 1979).
 A real correlation should remove the redshift effect.  In doing so, we use the method by Padovani  (1992).
 If variables  $i$ and $j$ are correlated with the third variable
$k$, then the correlation between $i$ and $j$ should exclude the
effect of the third variable $k$. In the case of three variables
$i$, $j$ and $k$, if the correlation coefficients of the relation
between any two variables  are represented as $r_{ij}$,
$r_{ik}$, $r_{jk}$ respectively, then  the correlation
coefficient $r_{ij}$  excluded the effect of the third
variable $k$ is represented as ${r_{ij,k}} = ({r_{ij}} -
{r_{ik}}{r_{jk}})/\sqrt {(1 - r_{ik}^2)(1 - r_{jk}^2)}$. When the
method is adopted to discuss the correlations between any two
luminosities, the correlation coefficients subtracting the effect
of redshift are thus listed in Table \ref{T2} ($r_{LL, z}$ is the
correlation coefficient and $p_{LL,z}$ is the chance probability).

From Table \ref{T2}, it is noted that there is a good correlation
between $\gamma$-ray luminosity (${\rm {log}}\,L_\gamma$) and radio
1.4 GHz luminosity (${\rm {log}}\,L_{\rm R}$) for the whole sample and
 the subclasses with correlation coefficients  $r_{L_\gamma
L_{\rm R}, z} =$ 0.554, 0.420£¬ 0.613, 0.535, 0.628 and 0.536 for all
blazars sample and subclasses of FSRQs, BL Lacs, HBLs, IBLs and
LBLs respectively. For ${\rm {log}}\,L_\gamma$ and  ${\rm
{log}}\,L_{\rm O}$, the corresponding correlation coefficients are
$r_{L_\gamma L_{\rm O}, z} =$ 0.257, 0.325, 0.479, 0.663, 0.518 and
0.277.  While for ${\rm {log}}\,L_\gamma$ and ${\rm {log}}\,L_{\rm X}$,
correlation coefficients are $r_{L_\gamma L_{\rm X}, z} = -0.134$,
0.254, $-0.082$, 0.265, 0.202 and 0.103. It can be seen that the correlation
between $\gamma$-rays and radio and that
between $\gamma$-rays and optical bands exist even after removing the redshift effect.
 But the correlation between $\gamma$-rays and X-ray bands only exists for FSRQs and HBLs.  There is no more correlation for LBLs.
For FSRQs, their X-ray emissions are from the synchrotron emission tail and external Compton emission, the later is also strongly believed
 to contribute to $\gamma$-ray emission. Therefore external Compton emission can also make  contribution to
 the X-ray emissions:  the more contribution to the X-ray emissions the more contribution to the $\gamma$-ray emissions.  Thus it can be expected that a correlation exists between X-ray and $\gamma$-ray emissions. For BL Lacs, synchrotron self-Compton emissions are mainly responsible for the whole electromagnetic wavebands. Then for HBLs, their X-rays are from synchrotron emissions while their $\gamma$-rays are from self-Compton. So a correlation between X-ray and $\gamma$-ray can be expected; while for LBLs, their X-ray emissions are from the summation of synchrotron emission tail and self-Compton emission, their $\gamma$-rays are from self-Compton. In this case, the summation will dilute the correlation between X-ray and $\gamma$-ray emissions.

From our analysis, it is found that for
FSRQs,  after removing the redshift effect,  there is still strong
correlations between $\gamma$-ray and radio, between $\gamma$-rays and optical, and between $\gamma$-rays and X-ray
emissions  with the correlation coefficients decreasing from
$r_{L_{\gamma}L_{\rm R},z}$ to  $r_{L_{\gamma}L_{\rm X},z}$. While
for BL Lacs, the strongest correlation is between $L_{\gamma}$ and $L_{\rm R}$, but there is  almost no correlation between $L_{\gamma}$
and $L_{\rm X}$.

\begin{figure}
    \centering
    \resizebox{\hsize}{!}{\includegraphics*{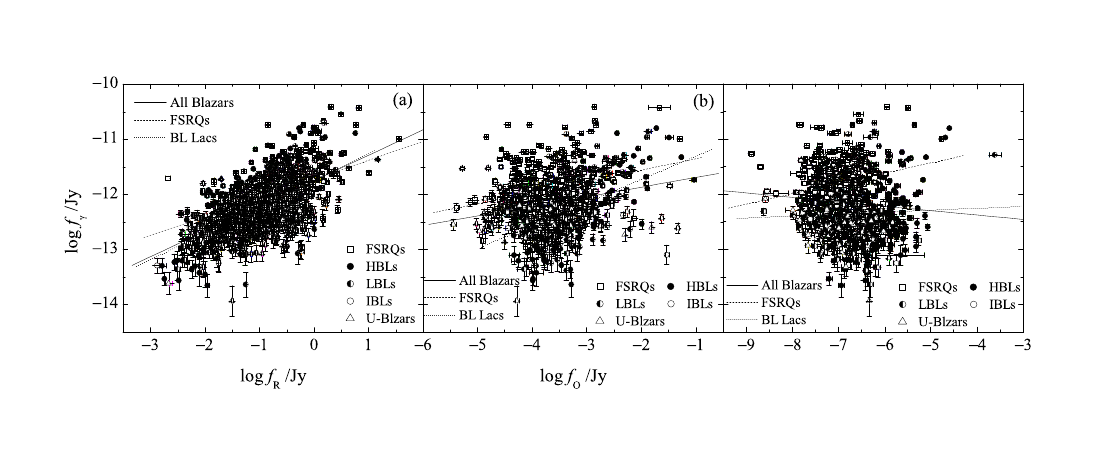}}
    \caption{The relationship between the $\gamma$-ray flux density at 1 GeV and the lower energy bands.
    ($a$) For $\gamma$-ray vs radio band (1.4 GHz);
    ($b$) For $\gamma$-ray vs optical band (R band); and
    ($c$) For $\gamma$-ray vs X-ray band at 1 KeV.
     }
    \label{fG-fROX}
\end{figure}

We now discuss the relationship between the $\gamma$-ray flux density and the lower energetic band flux densities by adopting the linear regression method. It is found that there exist correlations
between $\gamma$-ray and radio,
between $\gamma$-rays and optical, and
between $\gamma$-rays and X-ray emissions with chance probabilities being less than $10^{-4}$ for FSRQs.
For BL Lacs, there exist correlations
between $\gamma$-ray and radio,
between $\gamma$-rays and optical bands, but there is no correlation between $\gamma$-rays and X-ray emissions. However, there exists a  correlation between $\gamma$-rays and X-ray for HBLs. The linear regression analysis results are listed in Table \ref{T3} and shown in  Fig. \ref{fG-fROX}. It is interesting that the flux density-flux density analysis results are consistent with the luminosity-luminosity correlation results after the redshift effect is removed. It can be seen that the correlation between $\gamma$-rays and radio is the strongest.

For the correlation between radio and $\gamma$-ray emissions, similar analysis results are presented  in many literatures:
Abdo et al. (2009a) and Giroletti et al. (2010) found that there is a possible correlation between the 8.4 GHz
radio flux and the $\gamma$-ray flux.
Ackermann et al. (2011b) found that there is a correlation between $\gamma$-ray and 8.4GHz/15GHz radio fluxes.
Ghirlanda et al. (2010) found that there is  a strong correlation between the radio flux at 20 GHz ($F_r$) and the $\gamma$-ray flux above 100 MeV ($F_{\gamma}$).
Nieppola et al. (2011) found a significant correlation between both the flux densities and luminosities in $\gamma$-ray and 37 GHz radio bands.
 Our $\gamma$-ray vs radio correlation result is consistent with those reported. It is commonly found
   that the radio emission in blazars are strongly beamed, $\gamma$-ray emissions are also beamed (see Fan et al. 2014 and references therein). It is possible that the correlation between radio and $\gamma$-ray band is an apparent correlation caused by the beaming effect. If the $\gamma$-ray emissions are from an SSC process, then there is a correlation between $\gamma$-ray and radio emissions. It is also possible that the radio and $\gamma$-ray correlation implies that $\gamma$-ray radiation is
produced co-spatially with the radio emission in the jet (Nieppola et al. 2011).

Correlations between $\gamma$-ray luminosity and peak luminosity ( and integrated luminosity) are also discussed in this paper (Fig. \ref{LG-LpLbol}).
$\rm{ log\, L_{\gamma}\,}$ = $(1.05 \pm 0.04) \rm logL_{bol}$ -  $(2.26 \pm 1.80)$, and
$\rm{ log\, L_{\gamma}\,}$ = $ (1.00 \pm 0.04)\rm log L_{p}$ +  $(0.30 \pm 1.80)$ with chance probabilities  $p < 10^{-4}$ for FSRQs  and
$\rm{ log\, L_{\gamma}\,}$ = $ (1.11 \pm 0.03)\rm log L_{bol}$ - $ (5.66 \pm 1.18)$, and
$\rm{ log\, L_{\gamma}\,}$= $(1.08 \pm 0.03)\rm log L_{p}$ -  $(3.72 \pm 1.20)$ with  $p < 10^{-4}$ for BL Lacs.
 Their  correlations are strong, and the slopes of the relations for FSRQs are similar to those for BL Lacs. But the intercepts in the linear correlation for BL Lacs are smaller than those for FSRQs, which indicate that for the same peak luminosity (or integrated luminosity),
 FSRQs show more luminous $\gamma$-ray emissions than  BL Lacs (See also Table \ref{T3}). This is consistent with the claim  that the $\gamma$-ray emissions are mainly from EC in FSRQs, resulting in increased Compton dominance. For FSRQs, their radio emissions are strongly beamed and luminous while in the $\gamma$-ray, their increased Compton dominance also results in luminous emission. Therefore, such effects can produce a correlation between radio and $\gamma$-rays.

\subsubsection{Other correlations}

At the present work,  correlations are investigated with different parameters:
 Correlation between the peak frequency and monochromatic  luminosity ($\gamma$-ray, X-ray, optical, and Radio band) (Fig. \ref{vp-LROXG}),
correlation between the peak frequency and the integrated luminosity (and peak luminosity) (Fig. \ref{vp-LpLb}),
correlation between spectral curvature and integrated luminosity ( and peak frequency) (Fig. \ref{P1-vpLb}),
correlation between the peak frequency and the effective spectral indexes (Fig. \ref{vp-aRO-OX}), and
the correlation between effective spectral indexes (Fig. \ref{aRO-OX}). The linear regression analysis results are listed in  Tables \ref{T2} and \ref{T3}.

For peak frequency (${\rm {log\, \nu_{p}}}$) and monochromatic luminosity (${\rm {log}}\,L_{\rm \nu}$), ${\rm {log\, \nu_{p}}}$ vs ${\rm {log}}\,L_{\rm \nu}$  (Fig. \ref{vp-LROXG}); our analysis indicates  a strong
 anti-correlation between $\gamma$-ray luminosity ($ \rm log\, L_{\gamma}$) and peak frequency ($ \rm log\, \nu_{p}$),
 $\rm{ log\, L_{\gamma}\,=\,}$ -$(0.29\pm 0.02)\rm log \nu_{p}$  + $ (49.58 \pm 0.35)$ with a correlation coefficient $r = -0.32$ and a chance probability $p < 10^{-4}$, and that between radio luminosity ($ \rm log\, L_{R}$) and peak frequency ($ \rm log\, \nu_{p}$),
 $\rm{ log\, L_{R}\,=\,}$ -$(0.44\pm 0.03)\rm log \nu_{p}$ +  $ (48.70 \pm 0.38)$ with  $r = -0.45$ and  $p < 10^{-4}$ for the whole sample. A marginally positive correlation can be found between X-ray luminosity ($ \rm log\, L_{X}$) and peak frequency ($ \rm log\, \nu_{p}$), but there is no correlation between optical luminosity ($ \rm log\, L_{O}$) and peak frequency ($ \rm log\, \nu_{p}$). See Table \ref{T2} for details.
 It is possible that the strong anti-correlation is from the beaming effect as discussed in Nieppola et al. (2008). In their work, the Doppler factor is found to be larger at lowly peaked sources and  may be smaller at highly peaked sources, HBLs for instance. So, the lower peak sources have larger Doppler factors  resulting in a larger boosting, and higher peak sources have smaller Doppler factors resulting in a weaker boosting. Therefore there is  an anti-correlation between  the $\gamma$-ray luminosity and peak frequency and that between radio luminosity and peak frequency for BL Lacertae objects. For FSRQs, their $\gamma$-ray emissions are dominated by EC emissions, thus they are luminous as observations indicate. Their radio emissions are strongly beamed, in this sense, there is an anti-correlation between  the $\gamma$-ray luminosity and peak frequency and that between radio luminosity and peak frequency. For BL Lacs, the  X-ray emissions are from the synchrotron emission tail and self-Compton emissions for LBLs, but mainly from synchrotron emissions for HBLs, so that it results in a positive correlation between X-ray luminosity and peak frequency. For FSRQs, their X-ray emissions are from synchrotron emission tail, self-Compton emissions and some EC emissions, thus there is no clear correlation between X-ray luminosity and peak frequency ($p = 72.56\%$).
 In optical bands, it is quite complicated that the emissions may be beamed as radio, also they are  from different emission mechanisms. For  lowly peaked sources, when the peak frequency moves to lower frequency side,  its optical emissions will decrease, and the inverse Compton emission will also make  contributions
  to this band; For highly peaked sources, their intrinsic emissions are luminous as discussed in Nieppola et al. (2008). In addition, it is possible that there are emissions from host galaxies for BLs and accretion for FSRQs. All those effects  will dilute the correlation between the optical luminosity and the peak frequency.

For ${\rm {log}} \,\nu_{\rm p}$ vs ${\rm {log}}\,L_{\rm bol}$
(Fig. \ref{vp-LpLb}a): there is no correlation between synchrotron peak frequency (${\rm {log}}\,\nu_{\rm p}$)
  and integrated  luminosity (${\rm {log}}\,L_{\rm bol}$) for the
  whole sample ($r = -0.03$ and $p = 32.9\%$). But there are marginal correlations  for FSRQs ($r$ = 0.10 and $p = 3.1\%$) and BL Lacs ($r$ = 0.09 and $p = 2.9\%$).  For HBLs, we also have a positive correlation with $r$ = 0.19 and $p = 3.6 \times 10^{-3}$.

In 1996, Sambruna et al. calculated SEDs and obtained  effective
spectral indexes for a sample of blazars,  and found that the
primary correlations exist for $\alpha_{\rm RO}$ vs
$\alpha_{\rm XOX}$ ( = $\alpha_{\rm X} - \alpha_{\rm OX}$), $\alpha_{\rm RO}$ vs ${\rm {log}}L_{\rm bol}$, and $\alpha_{\rm
RO}$ vs redshift.
 From Fig. 5 of their paper,  it is found that
  there is a positive tendency for the integrated  luminosity to
  increase with peak frequency for XBLs while there is no such a
  tendency for FSRQs or RBLs.
  Our analysis shows that there is a    positive correlation for HBLs, which
   is consistent with the result for XBLs shown in Fig. 5 of Sambruna et
   al. (1996). As for FSRQs  and LBLs in our work, there is a marginal correlation for FSRQs but no clear correlation for LBLs ( $r = 0.08$ and $p = 28.7\%$). In fact, there is a strong  correlation between the integrated luminosity and the optical luminosity, $\rm log L_{bol} = (0.91 \pm 0.04) log L_{O} + 1.17 \pm 0.20$ from the present work. The effects discussed above for optical band will also cause the dilution for integrated luminosity for FSRQs and LBLs. From Fig. 1 of Nieppola et al. (2008), it can be seen that the Doppler factor is in the range of 1.5 to 28.5 for the lowly peaked sources, such that it is possible that some objects are strongly beamed while others are not so strongly beamed for FSRQs and LBLs. Such a difference  in Doppler factor will also dilute the correlation for FSRQs and BL Lacs. Thus there is no correlation for the whole sample.
 In this sense, our result is consistent with that  by  Sambruna et al. (1996).

For ${\rm {log}}\,\nu_{\rm p}$ vs ${\rm {log}}\,L_{\rm p}$ (Fig. \ref{vp-LpLb}b), there is an anti-correlation for the whole sample with $r = -0.10$ and $p = 3 \times 10^{-4}$ resulting in that  peak luminosity decrease with increasing $\nu_p$.  However there is no correlation for FSRQs ($p = 81.6\%$) or BL Lacs ($p = 95.6\%$). But there is a marginal positive correlation for HBLs with $r = 0.13$ and $p = 4.4\%$.   For HBLs, brighter sources tend to show higher peak frequency while for LBLs and FSRQs, there is no clear tendency. In the work presented by Fossati et al. (1998), it can be seen  that $\nu_p$ decreases with increasing luminosity as shown  in Fig. 7 for the whole sample. But for XBLs, there is a sign
for luminosity to increase with increasing peak frequency. Our results are consistent with those by Fossati et al. (1998).

Nieppola, et al. (2008) discussed the correlation of Doppler factor against the rest peak frequency and found the Doppler factors decrease with increasing rest peak frequencies  as shown  in  Fig. 1.
 They also investigated the correlation between $L_{\rm p}$ and $\nu_{\rm p}$ for a sample of blazars using their
corrected peak frequency and peak luminosity, and found that there
is an intrinsically  positive correlation between them.  Our  work
 shows an anti-correlations between luminosity ($\rm {log \, L_{bol}}$ and $\rm {log \, L_{p}}$) and peak frequency ($\rm {log \, \nu_p}$) for the whole sample, but there is a marginal correlation between  luminosity and  peak frequency for HBLs with $r\,=0.13$ and $p\,=\,4.4\%$. As investigated by Nieppola et al. (2008), it is found that the beaming factor (Doppler factor) decreases with peak frequency and high frequency sources may be weakly beamed, and the anti-correlation of peak luminosity and peak frequency is caused by a beaming effect.
 Since  Doppler boosting decreases with peak frequency and the Doppler factor may be small for highly peaked sources
  as shown in Fig. 1 by  Nieppola et al. (2008),
  then  we can say that  our HBLs  have weak beaming effect.  Therefore  their observed luminosity can be taken as  intrinsic one, and the marginally positive correlation for HBLs shows the intrinsic situation. For HBLs, if there are more  particles being accelerated to higher energy, one can expect more luminous luminosity in higher frequency so that there is  a positive correlation between the peak frequency and the peak luminosity.
  In the case of HBLs, the observed emissions do not need Doppler correction. We know from Fig. 1 by
   Nieppola et al. (2008) that difference in Doppler factor for FSRQs and LBLs will also dilute the correlation of  $L_{\rm p}$ and $\nu_{\rm p}$ for FSRQs and BL Lacs.

For peak frequency (${\rm {log}}\,\nu_{\rm p}$) and spectral curvature (P1), we investigate their relationship, ${\rm {log}}\,\nu_{\rm p}$ against ${\rm {log}}\,|P_1|$ (Fig. \ref{P1-vpLb}b). There is a clear anti-correlations between them,
${\rm log \nu_p\,=\, -(5.81 \pm 0.12)log(|P_1|)\, + (9.09 \pm 0.12)}$ with $r\,=\,-0.78$ and $ p \, < \, 10^{-4}$
for the whole sample.
  Clear anti-correlations exist for
  subclasses, FSRQs, BL Lacs, and the subclasses of BL Lacs with $p\, < \, 10^{-4}$ (See Table \ref{T3} for details).
 The SED of a blazar can be calculated using
$\rm log\, \nu f_{\nu} = - P_1 \rm (log\,\nu - P_2 )^2 + P_3 $. When multi-epochs of X-ray observations are considered for the spectra of Mkn 421, a log-parabolic law is found good for its spectra and anti-correlations between the  peak frequencies  and the curvatures show up (Massaro et al. 2004, Tramacere et al. 2007, 2009). Similar result was  presented for
 Mkn 501 (Massaro et al. 2006). Tramacere et al. (2007, 2009)  pointed out that the observed anti-correlation between the peak frequency  and the curvature  in the synchrotron SED  results from  a stochastic component in the acceleration process (see also Tramacere et al. (2011) for further physics consideration and numerical approach, and application to more BL Lacs).
Our statistical results confirm the anti-correlation (Massaro et al. 2004, Tramacere et al. 2007, 2009), and this kind of anti-correlation  is a strong signature of a stochastic component in the acceleration (Tramacere et al. 2011).

For the integrated luminosity and spectral curvature, there is a positive correlation for the whole sample, but such correlation does not exist
for FSRQs or BL Lacs. See Fig. \ref{P1-vpLb}a and Table \ref{T3} for details.

 For ${\rm {log}}\,\nu_{\rm p}$ vs $\alpha_{\rm RO}$ (and $\alpha_{\rm
OX}$ ), there are anti-correlations between them for the whole sample (Fig. \ref{vp-aRO-OX}a,b). It can be seen  that the sources follow the blazar sequence with LSP FSRQs occupying  the top left corner and HSP BLs at the bottom right corner, while ISP objects in the middle between LSPs and HSPs.
 From Fig. 8 by Fossati et al. (2008), it can be seen  that there is also an anti-correlation between  ${\rm
  {log}}\,\nu_{\rm p}$ and $\alpha_{\rm RO}$ for the whole sample. Our result is consistent with  that reported.
 But if we consider the sub-samples separately,  an anti-correlation  for  BL Lacs is found but there is a marginally positive correlation for FSRQs  between  ${\rm {log}}\,\nu_{\rm p}$ and $\alpha_{\rm OX}$ (See Table \ref{T3} and Fig. \ref{vp-aRO-OX}a).
For  HBLs,   an anti-correlation between  ${\rm {log}}\,\nu_{\rm p}$ and $\alpha_{\rm OX}$ is found but there is no correlation  between  ${\rm {log}}\,\nu_{\rm p}$ and  $\alpha_{\rm RO}$ (See Table \ref{T3}).
  For HBLs, when their $\nu_p$ moves to the higher frequency, their  radio and optical emissions are from synchrotron emission following the same power law, thus the ratio (i.e. $\alpha_{RO}$) of radio to optical emissions is fixed so that there is no correlation between $\alpha_{RO}$ and $\rm log \nu_p$. However, when their $\nu_p$ moves to the higher frequency, X-ray emissions increase and optical emissions decrease, so the ratio (i.e. $\alpha_{OX}$) of optical to the X-ray emissions will become smaller and that results in an anti-correlation for $\alpha_{OX}$ against
  $\rm log \nu_p$.
  For low peak frequency FSRQs, when their peak moves to the lower frequency, their optical emissions will become weaker, but their X-rays are dominated by inverse Compton emission, thus the ratio (i.e. $\alpha_{OX}$) of optical to X-rays will become smaller and that
   results in a positive correlation between $\alpha_{OX}$ and $\rm log \nu_p$. For LBLs, there are only 28 sources, linear regression analysis gives $r = $ 0.26 and $p = 18.7\%$ for $\alpha_{OX}$ against $\rm log \nu_p$. It is similar to the case of FSRQs.

For $\alpha_{\rm RO}$ vs $\alpha_{\rm OX}$ (Fig. \ref{aRO-OX}): Although
there are substantial scattering  in Fig. \ref{aRO-OX}, it is clear that HBLs and
FSRQs occupy different regions in the panel with IBLs and LBLs
being bridge between them. There are  anti-correlations between
$\alpha_{\rm RO}$ and $\alpha_{\rm OX}$ for FSRQs, HBLs and LBLs
with correlation coefficients being
$r\,=\,-0.61$ ($p\,<\, 10^{-4}$), $-0.54$ ($p\,<\, 10^{-4}$) and $-0.38$ ($p\,=\,4.4\%$) respectively.

 Abdo et al. (2010a) investigated the correlation between $\alpha_{\rm RO}$ and $\alpha_{\rm OX}$ for Fermi blazars and found that Fermi FSRQs are located along the top-left/bottom-right band, while Fermi BL Lacs are in all parts of the plane. From Fig. 27 as in the reference, it can be seen  that the HBLs also show a top-left/bottom-right tendency. Our anti-correlation results are  consistent with  those reported.  In Fig. 7
 by Ackermann et al. (2011b), it can be clearly seen that there are anti-correlations for LSP, ISP and HSP subclasses, and that LSP and HSP occupy different regions with ISP being the bridge connecting them. Our results are also consistent with those reported
  (Abdo et al. 2010a, Ackermann et al. 2011b).

\section{Conclusions}

In this paper, we compile multi-wavelength data for  a sample of 1425 Fermi blazars to calculate their SEDs, and successfully achieve  SEDs for 1392 blazars (461  FSRQs, 620  BLs and 311 BCUs).  999 objects (463 BLs, 461 FSRQs, and 75 BCUs) out of the 1392 blazars have available redshift. Synchrotron peak frequency ($\rm log \nu_p$), spectral curvature ($\rm P_1$),  peak flux ($\rm \nu_p F_{\nu_p}$), and integrated flux ($\rm \nu F_{\nu}$)  are obtained by fitting SED using a parabola function, ${\rm log}(\nu F_{\nu}) = P_1({\rm log}\nu - P_2)^2 + P_3$.  Monochromatic
luminosity at radio 1.4 GHz, optical R band, X-ray at 1 keV and
$\gamma$-ray at 1 GeV, peak luminosity, integrated luminosity and effective spectral indexes of radio to
optical ($\alpha_{\rm RO}$), and optical to X-ray ($\alpha_{\rm OX}$) are  calculated. We adopted the "Bayesian classification" to  log$\nu_{\rm p}$ in the rest frame for 999 blazars with available redshift and the result shows that 3 components are the best to fit the log$\nu_{\rm p}$ distribution. Based on the proposed method we classified the subclasses of blazars using the acronyms of Abdo et al. (2010a). Some mutual correlations are also studied.  Conclusions are drawn as follows:

 1. Based on the "Bayesian classification", blazars can be classified into 3 subclasses and there is no ultra high peaked subclass.
 According to the acronyms of Abdo et al. (2010a), blazars can be classified as follows:
 LSPs: ${\rm log \ \nu_p (Hz)} \leq 14.0$,
 ISPs: 14.0 Hz $ \leq {\rm log \ \nu_p (Hz)} \leq 15.3$, and
 HSPs:  ${\rm log \ \nu_p (Hz)} \geq 15.3$.
  When averaged redshifts are adopted to BLs and UCBs without redshifts,  we have
 9 HSPs, 180 ISPs and 272 LSP for 461 FSRQs;
 235 HSPs, 271 ISPs, and 114 LSPs for 620 BLs; and
 106 HSPs, 107 ISPs, and 98 LSPs for 311 BCUs for the 1392 sources.

2.  The $\gamma$-ray emissions are closely correlated with radio
 emissions, due to the facts that  radio and $\gamma$-ray emissions are both beamed and that  radio and $\gamma$-ray emissions are co-spatially produced in the jet.
 For BL Lacs, the fact that the emissions for the whole electromagnetic wavelength are from SSC  results
  in the correlation between $\gamma$-ray and radio emissions, while for FSRQs, their $\gamma$-rays are from SSC and EC with the EC making main contribution, their radio emissions are strongly beamed, therefore, there is an apparent correlation between $\gamma$-rays and radio emissions.
 $\gamma$-ray luminosity is also correlated with both synchrotron peak luminosity and integrated luminosity.

 3.  The spectral curvature is anti-correlated with the synchrotron peak frequency, which suggests a stochastic  acceleration.

4.   For synchrotron peak frequency and peak luminosity, we found an anti-correlation for the whole sample but a positive correlation for HBLs.
The anti-correlation is attributed to the fact that the lowly peaked sources are strongly beamed, and the highly peaked sources are weakly beamed. For  HBLs, their boosting effects are weak, the peak frequency and peak luminosity correlation is intrinsic. If  more  particles are accelerated to higher energy, one can expect more luminous luminosity in higher frequency so that there is  a positive correlation between the peak frequency and the peak luminosity.

5.    There is  a positive dependence of optical to X-ray spectral index ($\alpha_{ox}$) on peak frequency for FSRQs and there is no correlation between radio to optical index ($\alpha_{ro}$) and peak frequency for HBLs.
For FSRQs, when its peak  moves to the lower frequency side, the optical emissions decrease and the X-ray emissions will increase because of  more contribution from inverse Compton emission and EC emissions, therefore the ratio of optical to X-ray emissions ($\alpha_{ox}$) decreases and a positive tendency can be expected.
For HBLs, when the peak moves to the higher frequency side, the radio and optical emissions follow the same power law,  therefore the ratio of radio to optical emissions ($\alpha_{ro}$) is a constant and  there is no correlation between peak frequency  (log $\nu_{\rm p}$) and effective spectral index ($\alpha_{\rm RO}$).

\begin{acknowledgements}

 We thank the anonymous referee for the constructive and useful comments and suggestions, which have helped us to improve the manuscript.
This work is supported by the National Natural Science Foundation of China (U1531245, U1431112, 11203007, 11403006, 10633010, 11173009), and the Innovation Foundation of Guangzhou University (IFGZ),
Guangdong Province Universities and Colleges Pearl River Scholar
Funded Scheme(GDUPS)(2009), Yangcheng Scholar Funded
Scheme(10A027S), and supported for Astrophysics  Key Subjects of Guangdong Province and Guangzhou City.
\end{acknowledgements}

\begin{deluxetable}{ccccccccccccc}
\tabletypesize{\scriptsize}
 \rotate
  \tablecaption{Sample for blazars}
  \tablewidth{0pt}
 \tablehead{
  \colhead{3FGL name} &
   \colhead{$z$} &
  \colhead{C} &
  \colhead{$L_{\rm R}/\sigma_{L_{\rm R}}$} &
  \colhead{$L_{\rm O}/\sigma_{L_{\rm O}}$} &
  \colhead{$L_{\rm X}/\sigma_{L_{\rm X}}$} &
  \colhead{$L_\gamma/\sigma_{L_{\rm \gamma}}$} &
  \colhead{$\alpha _{\rm RO}/\sigma_{\alpha}$} &
  \colhead{$\alpha _{\rm OX}/\sigma_{\alpha}$} &
\colhead{$P_1/\sigma_{P_1}$} &
  \colhead{$\nu_{\rm p}/\sigma_{\nu_{\rm p}}$} &
  \colhead{$L_{\rm p}/\sigma_{L_{\rm p}}$} &
  \colhead{$L_{\rm bol}/\sigma_{L_{\rm bol}}$}
  }
 \startdata
(1)&(2)&(3)&(4)&(5)&(6)&(7)&(8)&(9)&(10)&(11)&(12)&(13)\\
\hline
J0001.2-0748    &       &   IB &   42.36/0.01    &   45.39/0.02    &               &   45.23/0.06    &   0.45/0.01    &               &      -0.12/0.01       &   14.37/0.12    &   45.35/0.03    &   45.71/0.05    \\
J0001.4+2120    &   1.106   &   HF  &   42.97/0.01    &               &               &   45.70/0.11    &               &               &      -0.05/0.00       &   16.79/0.28    &   45.70/0.03    &   46.32/0.04    \\
J0003.2-5246    &       &   HU  &               &               &   45.13/0.07    &   44.56/0.11    &               &               &      -0.05/0.01       &   17.89/0.81    &   45.15/0.14    &   45.76/0.14    \\
J0003.8-1151    &   1.310   &   LU  &   43.44/0.01    &   45.54/0.04    &               &   45.59/0.12    &   0.62/0.01    &               &      -0.12/0.01       &   13.06/0.14    &   45.57/0.11    &   46.01/0.15    \\
J0004.7-4740    &   0.880   &   IF  &               &   46.38/0.04    &   44.98/0.07    &   45.86/0.05    &               &   1.52/0.04    &      -0.12/0.01       &   14.14/0.09    &   46.20/0.06    &   46.59/0.09    \\
J0006.4+3825    &   0.229   &   IF  &   41.98/0.01    &   44.53/0.04    &   43.44/0.07    &   44.41/0.06    &   0.54/0.01    &   1.40/0.04    &      -0.11/0.01       &   14.03/0.12    &   44.65/0.10    &   45.08/0.14    \\
J0008.0+4713    &   0.280   &   IB &   41.18/0.01    &               &   43.51/0.07    &   44.87/0.03    &               &               &      -0.12/0.00       &   14.52/0.07    &   44.46/0.04    &   44.83/0.06    \\
\enddata
\label{T1-samp-all}
\end{deluxetable}

\begin{figure}
    \centering
    \resizebox{\hsize}{!}{\includegraphics*{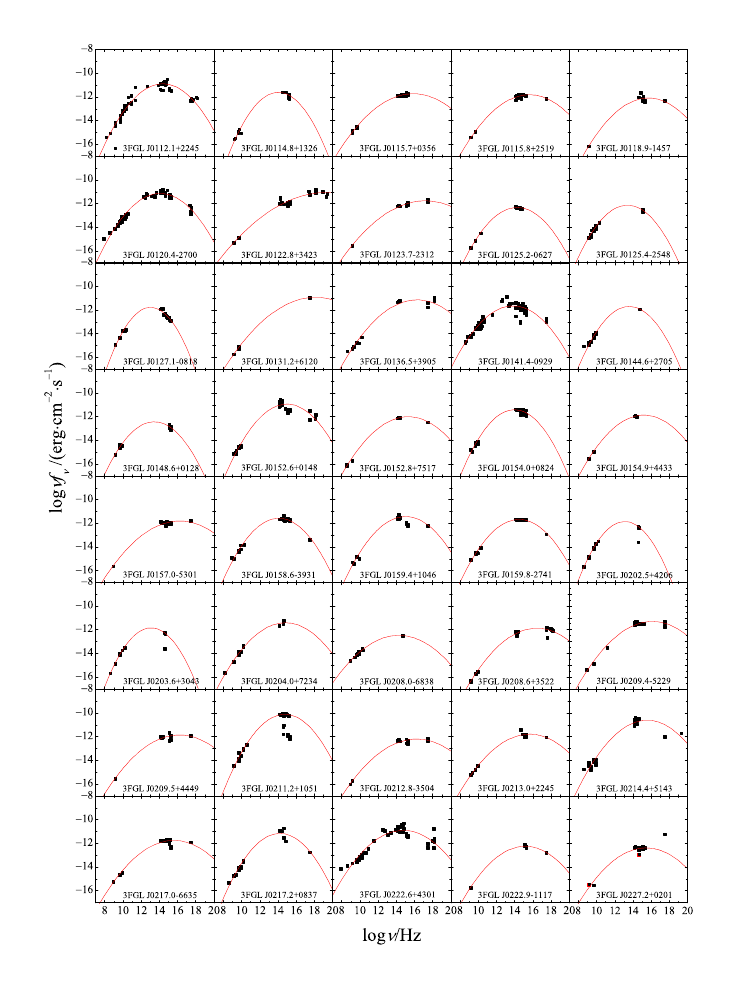}}
    \caption{Appendix: SED figures for BL Lacs.}
    \label{Fan-SED-Fig-B041-080}
\end{figure}

\begin{figure}
    \centering
    \resizebox{\hsize}{!}{\includegraphics*{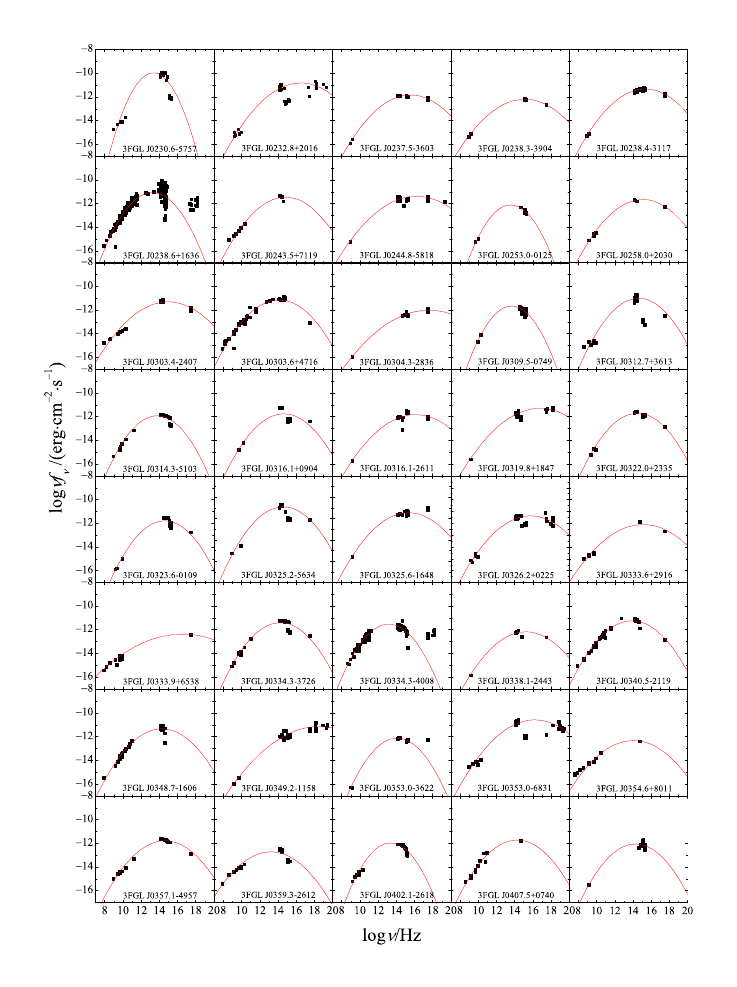}}
    \caption{Appendix: SED figures for BL Lacs.}
    \label{Fan-SED-Fig-B081-120}
\end{figure}

\begin{figure}
    \centering
    \resizebox{\hsize}{!}{\includegraphics*{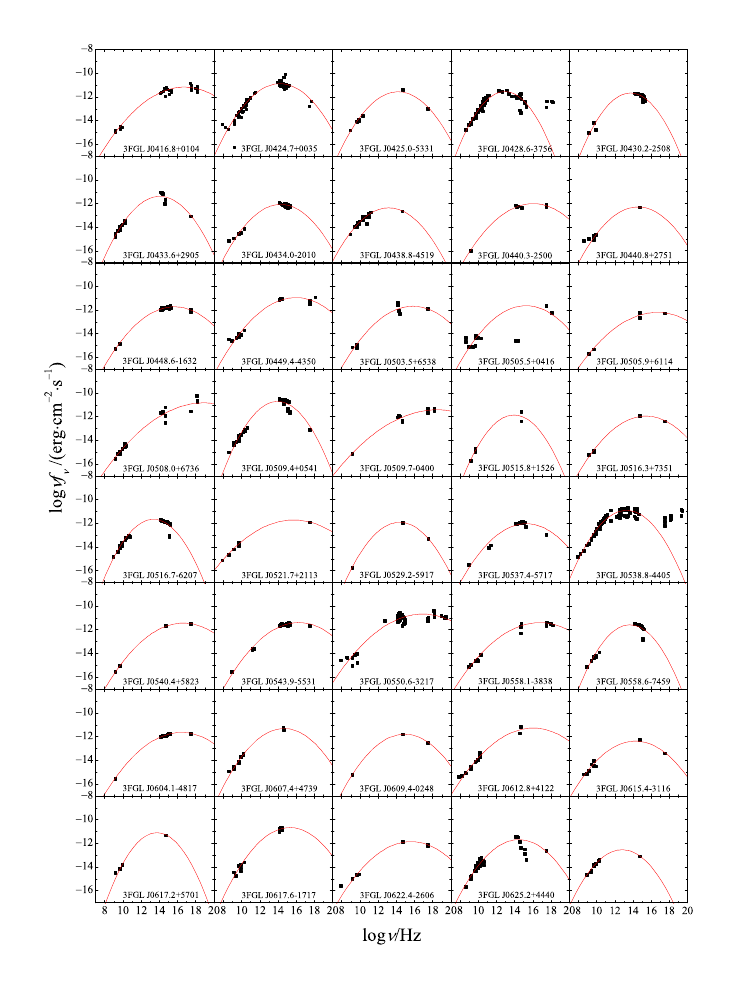}}
    \caption{Appendix: SED figures for BL Lacs.}
    \label{Fan-SED-Fig-B121-160}
\end{figure}

\begin{figure}
    \centering
    \resizebox{\hsize}{!}{\includegraphics*{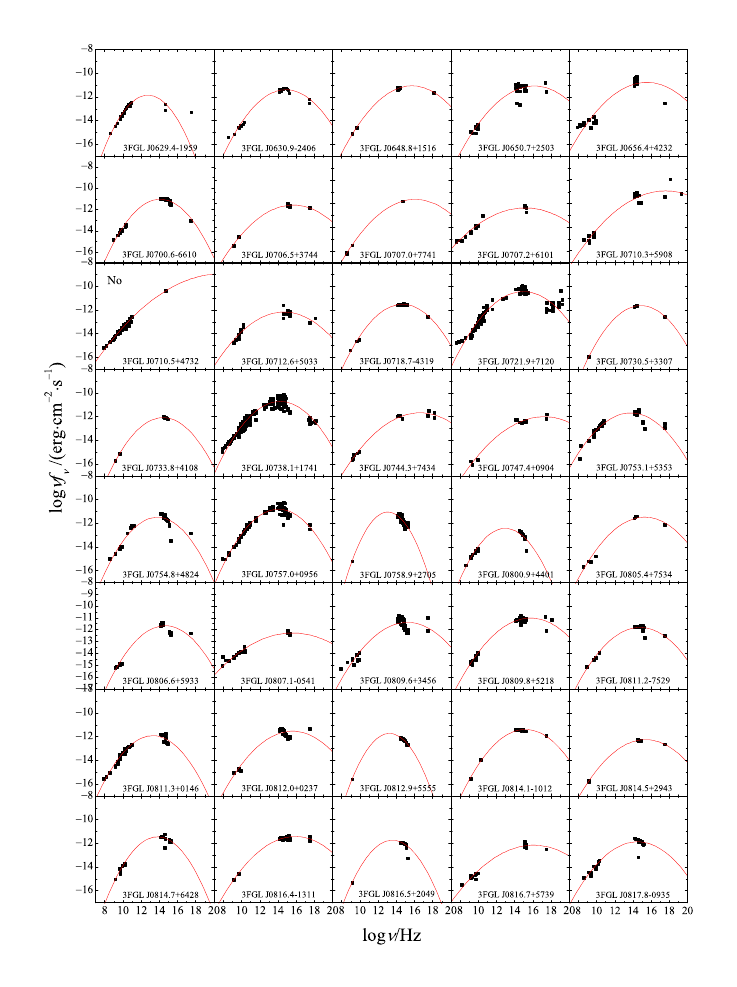}}
    \caption{Appendix: SED figures for BL Lacs.}
    \label{Fan-SED-Fig-B161-200}
\end{figure}

\begin{figure}
    \centering
    \resizebox{\hsize}{!}{\includegraphics*{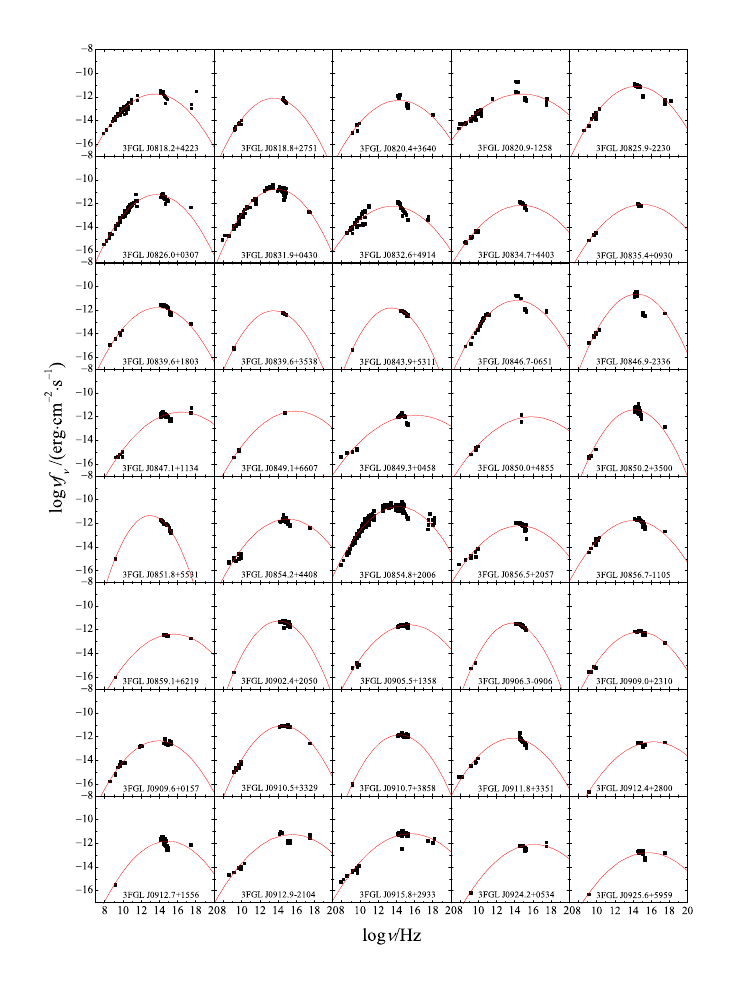}}
     \caption{Appendix: SED figures for BL Lacs.}
    \label{Fan-SED-Fig-B201-240}
\end{figure}

\begin{figure}
    \centering
    \resizebox{\hsize}{!}{\includegraphics*{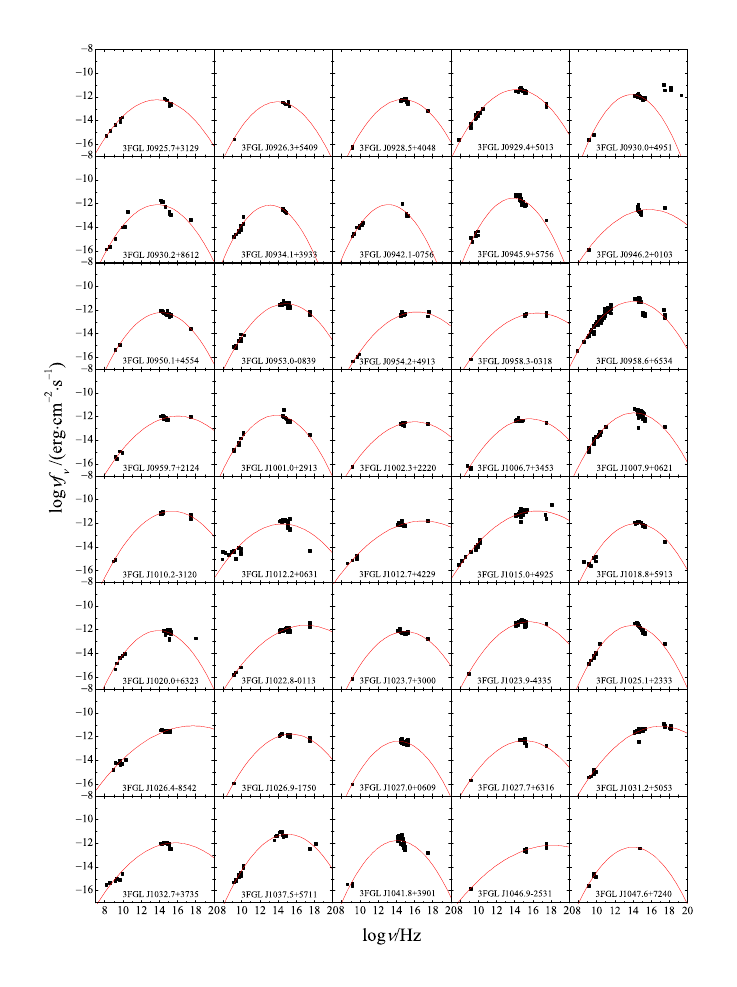}}
        \caption{Appendix: SED figures for BL Lacs.}
    \label{Fan-SED-Fig-B241-280}
\end{figure}

\begin{figure}
    \centering
    \resizebox{\hsize}{!}{\includegraphics*{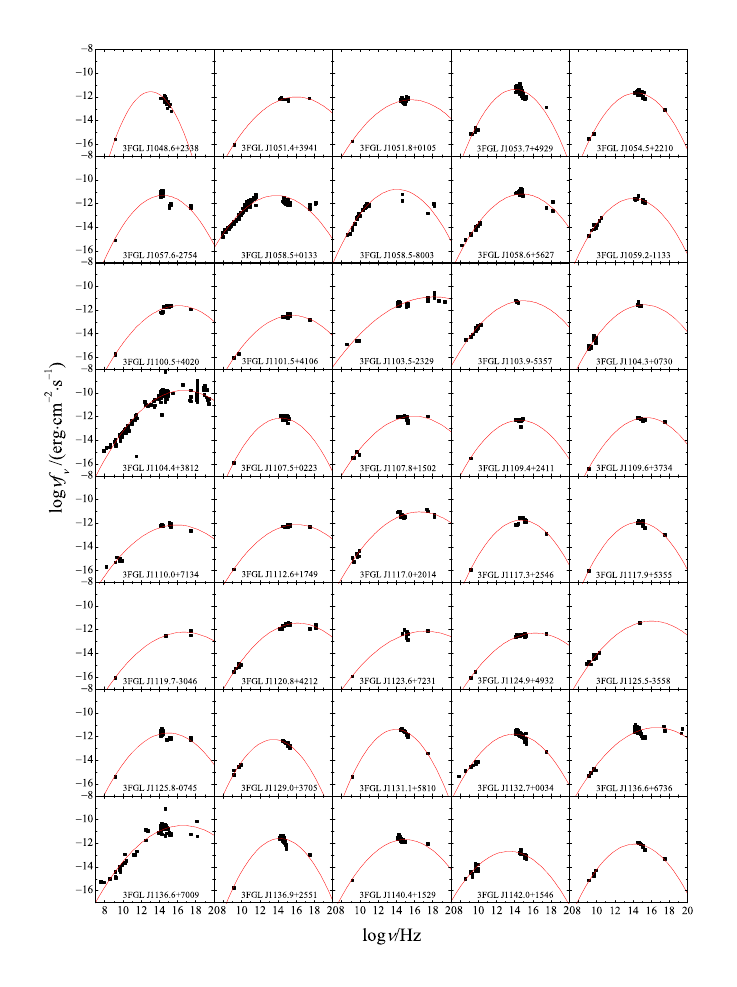}}
    \caption{Appendix: SED figures for BL Lacs.}
    \label{Fan-SED-Fig-B281-320}
\end{figure}

\begin{figure}
    \centering
    \resizebox{\hsize}{!}{\includegraphics*{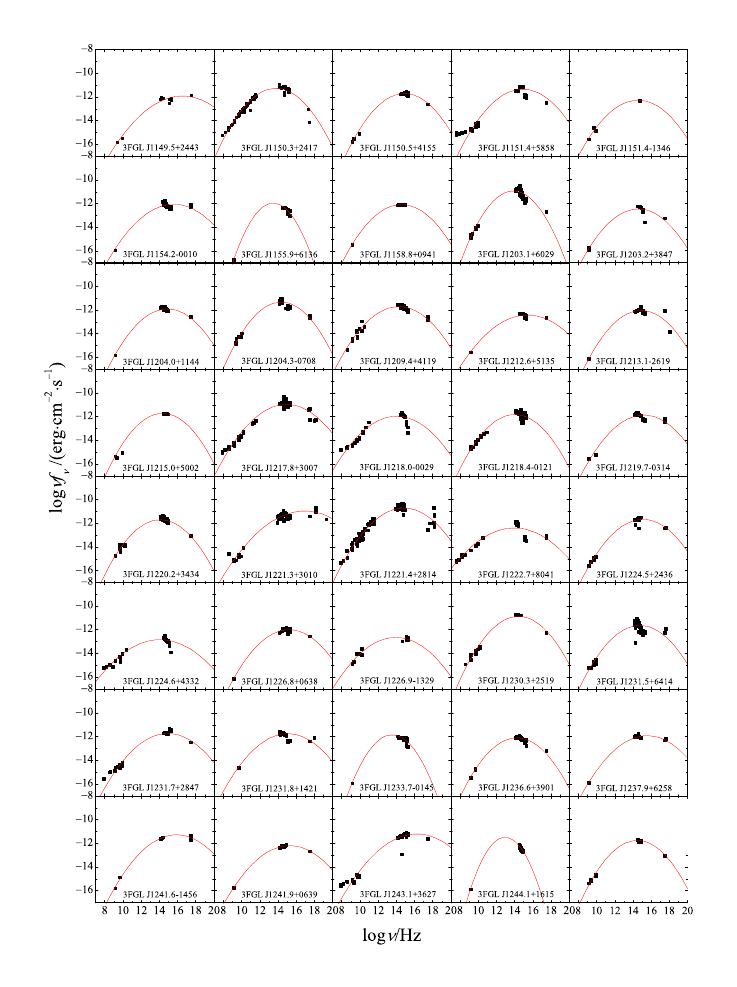}}
    \caption{Appendix: SED figures for BL Lacs.}
        \label{Fan-SED-Fig-B321-360}
\end{figure}

\begin{figure}
    \centering
    \resizebox{\hsize}{!}{\includegraphics*{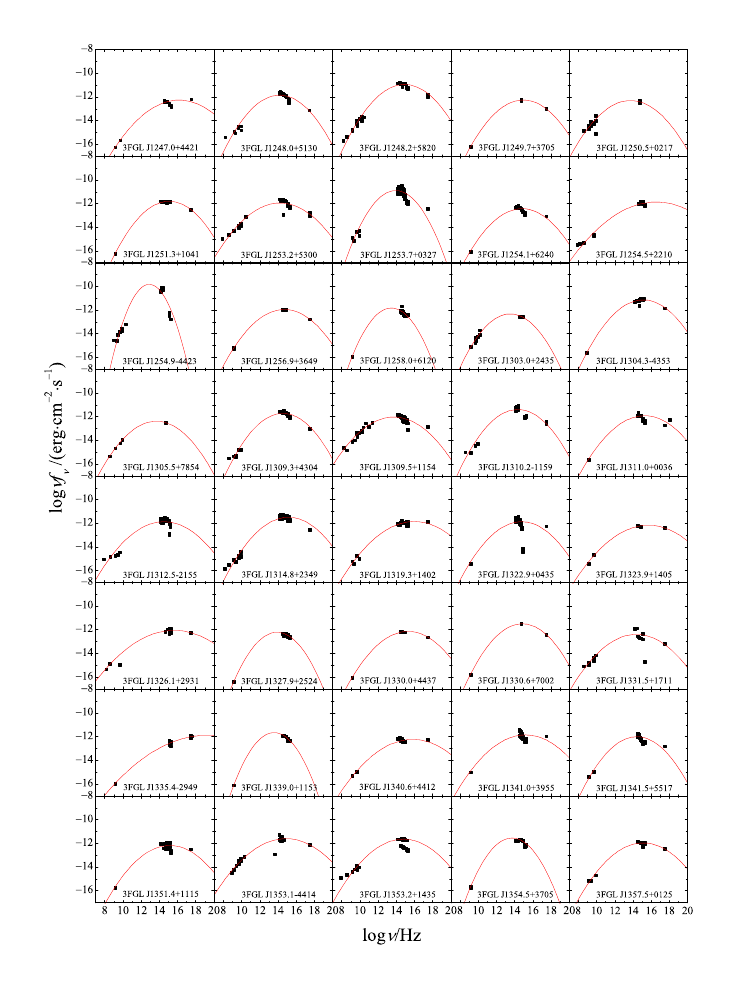}}
    \caption{Appendix: SED figures for BL Lacs.}
        \label{Fan-SED-Fig-B361-400}
\end{figure}

\begin{figure}
    \centering
    \resizebox{\hsize}{!}{\includegraphics*{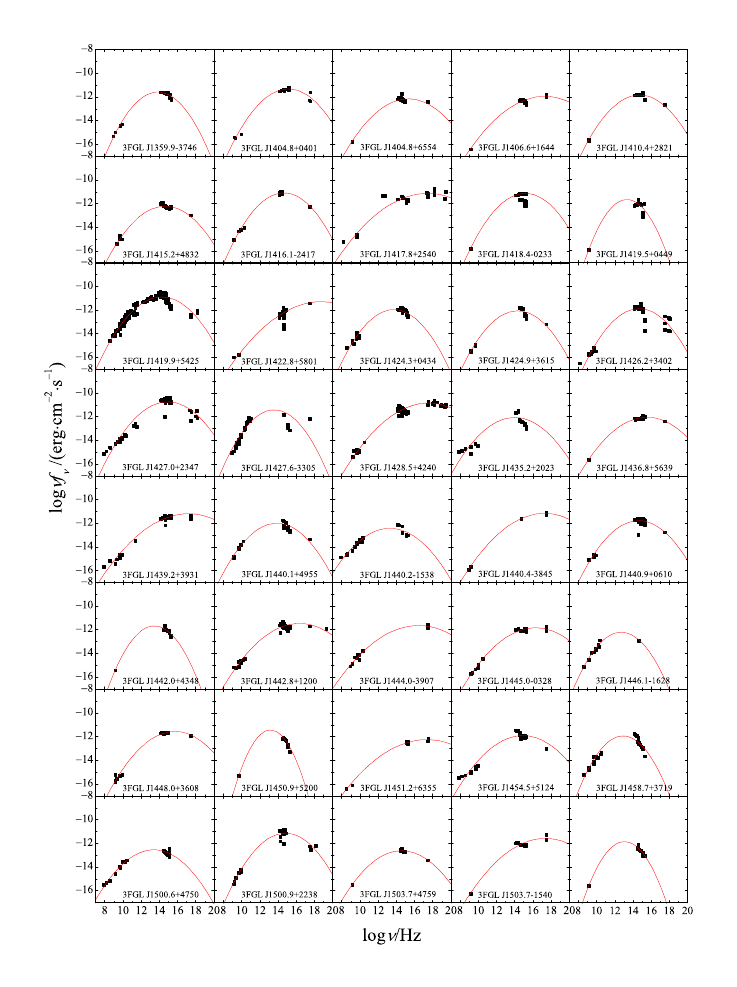}}
    \caption{Appendix: SED figures for BL Lacs.}
        \label{Fan-SED-Fig-B401-440}
\end{figure}

\begin{figure}
    \centering
    \resizebox{\hsize}{!}{\includegraphics*{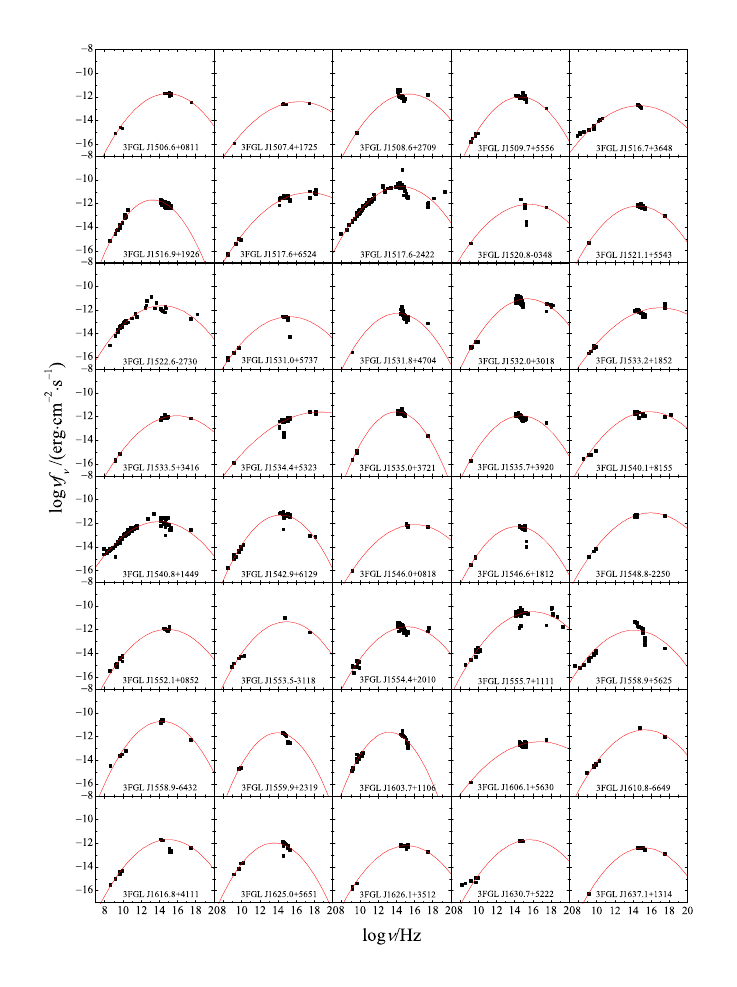}}
    \caption{Appendix: SED figures for BL Lacs.}
        \label{Fan-SED-Fig-B441-480}
\end{figure}

\begin{figure}
    \centering
    \resizebox{\hsize}{!}{\includegraphics*{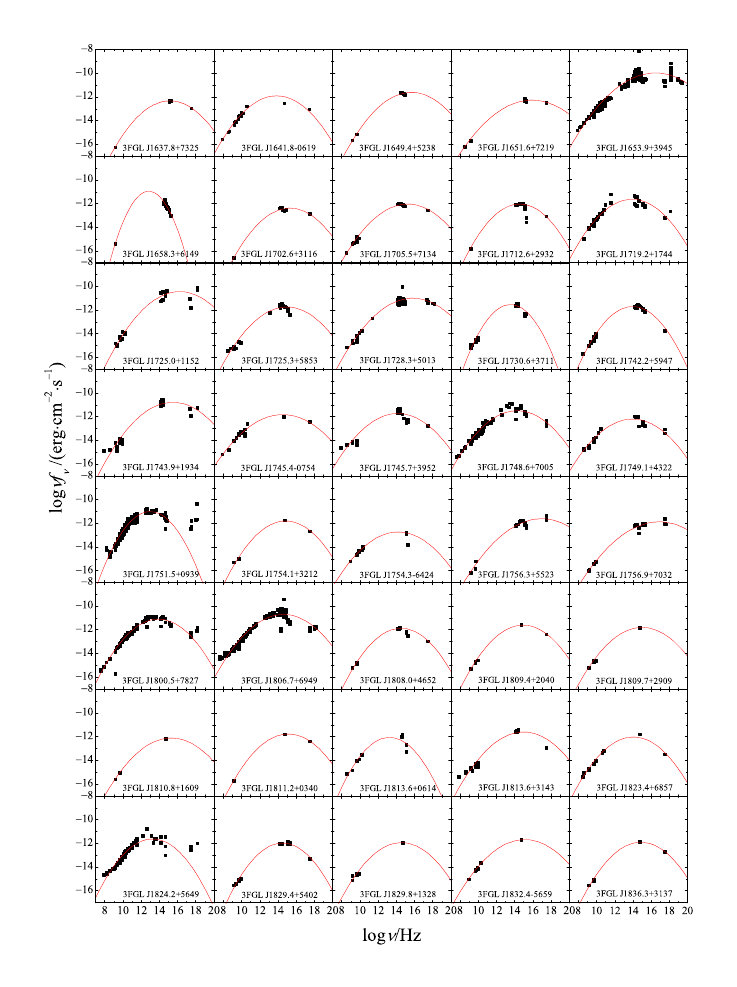}}
    \caption{Appendix: SED figures for BL Lacs.}
        \label{Fan-SED-Fig-B481-520}
\end{figure}

\begin{figure}
    \centering
    \resizebox{\hsize}{!}{\includegraphics*{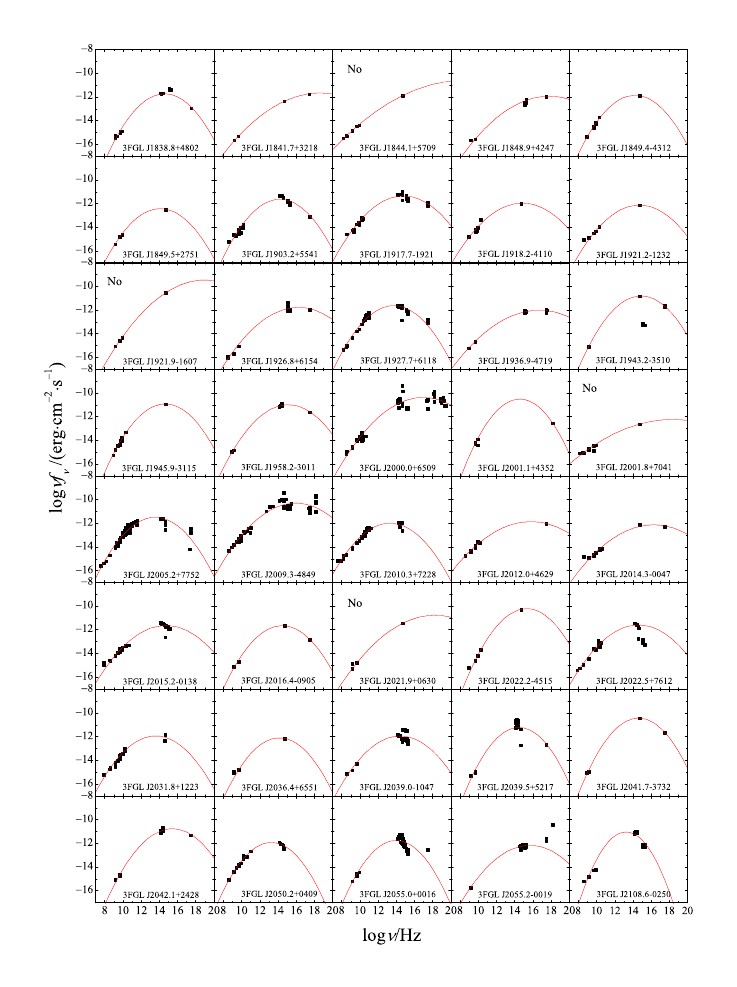}}
    \caption{Appendix: SED figures for BL Lacs.}
        \label{Fan-SED-Fig-B521-560}
\end{figure}

\begin{figure}
    \centering
    \resizebox{\hsize}{!}{\includegraphics*{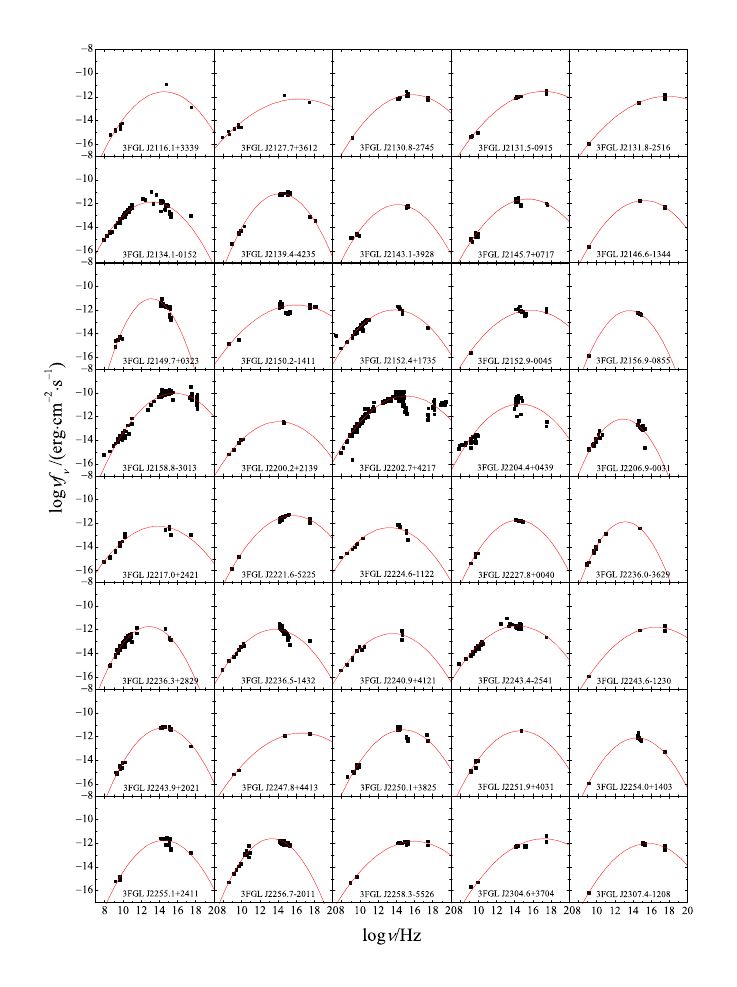}}
    \caption{Appendix: SED figures for BL Lacs.}
        \label{Fan-SED-Fig-B561-600}
\end{figure}

\begin{figure}
    \centering
    \resizebox{\hsize}{!}{\includegraphics*{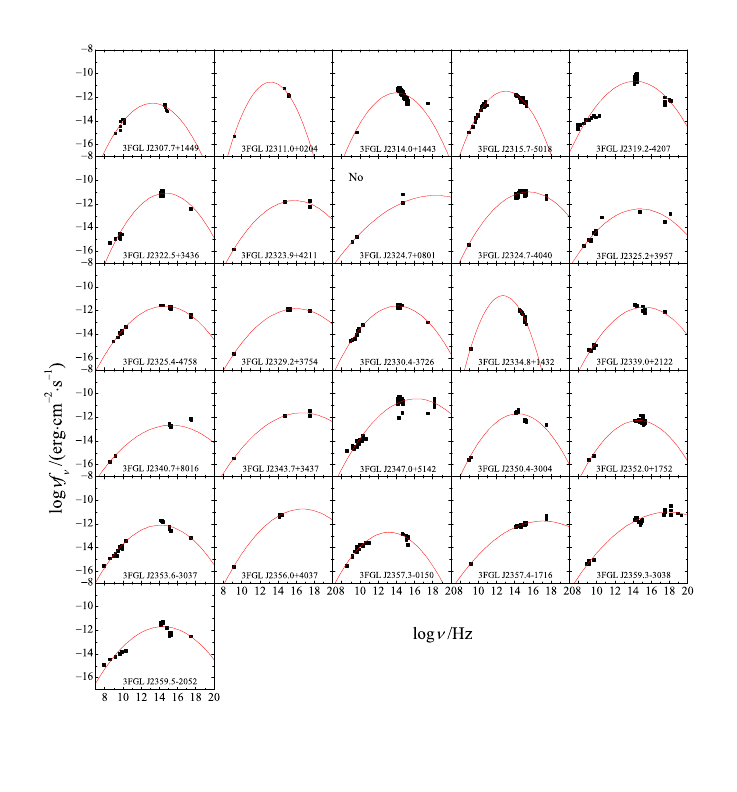}}
    \caption{Appendix: SED figures for BL Lacs.}
        \label{Fan-SED-Fig-B601-626}
\end{figure}


\begin{figure}
    \centering
    \resizebox{\hsize}{!}{\includegraphics*{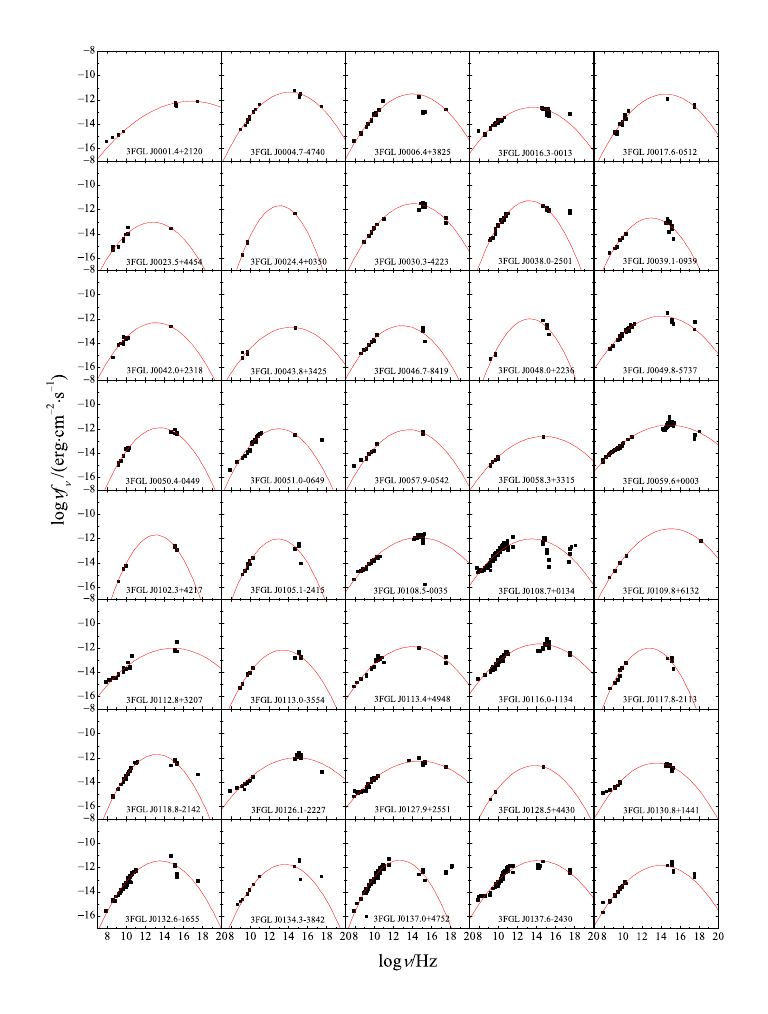}}
    \caption{Appendix: SED figures for FSRQs.}
    \label{Fan-SED-Fig-F001-040}
\end{figure}

\begin{figure}
    \centering
    \resizebox{\hsize}{!}{\includegraphics*{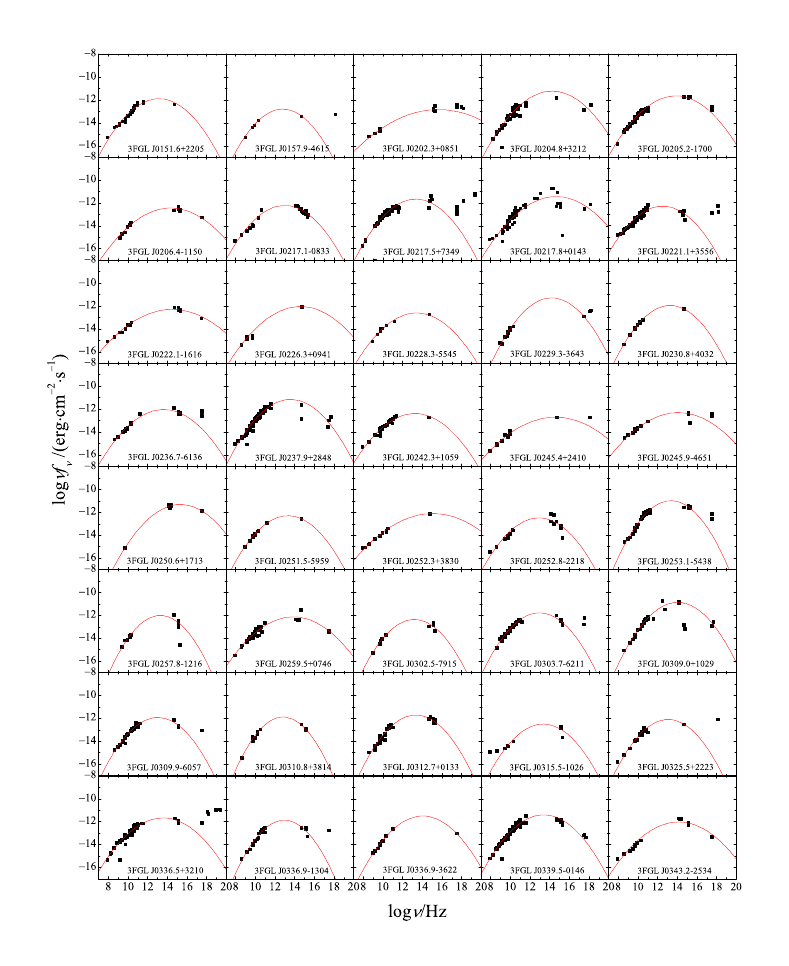}}
    \caption{Appendix: SED figures for FSRQs.}
    \label{Fan-SED-Fig-F041-080}
\end{figure}

\clearpage
\begin{figure}
    \centering
    \resizebox{\hsize}{!}{\includegraphics*{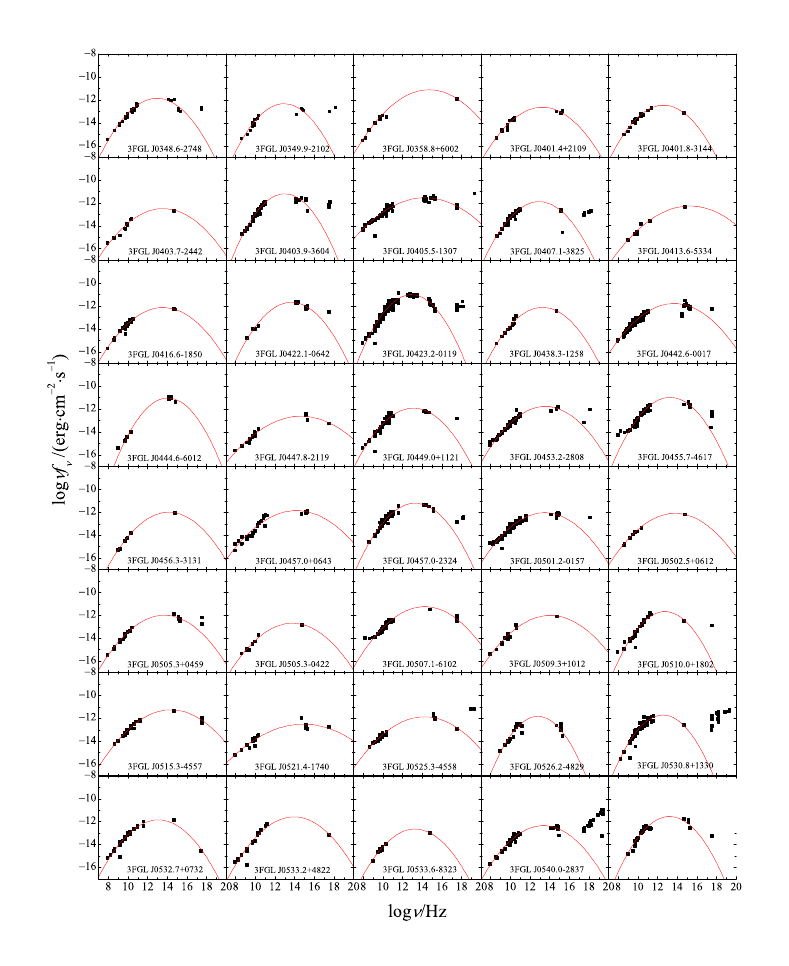}}
    \caption{Appendix: SED figures for FSRQs.}
    \label{Fan-SED-Fig-F081-120}
\end{figure}

\begin{figure}
    \centering
    \resizebox{\hsize}{!}{\includegraphics*{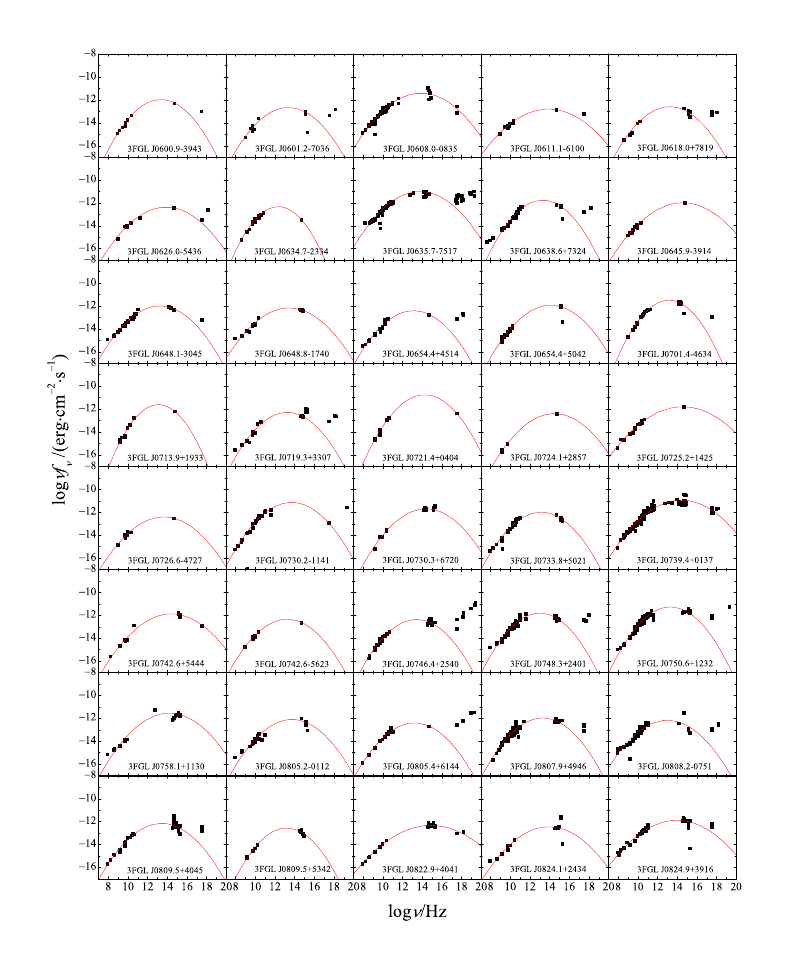}}
    \caption{Appendix: SED figures for FSRQs.}
    \label{Fan-SED-Fig-F121-160}
\end{figure}

\begin{figure}
    \centering
    \resizebox{\hsize}{!}{\includegraphics*{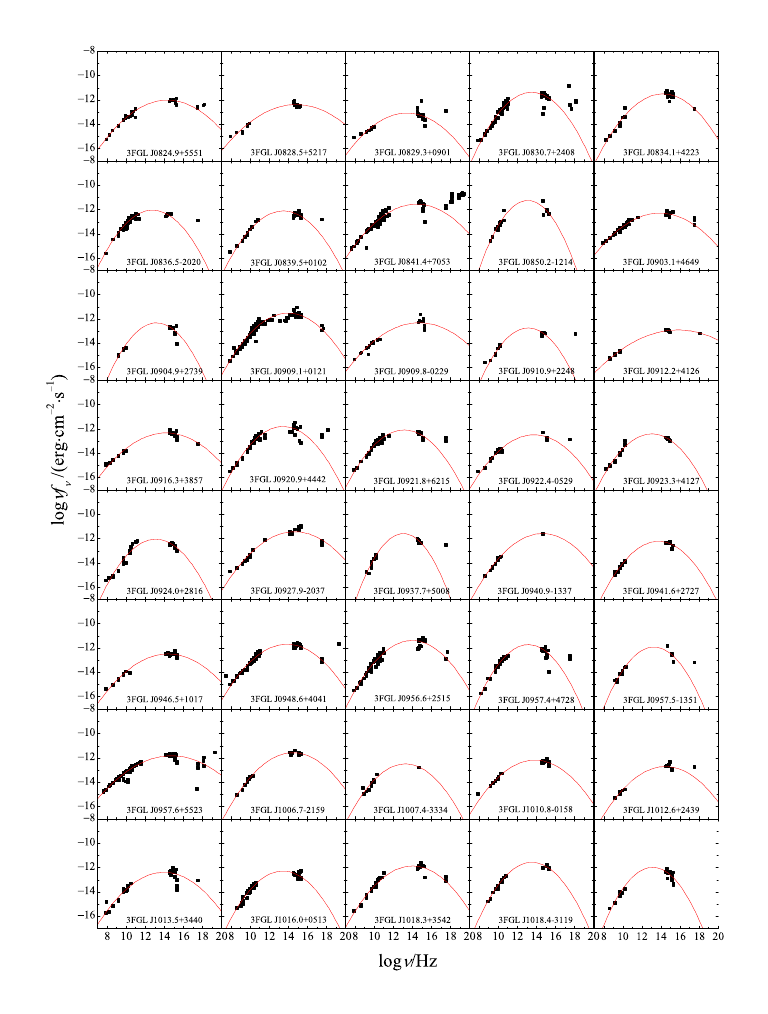}}
    \caption{Appendix: SED figures for FSRQs.}
    \label{Fan-SED-Fig-F161-200}
\end{figure}

\begin{figure}
    \centering
    \resizebox{\hsize}{!}{\includegraphics*{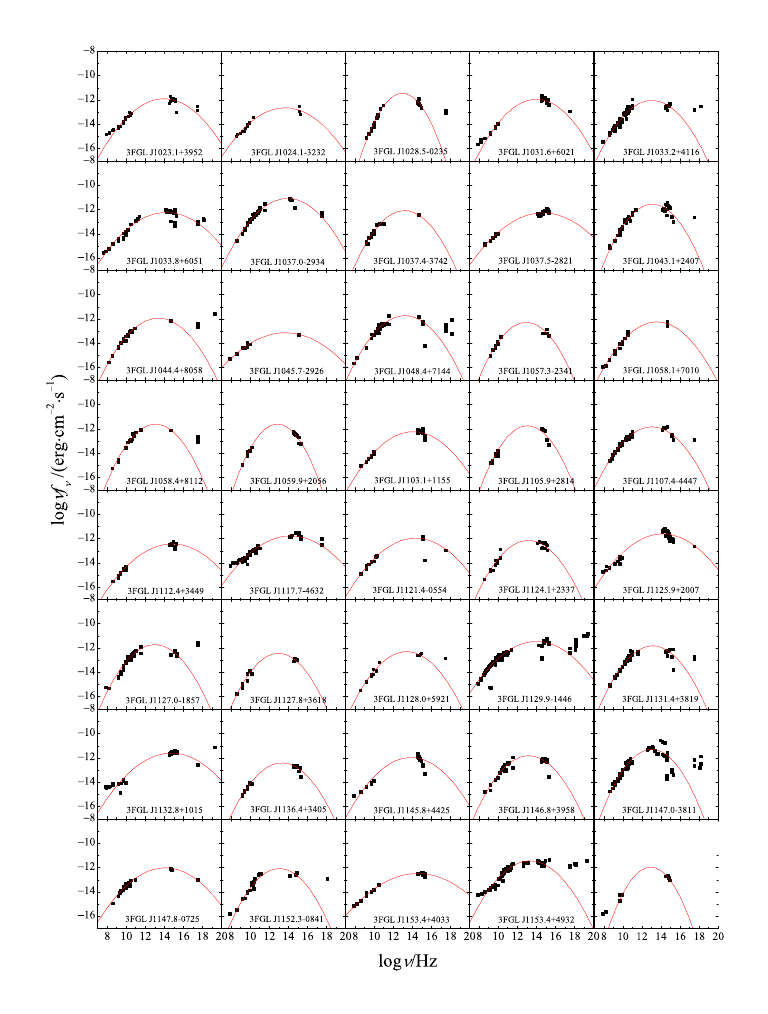}}
     \caption{Appendix: SED figures for FSRQs.}
    \label{Fan-SED-Fig-F201-240}
\end{figure}

\begin{figure}
    \centering
    \resizebox{\hsize}{!}{\includegraphics*{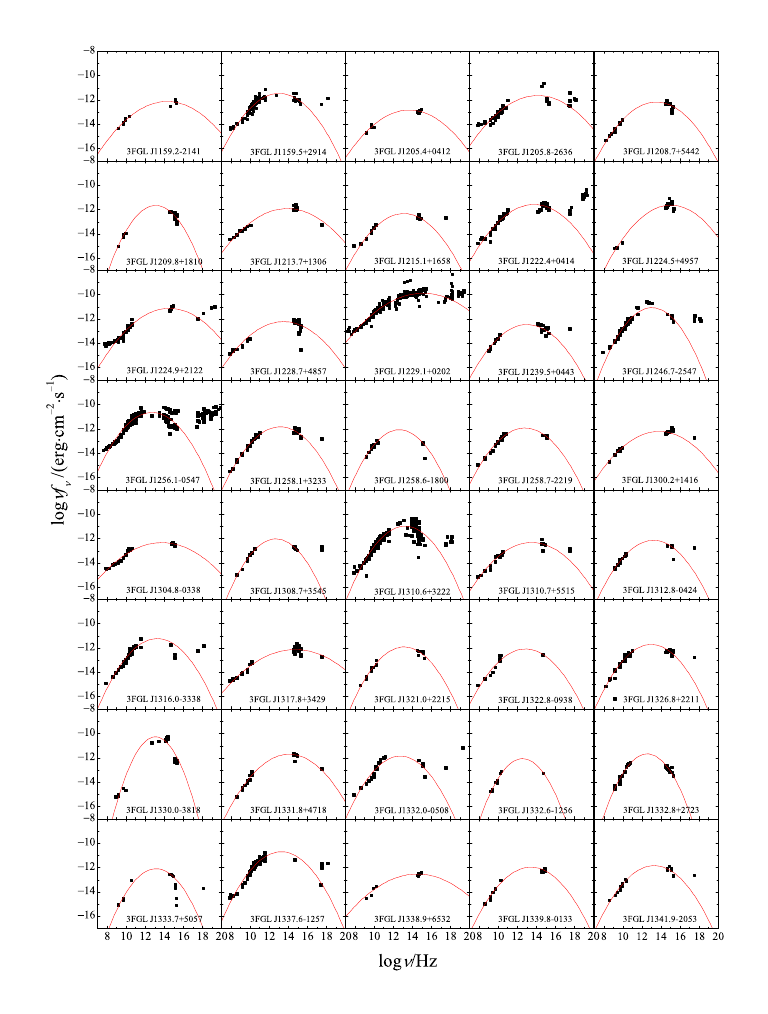}}
        \caption{Appendix: SED figures for FSRQs.}
    \label{Fan-SED-Fig-F241-280}
\end{figure}

\begin{figure}
    \centering
    \resizebox{\hsize}{!}{\includegraphics*{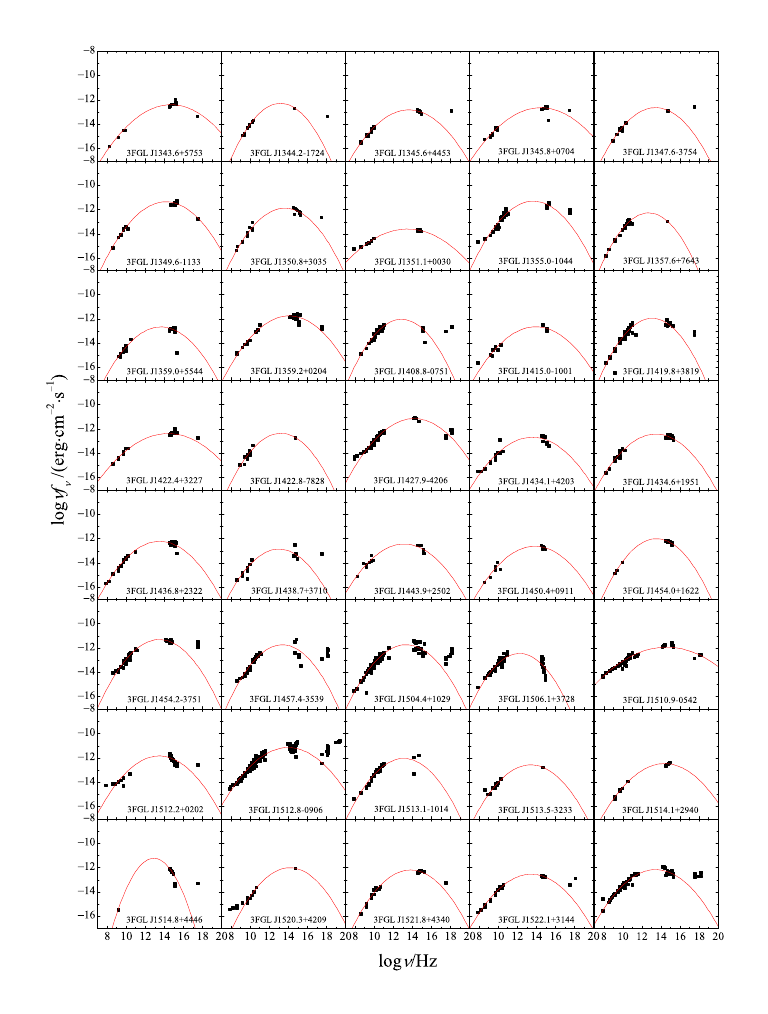}}
    \caption{Appendix: SED figures for FSRQs.}
    \label{Fan-SED-Fig-F281-320}
\end{figure}

\begin{figure}
    \centering
    \resizebox{\hsize}{!}{\includegraphics*{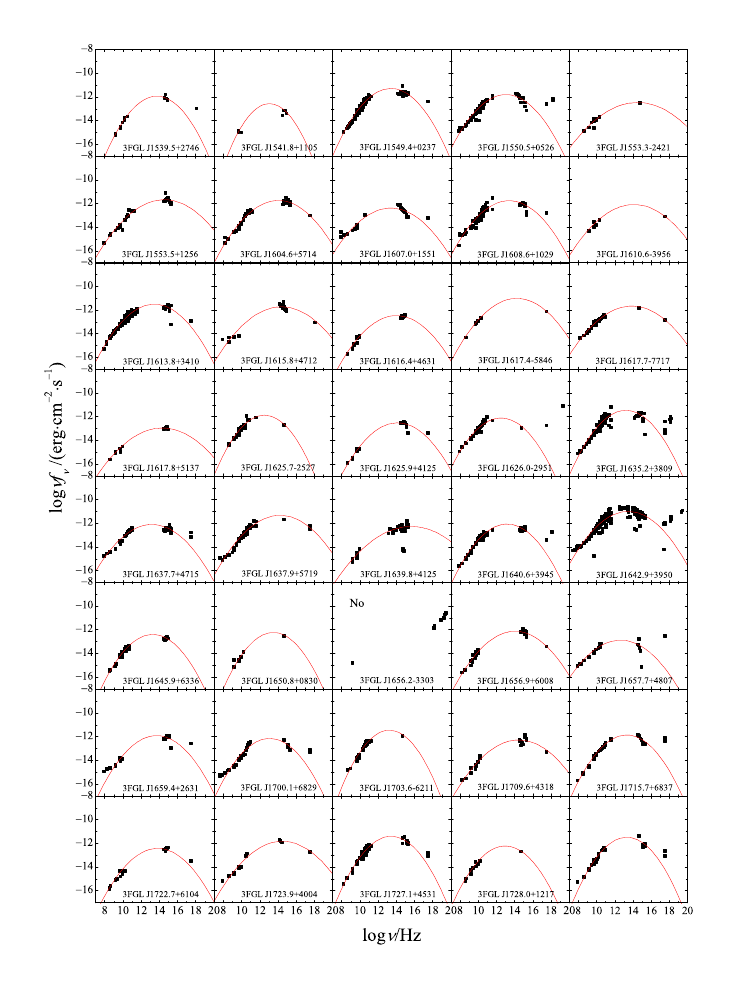}}
    \caption{Appendix: SED figures for FSRQs.}
        \label{Fan-SED-Fig-F321-360}
\end{figure}

\begin{figure}
    \centering
    \resizebox{\hsize}{!}{\includegraphics*{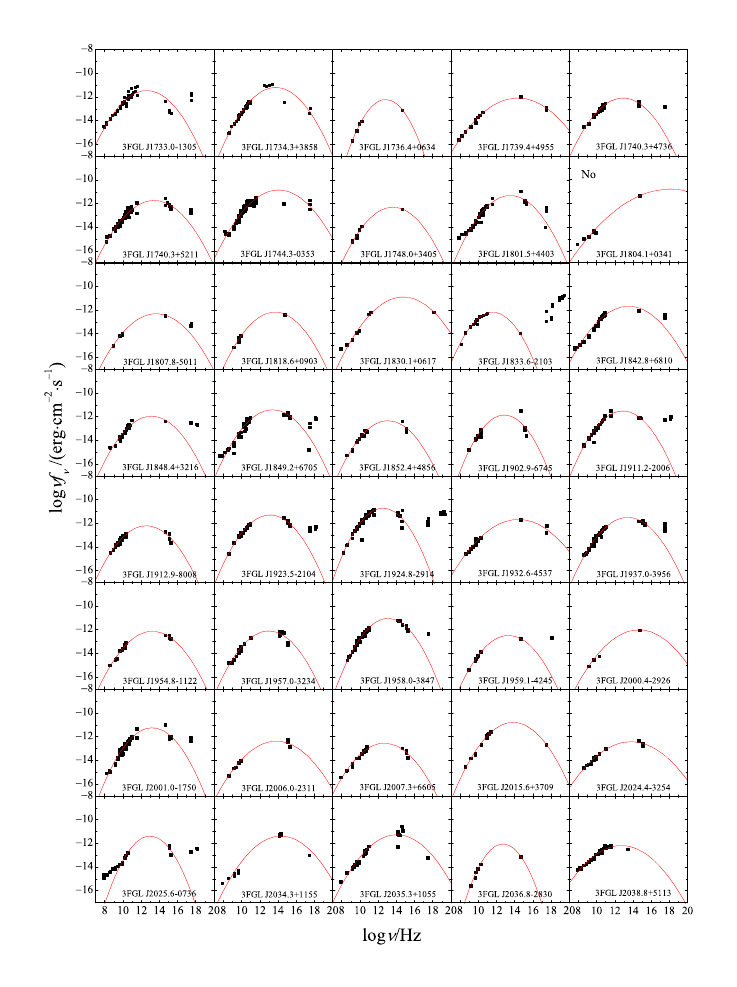}}
    \caption{Appendix: SED figures for FSRQs.}
        \label{Fan-SED-Fig-F361-400}
\end{figure}

\begin{figure}
    \centering
    \resizebox{\hsize}{!}{\includegraphics*{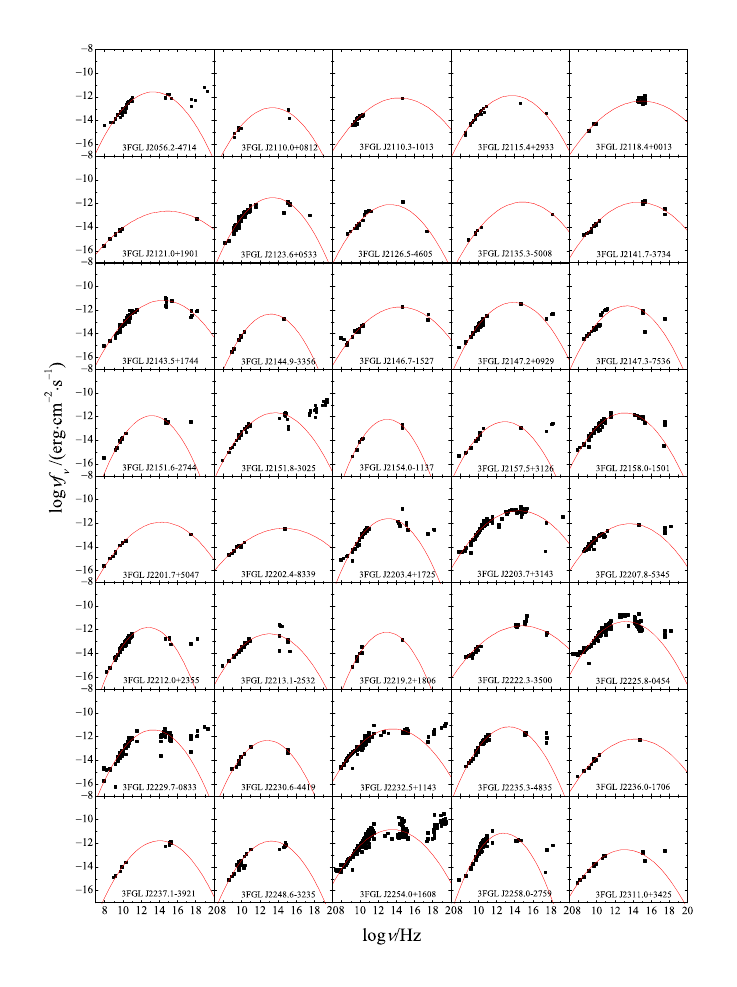}}
    \caption{Appendix: SED figures for FSRQs.}
        \label{Fan-SED-Fig-F401-440}
\end{figure}

\begin{figure}
    \centering
    \resizebox{\hsize}{!}{\includegraphics*{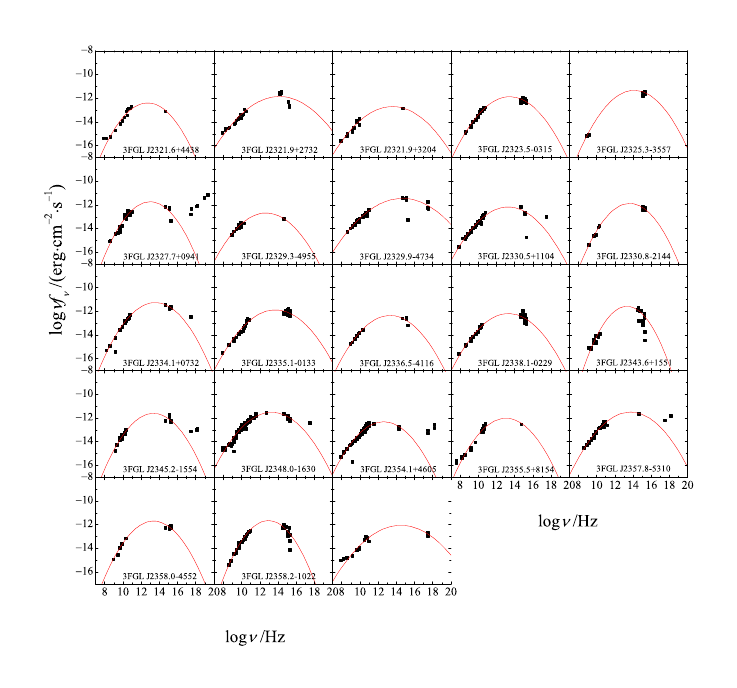}}
    \caption{Appendix: SED figures for FSRQs.}
        \label{Fan-SED-Fig-F441-463}
\end{figure}


\begin{figure}
    \centering
    \resizebox{\hsize}{!}{\includegraphics*{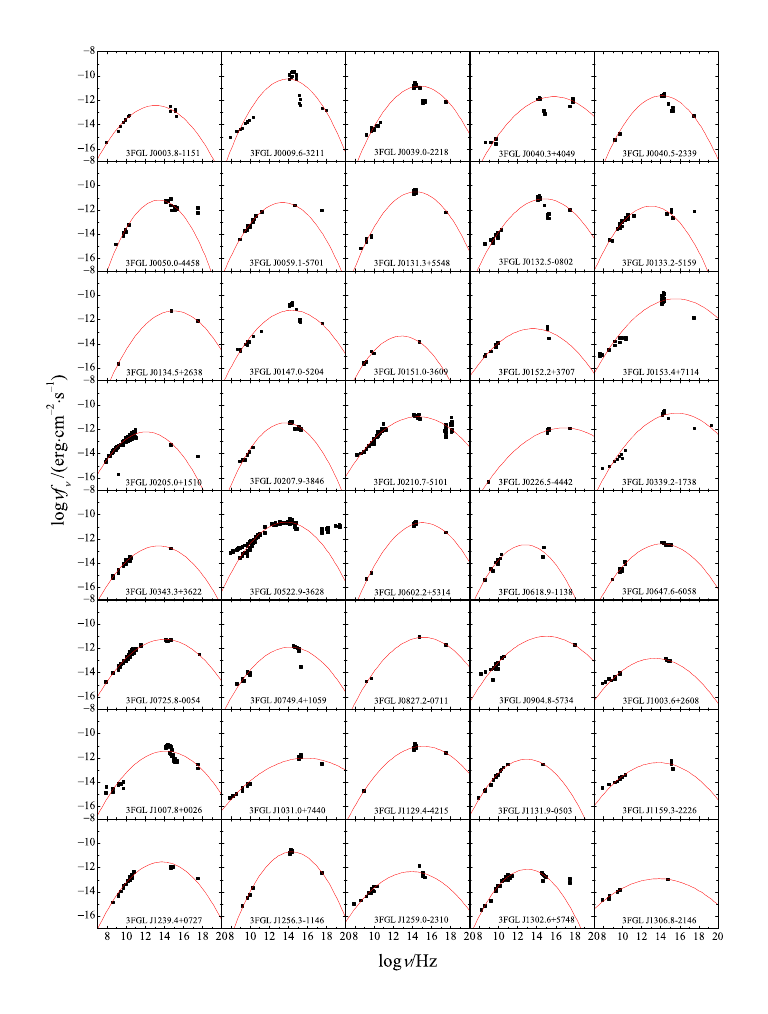}}
    \caption{Appendix: SED figures for UCBs.}
    \label{Fan-SED-Fig-U001-040}
\end{figure}

\begin{figure}
    \centering
    \resizebox{\hsize}{!}{\includegraphics*{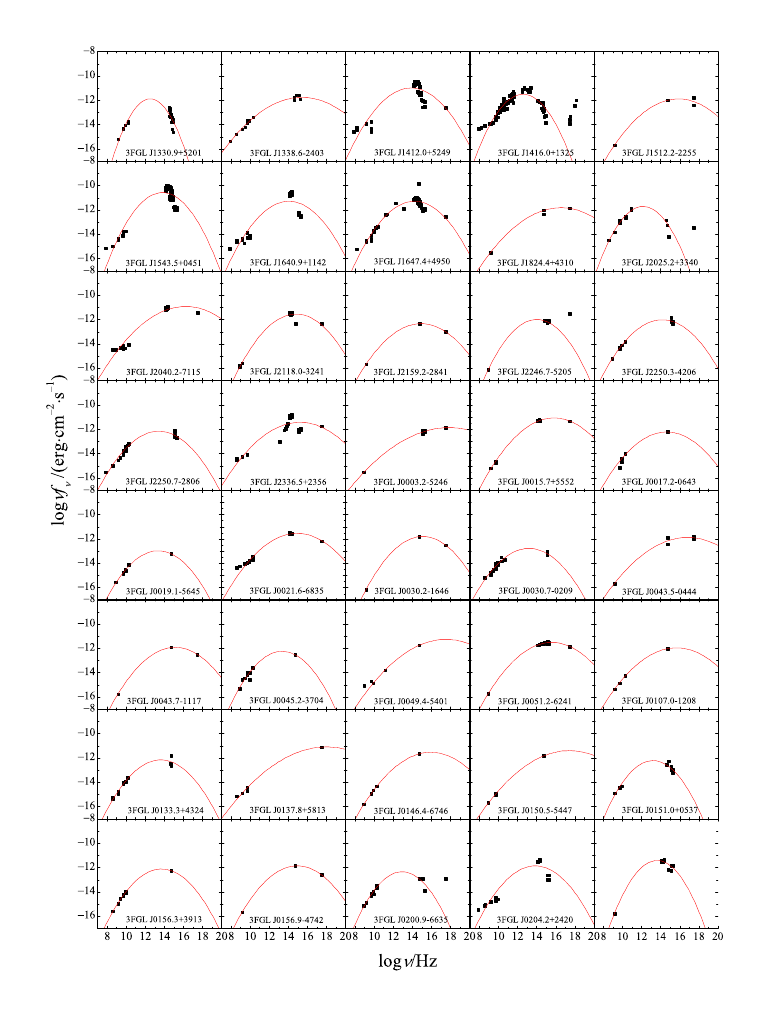}}
    \caption{Appendix: SED figures for UCBs.}
    \label{Fan-SED-Fig-U041-080}
\end{figure}

\clearpage
\begin{figure}
    \centering
    \resizebox{\hsize}{!}{\includegraphics*{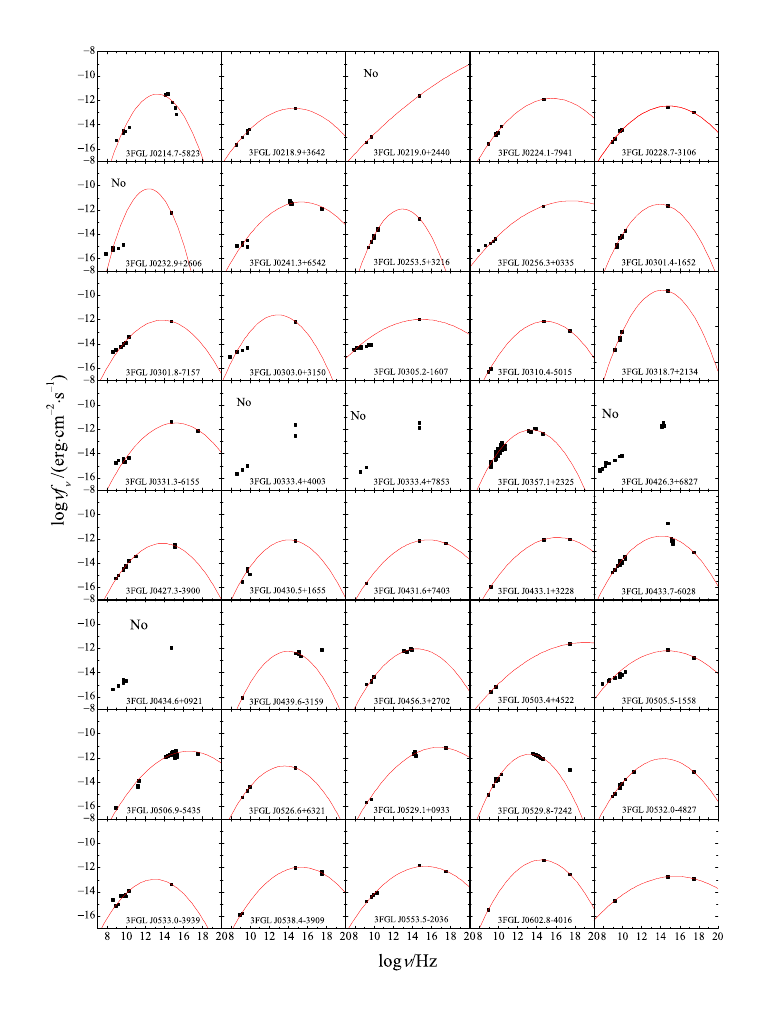}}
    \caption{Appendix: SED figures for UCBs.}
    \label{Fan-SED-Fig-U081-120}
\end{figure}

\begin{figure}
    \centering
    \resizebox{\hsize}{!}{\includegraphics*{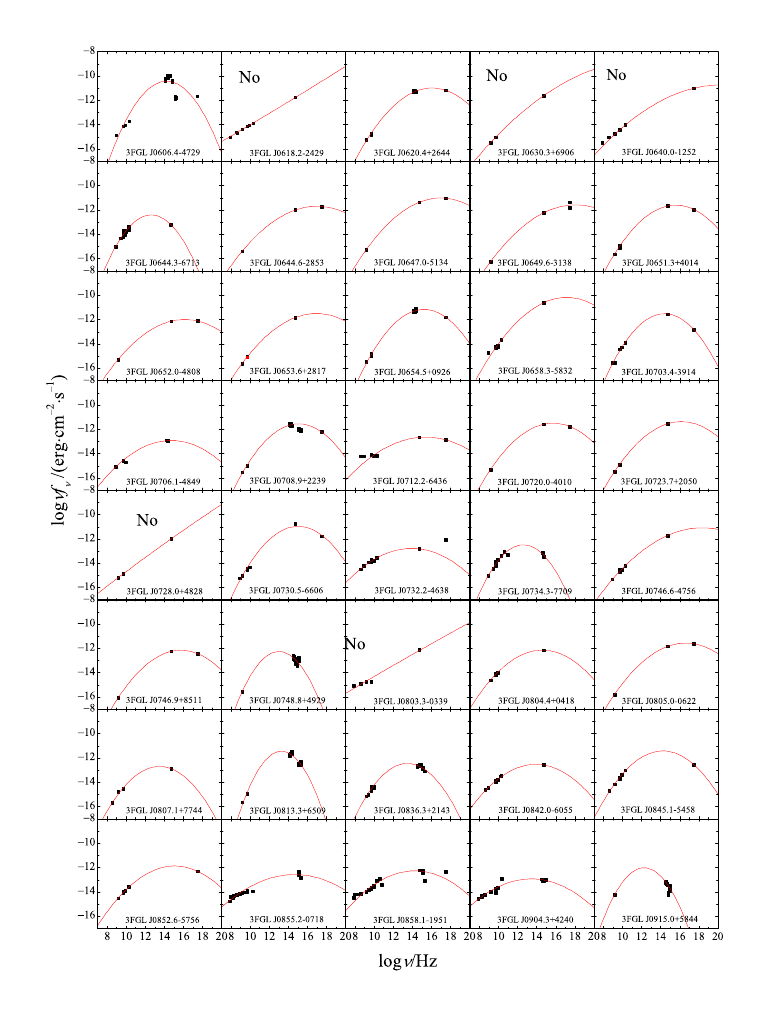}}
    \caption{Appendix: SED figures for UCBs.}
    \label{Fan-SED-Fig-U121-160}
\end{figure}

\begin{figure}
    \centering
    \resizebox{\hsize}{!}{\includegraphics*{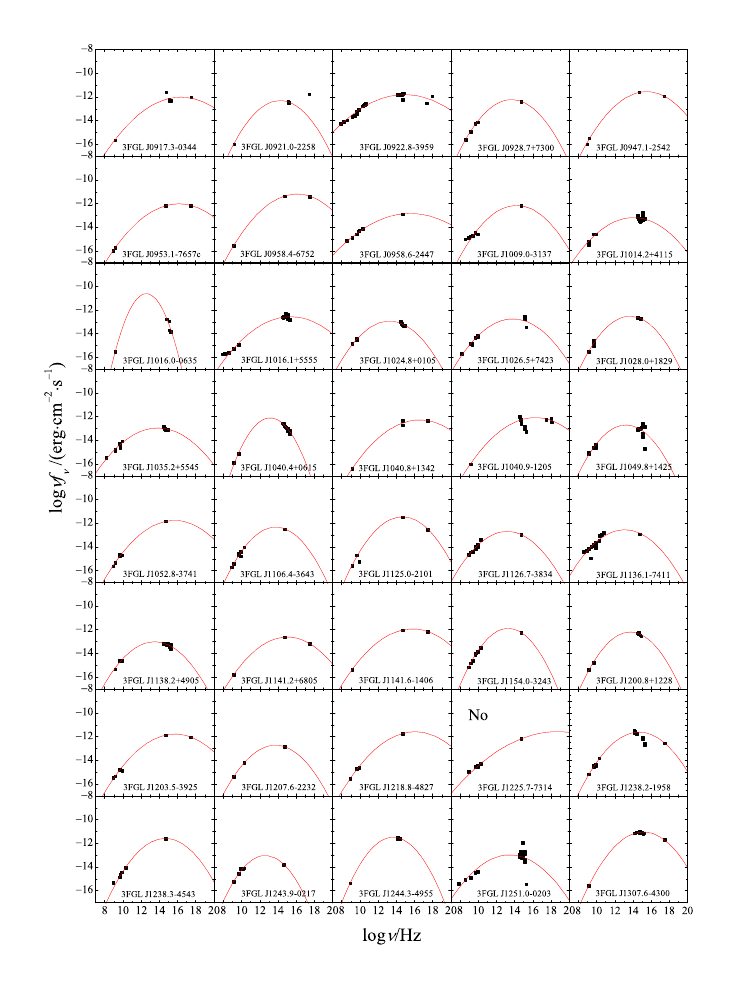}}
    \caption{Appendix: SED figures for UCBs.}
    \label{Fan-SED-Fig-U161-200}
\end{figure}

\begin{figure}
    \centering
    \resizebox{\hsize}{!}{\includegraphics*{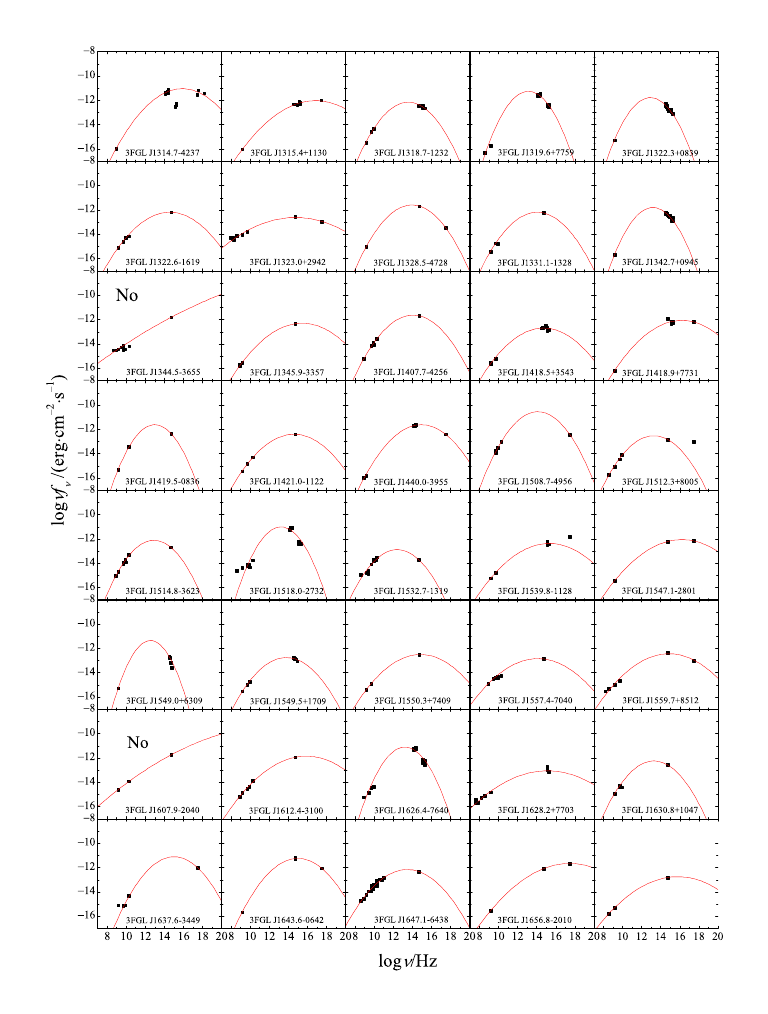}}
     \caption{Appendix: SED figures for UCBs.}
    \label{Fan-SED-Fig-U201-240}
\end{figure}

\begin{figure}
    \centering
    \resizebox{\hsize}{!}{\includegraphics*{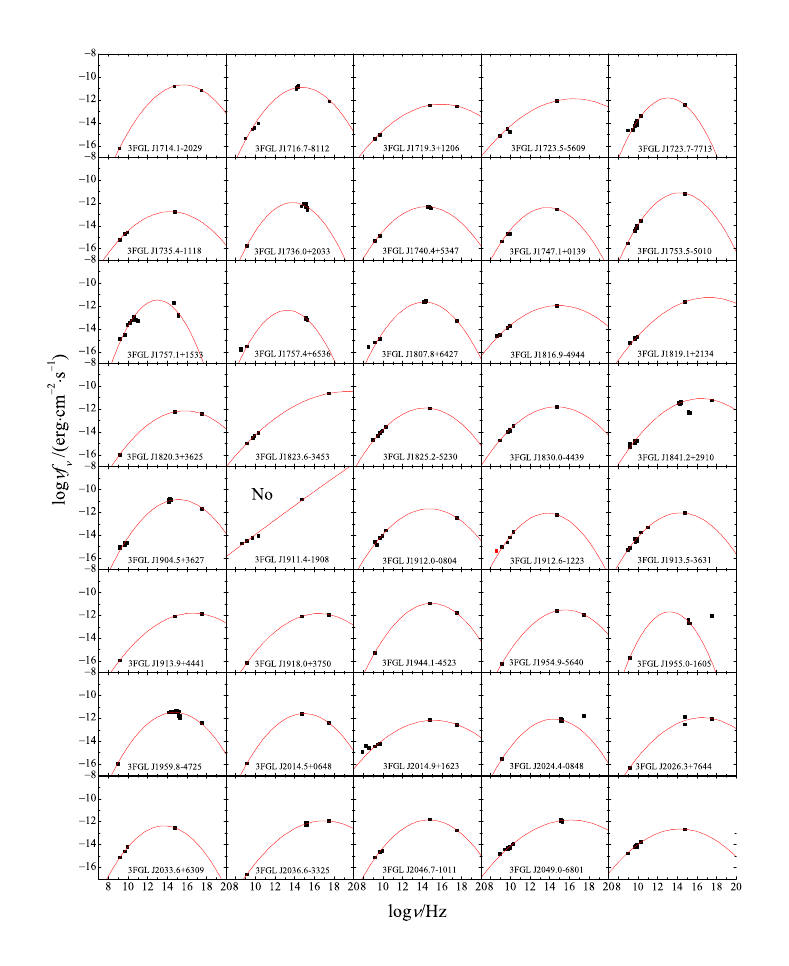}}
        \caption{Appendix: SED figures for UCBs.}
    \label{Fan-SED-Fig-U241-280}
\end{figure}

\begin{figure}
    \centering
    \resizebox{\hsize}{!}{\includegraphics*{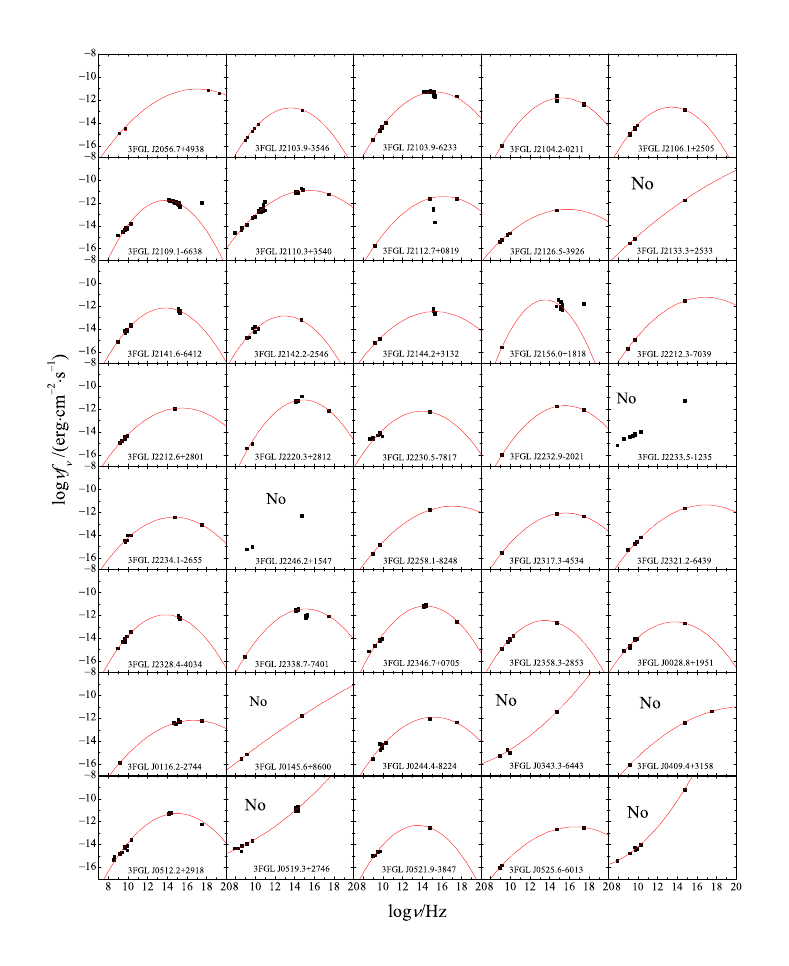}}
    \caption{Appendix: SED figures for UCBs.}
    \label{Fan-SED-Fig-U281-320}
\end{figure}

\begin{figure}
    \centering
    \resizebox{\hsize}{!}{\includegraphics*{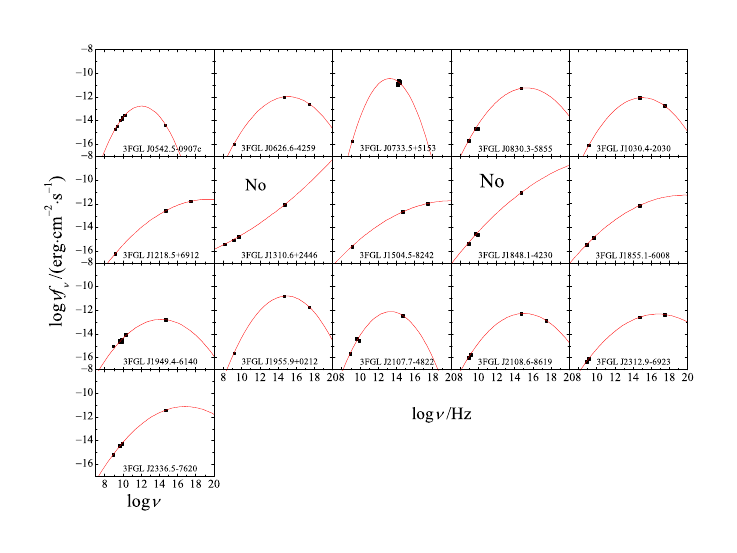}}
    \caption{Appendix: SED figures for UCBs.}
        \label{Fan-SED-Fig-U321-336}
\end{figure}

\end{document}